\documentclass[aps,prd,twocolumn,nofootinbib,floatfix,10pt]{revtex4-2}
\usepackage[T1]{fontenc}
\usepackage{lmodern}
\usepackage{amsmath,amssymb,amsthm,bm,graphicx}
\usepackage{booktabs}
\usepackage{placeins}
\usepackage[normalem]{ulem}
\usepackage[colorlinks=true,linkcolor=blue,citecolor=blue,urlcolor=blue]{hyperref}

\newtheorem{proposition}{Proposition}[section]

\newcommand{\dd}{\mathrm{d}}
\newcommand{\ii}{\mathrm{i}}
\newcommand{\e}{\mathrm{e}}
\newcommand{\Id}{\mathbf{1}}
\newcommand{\Dslash}{\mathcal{D}}
\newcommand{\cM}{c_{\mathrm{mix}}}
\newcommand{\cR}{c_{\rho}}
\newcommand{\cT}{c_{\theta}^{(s)}}
\usepackage{xcolor}

\begin{document}
\title{Geometry, Topology, and Confined Spectra Around Anisotropic Cosmic Strings}

\author{Gabriel G.\ Carvalho}
\email{gabriel.guimaraescarvalho@imec.be}
\affiliation{Interuniversity Microelectronics Centre, Kapeldreef 75, 3001 Leuven, Belgium}
\author{Eduardo Bittencourt}
\email{bittencourt@unifei.edu.br}
\affiliation{Institute of Physics and Chemistry, Federal University of Itajub\'a, Av.\ BPS 1303, Itajub\'a/MG, Brazil and Department of Mathematics and Applied Mathematics, University of Cape Town, Rondebosch 7700, Cape Town, South Africa}

\date{\today}

\begin{abstract}
We study scalar, Dirac, and electromagnetic test fields on a two-parameter family of static spacetimes, labeled by an angular winding number $k$ and an anisotropy parameter $\alpha$ that shortens one direction of the transverse metric relative to the other. Only the members with $k=1$ and $\alpha<1$ are conical, as for an ordinary straight cosmic string. The remaining members are nonconical and, except in the flat cases, curved and nonvacuum. A common spin-weighted connection Laplacian organizes the local equations, while the curvature coupling, global field conditions, and constraints are fixed separately for each physical field. Explicit frame transitions remove spurious winding dependence in the flat limit. For the nonconical members of the family, isothermal coordinates yield an exact factorization of the indicial equation and determine its monodromy: the indicial exponents are independent of the anisotropy parameter, although the finite-domain eigenvalues are not. We derive variational bounds and a weak-anisotropy selection rule for the scalar spectrum. The electromagnetic equations are obtained from the four-potential in Lorenz gauge, with gauge-invariant transverse-electric (TE) and transverse-magnetic (TM) reductions and an annular transverse-electromagnetic (TEM) mode. Finite-element computations on an annulus, validated against benchmarks, are consistent with the bounds, confirm the selection rule, give MIT bag Dirac and conducting-wall Maxwell spectra, and reproduce the invariance of the transverse-electromagnetic capacity under $k\mapsto2-k$, an exact consequence of a radial inversion of the Dirichlet energy.
\end{abstract}

\maketitle

\begin{center}
\textit{In memory of I. D. Soares and F. M. de Aguiar.}
\end{center}

\section{Introduction}
\label{sec:introduction}

The ordinary straight cosmic string has a locally flat exterior with a conical identification. For this geometry, the holonomy around the string is not determined by the curvature of the region where the fields live, so global and local gravitational effects on test fields can be distinguished. More general defect geometries can possess curvature away from the axis. The holonomy of a loop around the axis can then depend on the loop, and two such loops differ by the curvature enclosed between them.

As an effective geometry, this distinction between local curvature and global identification makes cosmic-string spacetimes useful as models of topological defects. The broader relation between cosmic strings and defects in condensed matter is reviewed in Ref.~\cite{HindmarshKibble}. Lensing effects in a nematic liquid crystal were studied in Ref.~\cite{moraes06}, and the parallel transport of light polarization around disclinations in Ref.~\cite{Carvalho2007}. In graphene, pentagonal and heptagonal rings act as conical defects with an angle deficit and an angle surplus, and their effect on the electrons has been modeled by the massless Dirac equation on a cosmic-string metric~\cite{Cortijo2007}. An analogy between hyperbolic metamaterials and the Milne universe was studied in \cite{PhysRevD.96.105012}. Here we consider the asymmetric family introduced by S\'atiro, Carvalho, and Moraes~\cite{Satiro2009}, whose transverse metric first appeared as an effective optical metric in Ref.~\cite{moraes06} and depends on an anisotropy parameter $\alpha$ and an angular winding number $k$. Only the members of this family with $k=1$ and $\alpha<1$ are conical, as for an ordinary straight cosmic string. The member $k=0$ is the Euclidean plane in stretched coordinates. The remaining members with $\alpha<1$ are nonconical, curved, and nonvacuum. Reference~\cite{Satiro2009} assigns a conical singularity at $\rho=0$ to the curved members as well, from the holonomy of coordinate circles. This is consistent for that family of loops, but for $k\notin\{0,1\}$ we show that its strength depends on the loops that shrink to the axis and that for $k\ne1$ there is no conical deficit in the conformal sense. We treat the metric as a prescribed or effective geometry for test fields, rather than assuming an ordinary positive-energy string source.

The question that organizes this paper is how the local geometric structure and a global boundary-value problem each enter the behavior of scalar, Dirac, and electromagnetic test fields on this background. Some properties of the fields are fixed by the local structure of the field equations, independently of any boundary condition. Others, such as spectra, are defined only once a domain and its boundary conditions are specified. We ask which properties belong to which class, how the anisotropy enters the spectra of confined fields, and which of the resulting parameter dependences can be attributed to curvature.

Answering these questions requires the physical field equations themselves. Although the equations of different fields can often be organized in a common geometric language, this organization does not by itself imply that an arbitrary spin-weighted equation is the equation of a physical field of the same spin. The Teukolsky construction~\cite{Teukolsky1972} illustrates the importance of both the field equations and the background assumptions in obtaining a master equation. Related work on the Dirac equation in the vacuum C-metric~\cite{bini2015} also shows how the geometry determines the available separations. For the background considered here, we derive the scalar, Dirac, and Maxwell equations directly, and retain the constraints that distinguish their physical solution spaces.

Two global issues are essential. First, components in a rotating orthonormal frame need not have the same angular periodicity as components in a single-valued reference frame. A spin structure must be specified independently of that choice of frame~\cite{Geroch1968}. Second, an exterior differential equation does not select its own boundary conditions at a singular core. The role of self-adjoint extensions in the ordinary conical Dirac problem is discussed in Ref.~\cite{DAncona2025}. A massive Dirac field obeying the MIT bag condition on a cylindrical shell coaxial with an ordinary cosmic string was studied in Ref.~\cite{BezerradeMello2008}, in the regions inside and outside the shell. We instead formulate finite-annulus problems with explicit boundary conditions, keeping the core-extension problem separate.

On the analytic side, we fix the angular transition conditions of each field component in a single-valued reference frame, including the spin structure of the Dirac field. We then carry out an exact indicial monodromy calculation, which specializes standard arguments to the present geometry. On an annulus, we impose Dirichlet or Neumann conditions on the scalar field, the MIT bag condition on the Dirac field, and perfectly conducting walls on the Maxwell field. For the scalar spectrum we derive eigenvalue bounds and, from a Fourier decomposition into invariant angular blocks, a perturbative selection rule. For Maxwell fields, we derive the master equation from the four-potential and reconstruct the physical fields, including the annular TEM sector, whose capacity obeys an exact inversion identity under $k\mapsto2-k$.

On the numerical side, we solve the physical boundary-value problems on an annulus with finite elements and validate the computations against analytic benchmarks. We compute the scalar and Maxwell cutoffs and the MIT bag Dirac spectrum over the sampled windings and anisotropies, check the implementation against the bounds, test the selection rule, reconstruct the physical fields, and compute the TEM capacity. Together, the analytic and numerical results distinguish the information contained in a local indicial equation from that contained in a global spectrum.

The paper is organized as follows. Section~\ref{sec:geometry} fixes the geometry: the metric and its curvature, the behavior of field components under frame rotations, and the angular transition conditions. Section~\ref{sec:fields} reduces the scalar, Dirac, and Maxwell equations to a common spin-weighted operator and identifies the curvature coupling of each field. Section~\ref{sec:indicial} analyzes the scale-invariant (indicial) problem in isothermal coordinates. Section~\ref{sec:spectra} specifies the boundary-value problems on an annulus and derives analytic properties of their spectra. Section~\ref{sec:numerics} solves these problems numerically, validates the computations, and compares them with the analytic results. Section~\ref{sec:discussion} summarizes what has been established and what remains open.

\section{Geometry, frames, and global conditions}
\label{sec:geometry}

\subsection{Metric and curvature}
We consider a spacetime $(\mathcal{M}, g)$ with signature $(+---)$ and units $c=\hbar=G=1$. In cylindrical coordinates \(x^\mu=(t,\rho,\theta,z)\), the line element of an anisotropic cosmic string is~\cite{Satiro2009}
\begin{equation}
 \dd s^2=\dd t^2-h-\dd z^2, \qquad h=(e^1)^2+(e^2)^2,
 \label{eq:metric}
\end{equation}
with transverse coframe
\begin{align}
 e^1&=\cos\xi\,\dd\rho+\rho\sin\xi\,\dd\theta,\nonumber\\
 e^2&=-\alpha\sin\xi\,\dd\rho +\alpha\rho\cos\xi\,\dd\theta,
 \label{eq:coframe}
\end{align}
where $\xi=(k-1)\theta$, $0<\alpha\leq1$, and $k\in\tfrac12\mathbb Z$. The dual orthonormal frame, defined by $e^a(e_b)=\delta^a{}_b$, is
\begin{align}
 e_1&=\cos\xi\,\partial_\rho+\frac{\sin\xi}{\rho}\partial_\theta, \nonumber\\
 e_2&=-\frac{\sin\xi}{\alpha}\partial_\rho +\frac{\cos\xi}{\alpha\rho}\partial_\theta.
 \label{eq:inverseframe}
\end{align}
The metric is single-valued under $\theta\mapsto\theta+2\pi$. For half-odd-integer $k$, the displayed coframe changes sign on this identification and must be interpreted with a frame transition. Since $e^1$ and $e^2/\alpha$ form an orthonormal coframe of the flat coordinate metric $\dd\rho^2+\rho^2\dd\theta^2$, rotated by the angle $\xi$, every member with $\alpha<1$ shortens lengths in one direction by the factor $\alpha$ relative to the coordinate plane and leaves the orthogonal direction unchanged. We call $\alpha$ the anisotropy parameter in this sense, as in the nematic setting of Ref.~\cite{moraes06}. For $k=1$, the number $\alpha$ controls the conical defect, with deficit angle, also known as the Frank angle \cite{Fumeron2021}, $2\pi(1-\alpha)$. For $k=0$, the metric is the Euclidean plane in stretched coordinates (see below). Only the members with $k\notin\{0,1\}$ and $\alpha<1$ have a curvature that depends on $\theta$, and no conical deficit in the conformal sense defined in Sec.~\ref{sec:isothermal}.

The spacetime is the product of the flat $(t,z)$ plane with the transverse surface of metric $h$. The coordinate fields $\partial_t$ and $\partial_z$ are therefore covariantly constant Killing fields~\cite{Nakahara2003}, and the metric is static. In adapted coordinates, no Christoffel symbol carries a $t$ or $z$ index, and the corresponding spin-connection components vanish in an orthonormal frame containing $\dd t$ and $\dd z$. Derivatives along $t$ and $z$ are thus ordinary partial derivatives. The field equations below, and the walls $\rho=\mathrm{const}$ introduced later, are invariant under translations in $t$ and $z$, so we work with modes proportional to $\e^{-\ii Et+\ii p_z z}$, which obey equations on the transverse surface alone. The dependence on $\rho$ and $\theta$ stays coupled, and the anisotropy, the angular transition conditions, and the boundary conditions all act on this two-dimensional problem. The spin structure, the gauge, and the boundary conditions are not fixed by the symmetries and are specified separately below. For later use, we define
\begin{align}
 B&=\cos^2\xi+\alpha^2\sin^2\xi,
 &&A=\sin^2\xi+\alpha^2\cos^2\xi,\nonumber\\
 C&=(1-\alpha^2)\sin\xi\cos\xi,
 &&AB-C^2=\alpha^2.
 \label{eq:ABC}
\end{align}
With the orientation in which $(\dd\rho\wedge\dd\theta)(\partial_\rho,\partial_\theta)>0$, the transverse metric and its area form are
\begin{equation}
 \begin{gathered}
 h=B\,\dd\rho^2+2\rho C\,\dd\rho\dd\theta        +\rho^2 A\,\dd\theta^2, \\
 \dd A_h=e^1\wedge e^2 =\alpha\rho\,\dd\rho\wedge\dd\theta.
 \end{gathered}
 \label{eq:metriccomponents}
\end{equation}
We also use $\dd A_h$ for the corresponding area measure.

We define the connection one-form $\omega$ (see Ref.\cite{Nakahara2003}, for example) by
\begin{equation}
 \dd e^1=\omega\wedge e^2,\qquad
 \dd e^2=-\omega\wedge e^1.
 \label{eq:connectionconvention}
\end{equation}
Direct substitution gives
\begin{equation}
 \omega=\frac{kC}{\alpha\rho}\,\dd\rho
               +\frac{kA}{\alpha}\,\dd\theta,
 \qquad
 \dd\omega=-K\,\dd A_h,
 \label{eq:connection}
\end{equation}
where the Gaussian curvature is
\begin{equation}
 K=\frac{k(k-1)(1-\alpha^2)}{\alpha^2\rho^2}
                  \cos[2(k-1)\theta].
 \label{eq:curvature}
\end{equation}

Since $K\propto\rho^{-2}$, the curvature decays at large $\rho$, but $\rho^2K$ does not. Except in the flat cases, the exterior never approaches the Euclidean plane or a cone. Our Riemann-tensor convention gives positive Gaussian curvature to a round two-sphere (see Ref.~\cite{Nakahara2003}). Accordingly, the transverse scalar curvature is $R_h=2K$ and the four-dimensional scalar curvature is $R_4=-2K$, in the smooth region $\rho>0$.

\begin{figure*}[tbp]
 \centering
 \includegraphics[width=\linewidth]{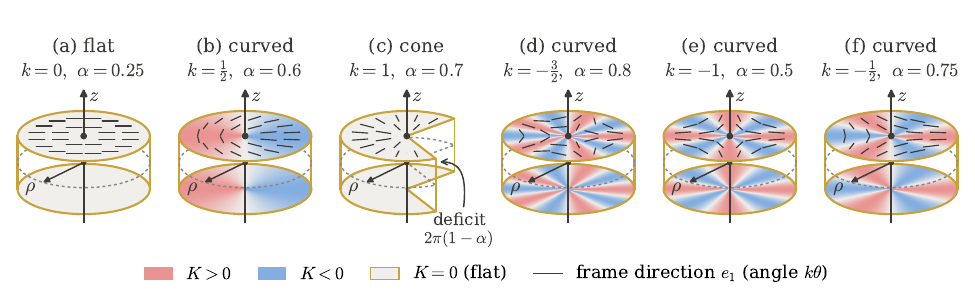}
 \caption{Transverse geometry $h$ of Eq.~\eqref{eq:metric} for six members of
 the family. The disks show the coordinate plane $(\rho\cos\theta,\rho\sin\theta)$,
 and the segments give the frame vector $e_1$ of Eq.~\eqref{eq:inverseframe}.
 Displacements along $e_1$ keep their coordinate length in $h$, while
 perpendicular ones are shortened by $\alpha$. For half-odd-integer $k$ the
 segments carry no orientation, since $e_1$ changes sign around the axis.
 The shading gives the sign of $K$ [Eq.~\eqref{eq:curvature}], which changes
 $4|k-1|$ times around the axis for $k\neq0,1$, with intensity proportional
 to $\rho^2K$ normalized in each panel separately. The members other than
 the cone have no conical deficit in the sense of Sec.~\ref{sec:isothermal}.
 The disks are coordinate pictures, not isometric embeddings.}
 \label{fig:transverse_geometry}
\end{figure*}

The metric is flat in three cases, which serve as references below. For $k=1$, $h=\dd\rho^2+\alpha^2\rho^2\dd\theta^2$ is the ordinary cone. For $\alpha=1$, the metric is Euclidean in polar coordinates for every $k$. For $k=0$, setting $x=\rho\cos\theta$ and $y=\rho\sin\theta$ gives $h=\dd x^2+\alpha^2\dd y^2$. This last case is also flat, but a coordinate annulus becomes an elliptical annulus after the Cartesian stretch $y\mapsto\alpha y$, so flatness alone does not imply a circular-domain spectrum. Figure~\ref{fig:transverse_geometry} shows the transverse geometry for representative values of $k$ and $\alpha$, with the frame direction $e_1$ winding by $2\pi k$ around the axis and the sign of the Gaussian curvature $K$ indicated by shading.

\subsection{Spin weight and the angular transition}
\label{sec:global}

The oriented orthonormal coframe $e=(e^1,e^2)$ is determined by $h$ only up to a local rotation. Before writing field equations in components, we must therefore know how the connection and the components change under such a rotation. 

Consider a local (transverse) rotation $e\mapsto e'=R(\chi)e$, with rotation matrix $R(\chi)$ given by
\begin{equation}
 R(\chi)=\begin{pmatrix}\cos\chi&\sin\chi\\
                    -\sin\chi&\cos\chi\end{pmatrix},
 \qquad \chi=\chi(\rho,\theta).
 \label{eq:rotation}
\end{equation}
Using Eq.~\eqref{eq:connectionconvention}, we obtain
\begin{align}
 \dd e'^1&=\dd(\cos\chi\,e^1+\sin\chi\,e^2)\nonumber\\
 &=-\sin\chi\,\dd\chi\wedge e^1+\cos\chi\,\omega\wedge e^2\nonumber\\
 &\quad+\cos\chi\,\dd\chi\wedge e^2-\sin\chi\,\omega\wedge e^1\nonumber\\
 &=(\omega+\dd\chi)\wedge(-\sin\chi\,e^1+\cos\chi\,e^2)\nonumber\\
 &=(\omega+\dd\chi)\wedge e'^2,
\end{align}
and in the same way $\dd e'^2=-(\omega+\dd\chi)\wedge e'^1$. The
connection of the rotated coframe is therefore
\begin{equation}
 \omega'=\omega+\dd\chi.
 \label{eq:connectiongauge}
\end{equation}
The connection one-form thus acts as a gauge potential for frame
rotations, with $\chi$ as the gauge function.

Frame components of a field also change under this rotation. Consider
a transverse one-form $a=a_1e^1+a_2e^2$. Since $a$ itself does not
depend on the frame, its components in the rotated frame are fixed by
$a=a'_1e'^1+a'_2e'^2$, which gives
\begin{equation}
 \begin{gathered}
 a'_1=\cos\chi\,a_1+\sin\chi\,a_2,\\
 a'_2=-\sin\chi\,a_1+\cos\chi\,a_2.
 \end{gathered}
\end{equation}
The two components mix, so an equation for one of them would not
remain an equation for that component alone in another frame. We
therefore look for combinations $a_1+c\,a_2$ that are only multiplied
by a factor under the rotation. Substituting the transformation, this
requires $\sin\chi\,(1+c^2)=0$ for all $\chi$, so $c=\mp\ii$. Writing
$c=-\ii s$ with $s=\pm1$,
\begin{equation}
 a'_1-\ii s\,a'_2=\e^{\ii s\chi}\,(a_1-\ii s\,a_2).
\end{equation}
The circular components $(a_1-\ii s\,a_2)/\sqrt2$ thus transform
independently, each by a phase. More generally, we say that a
component $\psi_s$ has signed spin weight $s$ if
$\psi_s\mapsto\e^{\ii s\chi}\psi_s$. Scalars have $s=0$, the circular
components of a one-form have $s=\pm1$, and the two transverse spinor
components have $s=\pm\tfrac12$, although these
labels are not, by themselves, helicity eigenvalues.

Equations for such components should keep this transformation
property, but the ordinary derivative does not, because $\chi$ depends
on position. In fact, $\partial_i(\e^{\ii s\chi}\psi_s)
=\e^{\ii s\chi}(\partial_i\psi_s+\ii s\,\partial_i\chi\,\psi_s)$.
The extra term is removed by the shift of the connection in
Eq.~\eqref{eq:connectiongauge}. The covariant derivative
\begin{equation}
 D_i^{(s)}=\partial_i-\ii s\omega_i,
\qquad i\in\{\rho,\theta\},
 \label{eq:covderivative}
\end{equation}
satisfies
\begin{equation}
 \bigl[\partial_i-\ii s(\omega_i+\partial_i\chi)\bigr]
 \bigl(\e^{\ii s\chi}\psi_s\bigr)
 =\e^{\ii s\chi}D_i^{(s)}\psi_s,
\end{equation}
and therefore transforms in the same way as $\psi_s$.

The next step is to determine which angular periodicity the field components must obey. Field components in the coframe~\eqref{eq:coframe} need not be periodic in $\theta$, because the coframe itself rotates as one goes around the axis. When $k$ is a half-odd integer, the coframe is not even single-valued. To determine the correct angular condition, we compare it with a frame that does not wind. Since $\dd\rho=\cos\theta\,\dd x+\sin\theta\,\dd y$ and $\rho\,\dd\theta=-\sin\theta\,\dd x+\cos\theta\,\dd y$, we have
\begin{equation}
 e=\operatorname{diag}(1,\alpha)\,R(k\theta)
   \begin{pmatrix}\dd x\\\dd y\end{pmatrix}.
\end{equation}
The coframe follows the direction at angle $k\theta$, which winds $k$ times around the axis. Thus, a convenient reference coframe removes this winding, that is
\begin{equation}
 \widetilde e=R(-k\theta)\,e =R(-k\theta)\operatorname{diag}(1,\alpha)R(k\theta)
  \begin{pmatrix}\dd x\\\dd y\end{pmatrix}.
 \label{eq:referenceframe}
\end{equation}
Explicitly, with $c_k=\cos k\theta$ and $s_k=\sin k\theta$, it can be written down as
\begin{equation}
 \begin{gathered}
 \widetilde e^1=(c_k^2+\alpha s_k^2)\,\dd x+(1-\alpha)\,c_k\,s_k\,\dd y, \\
 \widetilde e^2=(1-\alpha)\,c_k\,s_k\,\dd x+(s_k^2+\alpha c_k^2)\,\dd y.
 \end{gathered}
\end{equation}
These coefficients depend on $\theta$ only through $2k\theta$. Since $2k$ is an integer, $\widetilde e$ is single-valued for every allowed $k$. For $\alpha=1$ it reduces to the Cartesian coframe $(\dd x,\dd y)$. By Eq.~\eqref{eq:connectiongauge} with $\chi=-k\theta$, its connection is
\begin{equation}
 \widetilde\omega=\omega-k\,\dd\theta,
\end{equation}
and components in the two frames are related by
\begin{equation}
 \psi_s=\e^{\ii sk\theta}\,\widetilde\psi_s.
 \label{eq:framerelation}
\end{equation}

The periodicity of the components can now be fixed in the reference
frame, where it has a direct meaning. Scalar and vector fields have
single-valued components there. For spinors, a rotation by $2\pi$ acts
as $-1$, so the periodicity of the components is an additional choice,
the spin structure. The annulus has two, which we specify by
\begin{equation}
 \widetilde\psi_s(\theta+2\pi)
       =\e^{2\pi\ii\delta}\widetilde\psi_s(\theta),
 \qquad \delta\in\{0,\tfrac12\}.
 \label{eq:spinstructure}
\end{equation}
For $\alpha=1$, the choice $\delta=0$ is the spin structure that
extends over the disk, since the Cartesian frame does. Spin structures
exist for every $\alpha$ and $k$, because the noncompact exterior
$\rho>0$ carries the global orthonormal tetrad
$(\dd t,\widetilde e^{\,1},\widetilde e^{\,2},\dd z)$~\cite{Geroch1968}.
On the annulus they are classified by the cohomology group $H^1(\mathcal A;\mathbb Z_2)\cong\mathbb Z_2$~\cite{LawsonMichelsohn},
so these are the only two. The other choice, $\delta=\tfrac12$, does
not extend over the axis in this sense.
On the annulus it is equivalent to $\delta=0$ with a half quantum of
Aharonov--Bohm flux~\cite{AharonovBohm} through the hole. Indeed,
writing $\widetilde\psi_s=\e^{\ii\delta\theta}\psi'_s$ in
Eq.~\eqref{eq:spinstructure} makes $\psi'_s$ periodic and replaces
$\partial_\theta$ by $\partial_\theta+\ii\delta$. This is minimal
coupling, $\partial_\theta-\ii qA_\theta$, to the flat potential
$qA=-\delta\,\dd\theta$, whose flux $\oint qA=-2\pi\delta$ is half a
flux quantum, $2\pi$ in these units, for $\delta=\tfrac12$. We keep it to test how the confined Dirac spectrum depends on
this global choice (Sec.~\ref{sec:numerics}). For scalar and
vector fields we set $\delta=0$. Combining Eqs.~\eqref{eq:framerelation}
and \eqref{eq:spinstructure}, we get
\begin{equation}
 \begin{split}
 \psi_s(\theta+2\pi)
 &=\e^{\ii sk(\theta+2\pi)}\,\e^{2\pi\ii\delta}\,\widetilde\psi_s(\theta)\\
 &=\e^{2\pi\ii(sk+\delta)}\,\psi_s(\theta).
 \end{split}
\end{equation}
Components in the coframe~\eqref{eq:coframe} therefore satisfy
\begin{equation}
 \begin{gathered}
 \psi_s(\theta+2\pi)=\e^{2\pi\ii\chi_s}\psi_s(\theta),
 \\ \chi_s=sk+\delta\pmod1,
 \end{gathered}
 \label{eq:twist}
\end{equation}
and a Fourier mode $\e^{\ii m\theta}$ of $\psi_s$ has
$m\in\mathbb Z+\chi_s$. The field and its angular derivatives obey the
same transition. 

\begin{figure*}[tbp]
 \centering
 \includegraphics[width=\linewidth]{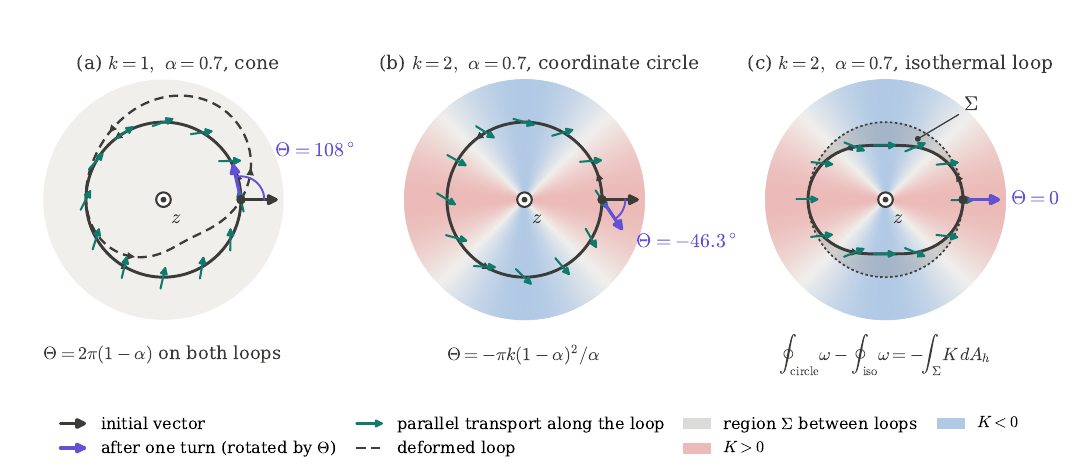}
 \caption{Holonomy of loops around the axis, in the coordinate plane
 $(\rho\cos\theta,\rho\sin\theta)$. A tangent vector is parallel transported
 counterclockwise with the Levi-Civita connection of $h$ from the base point
 $\theta=0$, and $\Theta$ is the angle, measured with $h$, by which it returns
 rotated. Since a parallel vector obeys $\dd(V^1+\ii V^2)=-\ii\omega(V^1+\ii V^2)$,
 $\Theta=-(\oint\omega-2\pi k)$ and
 $\operatorname{Hol}_s=\e^{-2\pi\ii\delta}\e^{-\ii s\Theta}$ [Eq.~\eqref{eq:holonomy}].
 (a) On the cone $\Theta$ is the same for every loop, since $K=0$ away from the
 axis. (b) Coordinate circle, Eq.~\eqref{eq:loopintegral}. (c) Isothermal loop
 through the same base point, with $\Theta=0$. The difference between (b) and (c)
 is the curvature in the signed region $\Sigma$ between the loops, which touch
 at $\theta=0,\pi$, so the coordinate-circle value characterizes the loop, not
 the defect. Numerical parallel transport reproduces all values of $\Theta$ to
 better than $10^{-6}$. Arrows have fixed coordinate length, so drawn angles
 differ from $\Theta$.}
 \label{fig:holonomy}
\end{figure*}

The same transition enters the holonomy \cite{Nakahara2003} of a field transported
around a loop. We compute it for the coordinate circles
$\rho=\mathrm{const}$. Parallel transport around such a circle must
include this transition. A parallel
component obeys $D^{(s)}_\theta\psi_s=0$, that is
$\partial_\theta\psi_s=\ii s\,\omega_\theta\psi_s$. After one turn,
$\psi_s(2\pi)=\e^{\ii s\oint\omega}\psi_s(0)$, and by
Eq.~\eqref{eq:framerelation}
$\widetilde\psi_s(2\pi)=\e^{\ii s(\oint\omega-2\pi k)}
\widetilde\psi_s(0)$. Comparing with the identification
\eqref{eq:spinstructure}, the holonomy in the reference trivialization
is
\begin{equation}
 \operatorname{Hol}_s
   =\e^{-2\pi\ii\delta}
     \exp\left\{\ii s\left(\oint\omega-2\pi k\right)\right\}.
 \label{eq:holonomy}
\end{equation}
The loop integral follows from Eq.~\eqref{eq:connection}. Since
$\omega_\theta=kA/\alpha$ and
$A=\tfrac12(1+\alpha^2)-\tfrac12(1-\alpha^2)\cos2\xi$, the cosine
integrates to zero over a period when $k\ne1$, whereas $A=\alpha^2$
when $k=1$. Hence
\begin{equation}
 \begin{gathered}
 \oint\omega=
 \begin{cases}
  2\pi\alpha,&k=1,\\
  \pi k(\alpha+\alpha^{-1}),&k\ne1,
 \end{cases}
 \\
 \oint\omega-2\pi k=
 \begin{cases}
  -2\pi(1-\alpha),&k=1,\\
  \pi k(1-\alpha)^2/\alpha,&k\ne1.
 \end{cases}
 \end{gathered}
 \label{eq:loopintegral}
\end{equation}
Both results are independent of $\rho$, as a consequence of the scale invariance of the metric, under which loops of the same shape give the same value. For $k\notin\{0,1\}$ and $\alpha\ne1$, however, loops of different shape can give different values. By Stokes' theorem \cite{Nakahara2003} and Eq.~\eqref{eq:connection}, two loops $\gamma_1,\gamma_2$ that each encircle the axis once counterclockwise, with $\gamma_2$ enclosing $\gamma_1$ and $\Sigma$ the region between them, satisfy $\oint_{\gamma_2}\omega-\oint_{\gamma_1}\omega=-\int_\Sigma K\,\dd A_h$. For $k\ne1$ the values in Eq.~\eqref{eq:loopintegral} therefore describe the coordinate circles rather than the defect. Carvalho et al.~\cite{Carvalho2007} computed the same coordinate-circle integral and, using the Gauss--Bonnet theorem, attributed the mismatch to a curvature singularity at $\rho=0$. S\'atiro et al.~\cite{Satiro2009} obtained an equivalent result, described the singularity as conical, and assigned it a linear mass density. This reading is consistent as long as only coordinate circles are used, because the positive and negative curvature sectors cancel on every coordinate annulus, so all such circles see the same point-like contribution at the axis. Near the axis, however, each sector carries infinite curvature, since for $k\notin\{0,1\}$ and $\alpha\ne1$ the quantity $|K|\,\dd A_h$ decreases only like $\dd\rho\,\dd\theta/\rho$. The cancellation then depends on the shape of the loops, and the isothermal loops, which also shrink to the axis, assign no contribution to it. A point-like curvature at $\rho=0$ is therefore not fixed by the exterior geometry. Figure~\ref{fig:holonomy} shows this loop dependence. For $k\ne1$ an isothermal loop through the same point has holonomy $\e^{-2\pi\ii\delta}$, with no geometric contribution, the difference being the curvature enclosed between the two loops.

The comparison of two loops provides a direct diagnostic of distributed curvature. For loops $\gamma_1$ and $\gamma_2$ with the same base point and winding, Eq.~\eqref{eq:holonomy} gives
\begin{equation*}
 \frac{\operatorname{Hol}_s(\gamma_2)}{\operatorname{Hol}_s(\gamma_1)}=\exp\left(-\ii s\int_\Sigma K\,\dd A_h\right),
\end{equation*}
where $\Sigma$ is oriented so that $\partial\Sigma=\gamma_2-\gamma_1$. The common spin-structure sign cancels. Thus the relative geometric phase, or the relative rotation of transported vectors, probes the curvature between the paths. For the ordinary cone this ratio is unity whenever $\Sigma$ lies entirely outside the axis whereas for the curved members it can be nontrivial.

Finally, the reference frame shows that the winding number $k$ has no physical effect in the flat case $\alpha=1$. The holonomy then reduces to the spin-structure sign $\e^{-2\pi\ii\delta}$. Moreover, in the reference frame the covariant derivative is $\partial_i$ and the transition is $\e^{2\pi\ii\delta}$, for every $k$. In the flat case, the operators and transition conditions written in the reference frame are therefore independent of $k$, and so are flat-space spectra with the same physical boundary conditions and spin structure.

\section{Physical field equations and a common local operator}
\label{sec:fields}

This section first defines a single spin-weighted operator with a curvature coupling $c_s$, and then shows how the scalar, Dirac, and Maxwell equations reduce to it. The factor $c_s$ is
not a free parameter and is derived separately for each field
and retains its own global conditions and constraints

\subsection{The spin-weighted operator}

Define the spatial (transverse) connection Laplacian and its curvature extension by
\begin{equation}
 \begin{gathered}
 \Delta_s\psi_s=
 \frac{1}{\alpha\rho}D_i^{(s)}
       \left(\alpha\rho h^{ij}D_j^{(s)}\psi_s\right),
 \\
 L_{s,c_s}=\Delta_s+c_sK,
 \qquad H_{s,c_s}=-L_{s,c_s}.
 \end{gathered}
 \label{eq:masterdef}
\end{equation}
Here $i,j\in\{\rho,\theta\}$ and $c_s$ is real. The common local
master equation is
\begin{equation}
 (L_{s,c_s}+\lambda^2)\psi_s=0.
 \label{eq:master}
\end{equation}
The physical values of $c_s$ will be derived below. The geometric
family at arbitrary $s,c_s$ is useful mathematically, but is not an
arbitrary-spin field theory.

For convenience, we introduce the auxiliary parameters
\begin{align}
 &\beta_1=A/\alpha^2,\qquad \beta_2=B/\alpha^2,\nonumber\\
 &\cM=-2C\alpha^2, \qquad\cR=k\beta_1-(k-1)\beta_2,\nonumber\\ &\cT=(k-1)\cM-\frac{2\ii sk}{\alpha}.
 \label{eq:coefficients}
\end{align}
Then the operator needed for a direct implementation is
\begin{align}
 L_{s,c_s}={}&\beta_1\partial_\rho^2
  +\frac{\beta_2}{\rho^2}\partial_\theta^2
  +\frac{\cM}{\rho}\partial_\rho\partial_\theta
  +\frac{\cR}{\rho}\partial_\rho
  +\frac{\cT}{\rho^2}\partial_\theta\nonumber\\
 &-\frac{s^2k^2\beta_1}{\rho^2}+c_sK.
 \label{eq:masterexpanded}
\end{align}
Useful checks on the expansion are
\begin{equation}
 \begin{gathered}
 h^{\rho\theta}=\frac{\cM}{2\rho},\qquad
 \omega^\rho=0,\\ \omega^\theta=\frac{k}{\alpha\rho^2},\qquad
 \rho^2K=k(k-1)(\beta_2-\beta_1).
 \end{gathered}
 \label{eq:operatorchecks}
\end{equation}
In this gauge the connection is divergence-free. An imaginary angular coefficient in Eq.~\eqref{eq:masterexpanded} is compatible with self-adjointness and does not by itself imply that the physical problem is non-Hermitian.

\subsection{Minimally coupled scalar field}

For a scalar field of mass $M_0$, the equation
$(\Box_g+M_0^2)\Phi=0$ and the mode
\begin{equation}
 \Phi=\e^{-\ii Et+\ii p_z z}\phi(\rho,\theta)
\end{equation}
give
\begin{equation}
 (\Delta_0+\lambda^2)\phi=0,
 \qquad \lambda^2=E^2-p_z^2-M_0^2.
 \label{eq:scalar}
\end{equation}
Thus $s=0$ and $c_0=0$ for the minimally coupled field assumed here. A nonminimal coupling $\zeta R_4\Phi$ would give $c_0=2\zeta$ in our sign conventions and could change the indicial result below.

\subsection{Dirac field and reconstruction}
\label{sec:dirac}

For a four-component Dirac spinor of mass $M$ we have
\begin{equation}
 (\ii\gamma^\mu\nabla_\mu-M)\Psi=0,
 \qquad \{\gamma^\mu,\gamma^\nu\}=2g^{\mu\nu}\Id.
 \label{eq:diracfirst}
\end{equation}
In the transverse spinor representation the connection is
$\partial_i-\ii\omega_i\sigma^3/2$. Define
\begin{align}
 \Dslash_h
  &=-\ii\sigma^a e_a{}^i
       \left(\partial_i-\frac{\ii}{2}\omega_i\sigma^3\right)
       \nonumber\\
  &=-\ii\left[
   \sigma^1\left(e_1+\frac{k\cos\xi}{2\rho}\right)+\sigma^2\left(e_2-\frac{k\sin\xi}{2\alpha\rho}\right)
       \right].
 \label{eq:dirac2d}
\end{align}
Here $a=1,2$, the $\sigma^a$ are Pauli matrices, and repeated transverse frame indices are summed regardless of their height. With the usual
Dirac representation, a stationary mode satisfies
\begin{equation}
 \begin{gathered}
 H_D\begin{pmatrix}\psi_{\mathrm u}\\\psi_{\mathrm d}\end{pmatrix}
 =E\begin{pmatrix}\psi_{\mathrm u}\\\psi_{\mathrm d}\end{pmatrix},
 \\
 H_D=\begin{pmatrix}M&T\\T&-M\end{pmatrix},
 \qquad T=\Dslash_h+p_z\sigma^3.
 \end{gathered}
 \label{eq:dirachamiltonian}
\end{equation}
Since $\Dslash_h$ contains only $\sigma^1$ and $\sigma^2$, both of which
anticommute with $\sigma^3$, one has $\{\Dslash_h,\sigma^3\}=0$. As a consequence, $T^2=\Dslash_h^2+p_z^2$ as a
local differential identity. The Lichnerowicz identity
\cite{LawsonMichelsohn} gives, on a component with $\sigma^3$ eigenvalue
$\sigma=\pm1$,
\begin{equation}
 \Dslash_h^2=-\Delta_{\sigma/2}+\frac{K}{2}.
 \label{eq:lichnerowicz}
\end{equation}
Consequently,
\begin{equation}
 \begin{gathered}
 \left(\Delta_{\sigma/2}-\frac{K}{2}+\lambda^2\right)\psi_\sigma=0,
 \\ \lambda^2=E^2-p_z^2-M^2.
 \end{gathered}
 \label{eq:diracsquared}
\end{equation}
Both signed components have $c_{+1/2}=c_{-1/2}=-1/2$.
For comparison with a direct expansion, the dimensionless potential is
\begin{equation}
 -\frac{k^2}{4}\beta_1-\frac{\rho^2K}{2}
 =\frac{k(k-2)}{4}\beta_1-\frac{k(k-1)}{2}\beta_2.
 \label{eq:diracpotential}
\end{equation}

The squared component equations are necessary but do not replace the
first-order field equation and its boundary domain. If $E+M\ne0$, the
lower two-spinor is reconstructed by
\begin{equation}
 \psi_{\mathrm d}=\frac{T\psi_{\mathrm u}}{E+M}.
 \label{eq:diracreconstruction}
\end{equation}
The exceptional value $E=-M$ must be treated with
Eq.~\eqref{eq:dirachamiltonian}. Independent component solutions of
Eq.~\eqref{eq:diracsquared} are not automatically independent physical
polarizations.

\subsection{Maxwell field in the four-potential formulation}
\label{sec:maxwell}

The electromagnetic field is described by a four-potential $A^\mu$, with source-free equation of motion given by
\begin{equation}
 \begin{gathered}
 \nabla_\nu F^{\mu\nu}=0,\quad\mbox{with}\quad F_{\mu\nu}=2\nabla_{[\mu}A_{\nu]}.
 \end{gathered}
 \label{eq:maxwellaction}
\end{equation}
In Lorenz gauge, we have
\begin{equation}
 \begin{gathered}
 \nabla_\mu A^\mu=0,\qquad
 \Box_g A^\mu-R^\mu{}_\nu A^\nu=0,
 \\ \Box_g=\nabla_\alpha\nabla^\alpha.
 \end{gathered}
 \label{eq:potentialwave}
\end{equation}
The Ricci term follows from commuting covariant derivatives and
must be retained on the present nonvacuum background
\cite{PoissonPoundVega2011}.

For the product metric~\eqref{eq:metric},
$R^t{}_\nu=R^z{}_\nu=0$ and $R^i{}_j=-K\delta^i{}_j$.
We decompose a mode through its covector:
\begin{align}
 A_\mu\dd x^\mu
 &=\e^{-\ii Et+\ii p_z z}
       \bigl(\varphi\,\dd t-a-a_z\,\dd z\bigr),\nonumber\\
 a&=h_{ij}a^j\dd x^i=a_1e^1+a_2e^2,\nonumber\\
 \lambda^2&=E^2-p_z^2.
 \label{eq:potentialansatz}
\end{align}
Thus $\varphi$, $a^i$, and $a_z$ are the amplitudes of $A^t$,
$A^i$, and $A^z$, respectively. The minus signs in the covector
reflect the spacetime signature.

The reduced equations are
\begin{equation}
 \begin{gathered}
 (\Delta_0+\lambda^2)\varphi=0,\qquad
 (\Delta_0+\lambda^2)a_z=0,\\
 (\Delta^{\rm H}_1-\lambda^2)a=0,
 \end{gathered}
 \label{eq:potentialreduced}
\end{equation}
where the positive transverse Hodge Laplacian \cite{Nakahara2003}, written $\Delta^{\rm H}_1$
to distinguish it from $\Delta_s$ at $s=1$, which has the opposite sign,
satisfies
\begin{equation}
 \Delta^{\rm H}_1=\dd_h\delta_h+\delta_h\dd_h
          =-\nabla^i\nabla_i+K.
 \label{eq:weitzenbock}
\end{equation}
Here $\dd_h$ is the transverse exterior derivative, $\delta_h$
is its codifferential, and the transverse indices are raised
with $h$.

Define the circular components of the transverse potential by
\begin{equation}
 A_s=\frac{a_1-\ii s a_2}{\sqrt2},\qquad s=\pm1.
 \label{eq:potentialcircular}
\end{equation}
Their connection Laplacian is precisely $\Delta_s$ in
Eq.~\eqref{eq:masterdef}, so that
\begin{equation}
 (\Delta_s-K+\lambda^2)A_s=0,\qquad s=\pm1.
 \label{eq:maxwellvector}
\end{equation}
Both signs therefore have $c_{+1}=c_{-1}=-1$.
The amplitudes $\varphi$ and $a_z$ are periodic in $\theta$,
whereas $A_s$ obeys Eq.~\eqref{eq:twist} with
$\chi_s=sk\pmod1$. The spin-zero amplitudes $\varphi$ and $a_z$
belong to the same electromagnetic potential and do not
represent additional scalar particles.

The component equations are supplemented by the Lorenz constraint
\begin{equation}
 \begin{gathered}
 -\ii E\varphi+\operatorname{div}_h a^\sharp
                     +\ii p_z a_z=0,\\
 \operatorname{div}_h a^\sharp
 =\frac{1}{\alpha\rho}\partial_i(\alpha\rho\,a^i),
 \end{gathered}
 \label{eq:potentialconstraint}
\end{equation}
where $a^\sharp$ is obtained by raising the index with $h$. There remains the gauge freedom
\begin{equation}
 A_\mu\longmapsto A_\mu+\nabla_\mu\Lambda,\qquad \Box_g\Lambda=0.
 \label{eq:gaugefreedom}
\end{equation}
For $\Lambda=\e^{-\ii Et+\ii p_z z}\ell(\rho,\theta)$, the residual gauge function obeys $(\Delta_0+\lambda^2)\ell=0$ and is single-valued. Potentials related by admissible residual gauge transformations represent the same electromagnetic field. The signed components $A_s$ are consequently not independent gauge-invariant photon polarizations.

The physical fields are recovered from $F_{\mu\nu}$. In particular, the gauge-invariant longitudinal amplitudes are
\begin{equation}
 u=E_z=\ii E a_z-\ii p_z\varphi,\qquad
 v=B_z=*_h\dd_h a,
 \label{eq:maxwelllongitudinal}
\end{equation}
with $*_h e^1=e^2$ and $*_h e^2=-e^1$. The TE sector has $u=0$, and the TM sector has $v=0$. Their scalar equations and the reconstruction of the physical fields are given in Appendix~\ref{app:maxwell}. For $\lambda^2>0$, these variables provide an equivalent description of the propagating modes after the gauge constraint and the physical boundary conditions have been imposed. The TEM sector, $\lambda=0$, is treated separately in Sec.~\ref{sec:tem}.

\begin{table}[htbp]
\centering
\caption{Physical uses of the common local operator. The signed spin
weight specifies a frame component, not an independent physical
degree of freedom. The scalar is minimally coupled.}
\label{tab:fields}
\small
\setlength{\tabcolsep}{4pt}
\begin{tabular}{@{}lccl@{}}
\toprule
Variable & $s$ & $c_s$ & Conditions\\
\midrule
Scalar & $0$ & $0$ & Scalar boundary domain\\
Dirac component & $\pm\tfrac12$ & $-\tfrac12$ & First-order reconstruction\\
Transverse $A_s$ & $\pm1$ & $-1$ & Gauge constraint, quotient\\
$\varphi$, $a_z$ & $0$ & $0$ & Same gauge constraint\\
$E_z$, $B_z$ & $0$ & $0$ & TE/TM reconstruction\\
\bottomrule
\end{tabular}
\end{table}

Within the minimally coupled metric description used here, the physical field equations share the characteristic null cone of $g$. For the scalar and Maxwell equations, this follows from their common principal part. For the Dirac equation, it follows from the Clifford relation in Eq.~\eqref{eq:diracfirst}. The spin connection and curvature couplings affect finite-wavelength solutions without introducing different characteristic cones for the field components.

\subsection{The spin-two obstruction}
\label{sec:spintwo}

For the prescribed product metric, the mixed Einstein tensor in the region $\rho>0$ is \cite{Satiro2009}
\begin{equation}
 \begin{gathered}
 G^\mu{}_{\nu}=\operatorname{diag}(K,0,0,K),\\
 \text{in the order }(t,\rho,\theta,z).
 \end{gathered}
 \label{eq:einsteintensor}
\end{equation}
Except in the flat cases, the background is not a vacuum solution. If interpreted as an Einstein solution with vanishing cosmological constant, it has an effective energy density proportional to \(K\). For \(\alpha\ne1\) and \(k\notin\{0,1\}\), this quantity changes sign with the angle, and a transverse null vector yields a negative null-energy contraction in the negative-curvature sectors. We therefore treat the metric as a prescribed or effective geometry rather than assuming an ordinary positive-energy string source. Effective geometries of this kind, with negative mass density as gravitational sources, also arise for heptagonal defects in graphene~\cite{Cortijo2007}.

A gravitational perturbation must be derived together with the perturbation
of whatever matter supports it. Setting the matter perturbation to
zero without a consistent matter theory is not, in general, a
gauge-invariant prescription. The formal choice $s=\pm2$ in
Eq.~\eqref{eq:masterdef} defines an associated-bundle operator, but
does not establish an equation for physical gravitons. We restrict the
physical comparisons to the fields in Table~\ref{tab:fields}.

\section{Exact indicial structure}
\label{sec:indicial}

This section asks how solutions of the master
equation~\eqref{eq:master} scale with $\rho$ when the $\lambda^2$ term
is dropped. Solutions of the form $\rho^p$ times an angular function
are characterized by indicial exponents $p$. The annulus problems of
Secs.~\ref{sec:spectra} and~\ref{sec:numerics} have an inner wall at
$a>0$ and do not use these exponents. The analysis is included because
it describes the local structure of the equations independently of any
boundary condition, and because its isothermal coordinates show that
for $k\ne1$ there is no conical deficit in the conformal sense and
supply the loops used in Sec.~\ref{sec:global}. We first construct
isothermal coordinates, in which the metric is conformally flat
(Sec.~\ref{sec:isothermal}). In these coordinates the indicial
operator factorizes for the physical curvature couplings, and its
monodromy can be computed exactly (Sec.~\ref{sec:factorization}).

The monodromy records how a solution changes after one turn around
the axis. A field component must return multiplied by the phase fixed
by its transition rule~\eqref{eq:twist}, so only exponents whose
monodromy reproduces that phase are admissible. For $k\ne1$ the
admissible exponents are $p\in\pm(\mathbb Z+\delta)$
(Proposition~\ref{prop:indicial}). They depend on the spin structure
but not on $\alpha$ or $k$, in contrast with the $\alpha$-dependent
Bessel orders of the cone [Eq.~\eqref{eq:conebessel}]. The
finite-annulus eigenvalues do depend on $\alpha$
(Secs.~\ref{sec:spectra} and~\ref{sec:numerics}), so the local
exponents do not determine the spectrum. They are the input for the
classification of core domains in the region $\rho<a$, which we leave
open (Sec.~\ref{sec:discussion}).

\subsection{Isothermal coordinates for the nonconical cases}
\label{sec:isothermal}

Coordinates $(u,v)$ on a surface are called isothermal if the metric
takes the conformally flat form
\begin{equation}
 h=\Omega^2\left(\dd u^2+\dd v^2\right),\qquad \Omega>0.
 \label{eq:isothermaldef}
\end{equation}
Such coordinates exist locally on any surface \cite{docarmo2016differential}. For the present metric
they can be found explicitly. Completing the square in
Eq.~\eqref{eq:metriccomponents} and using $AB-C^2=\alpha^2$ gives
\begin{equation}
 h=\rho^2B\left[\left(\frac{\dd\rho}{\rho}+\frac{C}{B}\,\dd\theta\right)^2
 +\left(\frac{\alpha}{B}\,\dd\theta\right)^2\right].
 \label{eq:completesquare}
\end{equation}
Henceforth, $\log$ denotes the natural logarithm. For $k\ne1$, the
identity $\dd B/\dd\theta=-2(k-1)C$ gives
$(C/B)\,\dd\theta=-\dd(\log B)/[2(k-1)]$, so the first one-form in the
brackets is exact on the annulus. The second, $(\alpha/B)\,\dd\theta$, is
closed but not exact, since its integral around the axis is $2\pi$
[Eq.~\eqref{eq:isothermalperiod}]. With a reference length $\rho_0>0$,
the two one-forms are $\dd u$ and $\dd v$, where
\begin{equation}
 u=\log\frac{\rho}{\rho_0}-\frac{\log B(\theta)}{2(k-1)},\qquad
 v=\int_0^\theta\frac{\alpha}{B(\vartheta)}\,\dd\vartheta.
 \label{eq:isothermalcoords}
\end{equation}
Equation~\eqref{eq:completesquare} then takes the form
\eqref{eq:isothermaldef} with $\Omega=\rho\sqrt B$.

Since $\dd v/\dd\theta=\alpha/B>0$, the map $\theta\mapsto v$ is
monotone. Because $\int_0^{2\pi}\dd\theta/B=2\pi/\alpha$ and $u$
differs from $\log(\rho/\rho_0)$ by a periodic function of $\theta$,
\begin{equation}
 \begin{gathered}
 v(\theta+2\pi)=v(\theta)+2\pi,\qquad
 \Omega=\rho_0\,\e^{u+H(v)},\\
 H(v)=\frac{k}{2(k-1)}\log B\bigl(\theta(v)\bigr),
 \end{gathered}
 \label{eq:isothermalperiod}
\end{equation}
with $H(v+2\pi)=H(v)$. The exterior $\rho>0$ is therefore the
cylinder $u\in\mathbb R$, $v\in[0,2\pi)$. Primes on $H$ denote
derivatives with respect to $v$. Since $\alpha^2\le B\le1$, the factor
$\e^{2H}=B^{k/(k-1)}$ is bounded between positive constants. With
$Z=\e^{u+\ii v}$, which is single-valued because
$v(\theta+2\pi)=v(\theta)+2\pi$, Eq.~\eqref{eq:isothermalperiod} gives
$h=\rho_0^2\e^{2H(v)}|\dd Z|^2$ on the whole exterior. We say that there
is no conical deficit in the conformal sense if, near the axis, the metric
is conformal to the flat metric of a punctured disk, of full angle $2\pi$,
with a conformal factor bounded above and below by positive constants. The isothermal coordinates
constructed above show that this is the case for every $k\ne1$. The bound on
the factor is essential, since for $k=1$,
$r=\rho^{1/\alpha}$ gives
$\dd\rho^2+\alpha^2\rho^2\dd\theta^2=\alpha^2r^{2(\alpha-1)}(\dd r^2+r^2\dd\theta^2)$,
which is also conformal to a punctured plane, but with a factor unbounded
as $r\to0$ for $\alpha<1$. A conformally flat metric of this type, with one logarithmic term in the
exponent for each defect, is used for pentagonal and heptagonal defects in
graphene~\cite[Eq.~(14)]{Cortijo2007}. For a single defect of total angle $2\pi\alpha$ it reduces to
$r^{2(\alpha-1)}(\dd r^2+r^2\dd\theta^2)$, the form above up to a constant factor. For $k=1$, $v$ instead has period $2\pi\alpha$, which
is the cone (see below). Figure~\ref{fig:isothermal}(a)--(c) shows the isothermal grid in the plane and the exterior as the cylinder $(u,v)$, on which the walls $\rho=a,b$ become angle-dependent curves.
Panels (d)--(f) show the corresponding conformal picture of the
cone, in which the factor is unbounded at the axis.

\begin{figure*}[tbp]
 \centering
  \includegraphics[width=\linewidth]{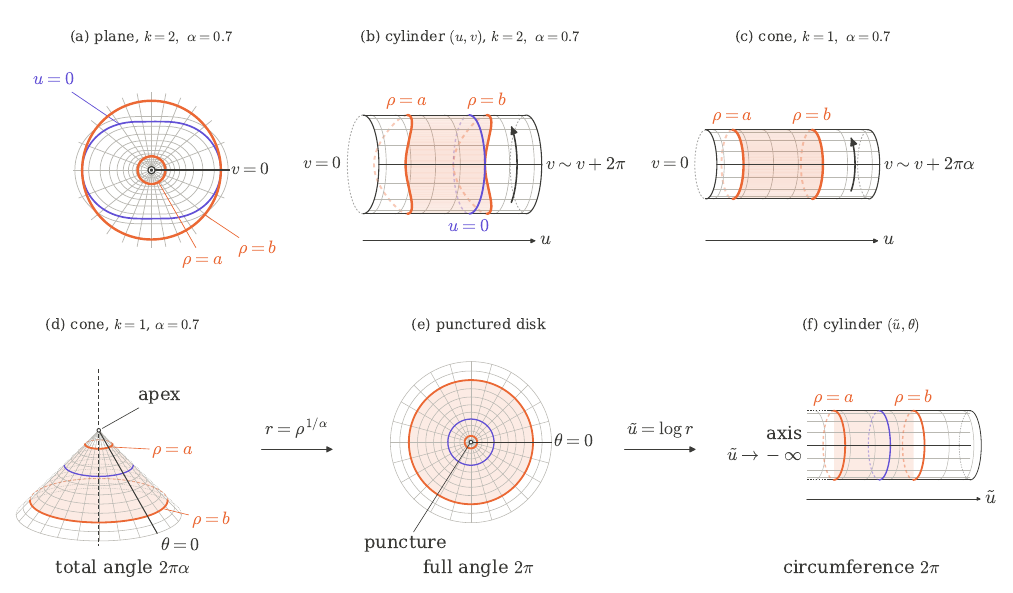}
 \caption{Conformal pictures of the exterior, with $a/b=0.2$ and $\rho_0=1$.
 (a)--(c) Isothermal coordinates of Eq.~\eqref{eq:isothermalcoords}.
 (a) Coordinate plane for $k=2$. The gray curves are the isothermal loops
 $u=\mathrm{const}$ and rays spaced uniformly in $v$, orthogonal in $h$, and
 the loop $u=0$ is that of Fig.~\ref{fig:holonomy}(c). (b) The same exterior as
 the cylinder $v\sim v+2\pi$ [Eq.~\eqref{eq:isothermalperiod}], with the axis
 at $u\to-\infty$. The isothermal loops become its circles and the walls the
 angle-dependent curves $u=\log(\rho_{\rm wall}/\rho_0)-\log B/[2(k-1)]$.
 (c) For the cone, $u=\log(\rho/\rho_0)$ and $v=\alpha\theta$, so the cylinder
 has circumference $2\pi\alpha$ and circular walls. Panels (b) and (c) share
 the same scale. (d)--(f) The cone drawn in $\mathbb R^3$, where the dashed line
 is the symmetry axis of the embedding and not $z$, is mapped by $r=\rho^{1/\alpha}$
 to a punctured disk and by $\tilde u=\log r=\alpha^{-1}\log\rho$ to a cylinder
 of circumference $2\pi$, which is (c) rescaled by $1/\alpha$. All maps are
 conformal. Relative to the flat disk the metric is
 $\alpha^2r^{2(\alpha-1)}(\dd r^2+r^2\dd\theta^2)$, whose factor is unbounded as
 $r\to0$. This is the conical deficit in the conformal sense, and for $k\ne1$
 the factor stays bounded. Walls $\rho=a,b$ are orange, the loop $\rho=0.5$ is
 purple, the seam is black, and hidden curves are dashed.}
 \label{fig:isothermal}
\end{figure*}

The coframe $(\Omega\,\dd u,\Omega\,\dd v)$ has connection
$-H'\dd u+\dd v$. The original coframe~\eqref{eq:coframe} is obtained
from it by a rotation through the angle $(k-1)v$. The two coframes
agree at $\theta=0$. Consequently,
\begin{equation}
 \omega=-H'(v)\,\dd u+k\,\dd v,\qquad
 K=-\Omega^{-2}H''(v).
 \label{eq:isothermalmetric}
\end{equation}
This is the connection of the original coframe expressed in the new
coordinates, so the field transition is unchanged.

In isothermal coordinates $\sqrt h\,h^{ij}=\delta^{ij}$, so
$\Delta_s=\Omega^{-2}[(\partial_u-\ii s\omega_u)^2
+(\partial_v-\ii s\omega_v)^2]$. Multiplying Eq.~\eqref{eq:master}
by $\Omega^2$ gives
\begin{equation}
 \begin{split}
 &\left[(\partial_u+\ii sH')^2+(\partial_v-\ii sk)^2\right.\\
 &\quad\left.-c_sH''+\lambda^2\rho_0^2\e^{2u+2H}\right]\psi_s=0.
 \end{split}
 \label{eq:isothermalmaster}
\end{equation}
This form isolates the scale-invariant indicial problem from the term
that couples $u$ and $v$ at nonzero $\lambda$. That term carries the
factor $\e^{2u}$, and setting $\lambda=0$ removes it.

For $k=1$ the same construction gives $u=\log(\rho/\rho_0)$ and
$v=\alpha\theta$, so $v$ has period $2\pi\alpha$. This is the ordinary
cone, which is separable and is treated in Sec.~\ref{sec:separable}.
\subsection{Factorization and monodromy}
\label{sec:factorization}

The indicial equation is obtained by setting $\lambda=0$ and seeking
$\psi_s=\e^{pu}w(v)$. Since the difference between $u$ and
$\log(\rho/\rho_0)$ is periodic in angle, $p$ is also the exponent
in an ansatz $\rho^p$ times an angular function. Define
\begin{equation}
 \begin{gathered}
 D_v=\partial_v-\ii sk,\\
 A_p=D_v+sH'-\ii p,\qquad B_p=D_v-sH'+\ii p.
 \end{gathered}
 \label{eq:factoroperators}
\end{equation}
Substituting $\psi_s=\e^{pu}w(v)$ into Eq.~\eqref{eq:isothermalmaster}
with $\lambda=0$ replaces $\partial_u$ by $p$ and gives
$P_{s,c}(p)w=0$, where $P_{s,c}(p)$ is an ordinary differential operator
in $v$ whose coefficients are quadratic polynomials in $p$. A polynomial
family of operators of this kind is called an operator
pencil~\cite{KozlovMazyaRossmann}. For the two curvature couplings it factors as
\begin{align}
 P_{s,+s}(p)&=(p+\ii sH')^2+D_v^2-sH''=A_pB_p,\nonumber\\
 P_{s,-s}(p)&=(p+\ii sH')^2+D_v^2+sH''=B_pA_p.
 \label{eq:factorization}
\end{align}
Both factorizations are relevant. The physical Dirac and Maxwell
couplings equal $+s$ for negative signed spin weight and $-s$ for
positive signed spin weight.

\begin{proposition}[Indicial multipliers]
\label{prop:indicial}
Let $k\ne1$ and $c_s\in\{s,-s\}$. Call $p$ an indicial exponent if
the equation with $\lambda=0$ has a solution
$\psi_s=\e^{pu}w(v)$ obeying the transition~\eqref{eq:twist}.
\begin{enumerate}
 \item[(i)] Over one period $v\mapsto v+2\pi$, the angular equation
 for $w$ has the two monodromy multipliers
 \begin{equation}
  \mu_\pm(p)=\exp[2\pi\ii(sk\pm p)].
  \label{eq:indicialmultipliers}
 \end{equation}
 \item[(ii)] For a twist $\chi_s$, the indicial exponents are
 \begin{equation}
  p=\pm(m-sk),\qquad m\in\mathbb Z+\chi_s.
  \label{eq:indicialroots}
 \end{equation}
 \item[(iii)] For the physical twist $\chi_s=sk+\delta$, this set is
 $p\in\pm(\mathbb Z+\delta)$.
\end{enumerate}
\end{proposition}

\begin{proof}
The assumption $k\ne1$ ensures that the isothermal coordinates of
Eq.~\eqref{eq:isothermalcoords} are defined, with $v$ of period $2\pi$ and
$H$ periodic. For $k=1$ isothermal coordinates also exist, but
$v=\alpha\theta$ has period $2\pi\alpha$, and the cone is treated separately
in Sec.~\ref{sec:separable}. The assumption
$c_s\in\{s,-s\}$ ensures that the indicial equation factorizes as in
Eq.~\eqref{eq:factorization}.

(i) Take $c_s=s$, so the equation is $A_pB_pw=0$, and set $g=B_pw$.
The equation is equivalent to
\begin{equation}
 \begin{gathered}
 w'=\bigl[\ii(sk-p)+sH'\bigr]w+g,\\
 g'=\bigl[\ii(sk+p)-sH'\bigr]g,
 \end{gathered}
 \label{eq:triangular}
\end{equation}
where the first equation is the definition of $g$ and the second is
$A_pg=0$. The equation for $g$ does not involve $w$. Since $H$ is
periodic, $\int_0^{2\pi}H'\,\dd v=0$, so after one period $g$ is
multiplied by $\mu_+$. The equation for $w$ has the homogeneous factor
$\mu_-$. For $c_s=-s$, the equation is $B_pA_pw=0$, and the choice
$g=A_pw$ gives the same result with $\mu_+$ and $\mu_-$ interchanged.

(ii) The coordinate $u$ is periodic in $\theta$, and
$v(\theta+2\pi)=v(\theta)+2\pi$. Hence $\psi_s=\e^{pu}w(v)$ obeys the
transition~\eqref{eq:twist} exactly when
$w(v+2\pi)=\e^{2\pi\ii\chi_s}w(v)$. Such a solution exists if and only
if $\e^{2\pi\ii\chi_s}$ equals $\mu_+$ or $\mu_-$, that is, if
$sk\pm p=\chi_s+n$ for some $n\in\mathbb Z$. Writing $m=\chi_s+n$
gives $p=\pm(m-sk)$ with $m\in\mathbb Z+\chi_s$.

(iii) If $\chi_s=sk+\delta$, then $m\in\mathbb Z+sk+\delta$, so
$m-sk\in\mathbb Z+\delta$.
\end{proof}

\begin{figure*}[tbp]
 \centering
 \includegraphics[width=\linewidth]{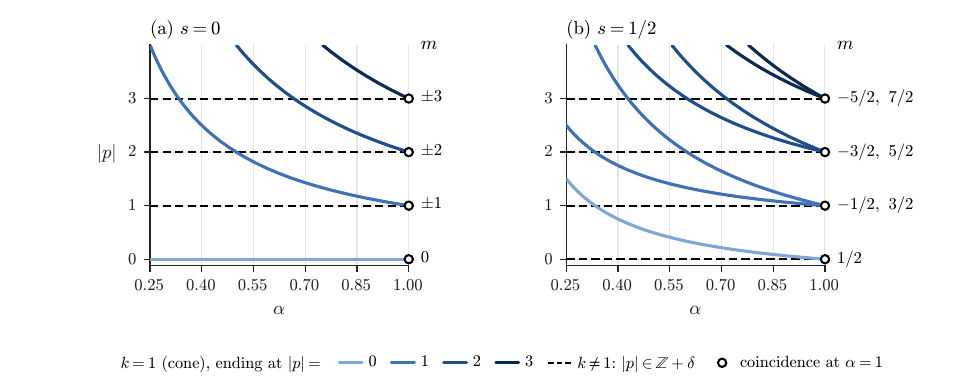}
 \caption{Indicial exponents as functions of $\alpha$ for $\delta=0$,  (a) for the scalar, $s=0$, and (b) for a spinor component, $s=\tfrac12$. Solid curves: for the cone, $k=1$, the exponents $\nu_m=|m/\alpha-s|$ of Eq.~\eqref{eq:conebessel}, $m\in\mathbb Z+s+\delta$, which move with $\alpha$. Each curve is shaded by the value $|m-s|$ it reaches at $\alpha=1$, and the values of $m$ ending at each level are listed on the right. For $s=0$ the curves for $\pm m$ coincide. For $s=\tfrac12$, two curves end at each nonzero level, the steeper one having $m>0$. Dashed lines: for every $k\ne1$, Proposition~\ref{prop:indicial} gives $|p|\in\mathbb Z+\delta$, independent of $\alpha$ and $k$. The two sets coincide at $\alpha=1$ (open circles), Eq.~\eqref{eq:flatbessel}.}
 \label{fig:indicial}
\end{figure*}

The proof combines the factorization of Eq.\eqref{eq:factorization}, which is specific to the present geometry, with the monodromy (Floquet) theory of linear first-order equations with periodic coefficients \cite[Sec.~3.6]{TeschlODE}. For $c_s=\pm s$ the indicial operator factorizes as in Eq.~\eqref{eq:factorization}. Because $H$ is periodic, the resulting triangular first-order system [Eq.~\eqref{eq:triangular}, and its analogue for $c_s=-s$] has the same monodromy multipliers as in the case $H\equiv0$. The proposition applies to the minimally coupled scalar and to the Dirac and Maxwell potential-component equations listed in Table~\ref{tab:fields}. It determines the set of component exponents rather than a decomposition into individual Fourier harmonics. In the electromagnetic case, physical solutions must additionally satisfy the Lorenz constraint and be considered modulo admissible residual gauge transformations, so a potential indicial root does not by itself define a physical photon mode. When multipliers coincide, the monodromy need not be diagonalizable, and logarithmic generalized solutions and their multiplicities require separate analysis. The proposition neither selects a self-adjoint core domain nor imposes the full set of physical constraints. Figure~\ref{fig:indicial} contrasts the $\alpha$-dependent exponents of the cone with the $\alpha$- and $k$-independent set of Proposition~\ref{prop:indicial}.

For $k\neq1$, the indicial result also controls candidate near-axis behavior at finite spectral parameter. Indeed, the term omitted from Eq.~\eqref{eq:isothermalmaster} is $\lambda^2\Omega^2=\lambda^2\rho^2B$, which vanishes quadratically
as $\rho\to0$ at fixed $\lambda$ and $\alpha>0$. For a scalar or orthonormal spinor component with pure-power behavior $\rho^p$ and a bounded angular profile that does not vanish identically, local square integrability requires
\begin{equation*}
 \int_0^\epsilon \rho^{2\operatorname{Re}p+1}\,\dd\rho<\infty \quad\Longleftrightarrow\quad \operatorname{Re}p>-1.
\end{equation*}
For example, the exponent $p=-\tfrac12$ allowed for spinor components when $\delta=\tfrac12$ describes a component that diverges pointwise but remains locally square integrable. The monodromy therefore constrains candidate core behavior. Selecting physical core solutions additionally requires the field constraints and a specified operator domain.

\subsection{Separable benchmarks and the problem at nonzero \texorpdfstring{$\lambda$}{lambda}}
\label{sec:separable}

For $k=1$, which was excluded from Eq.~\eqref{eq:isothermalcoords}, the
frame does not rotate, the connection~\eqref{eq:connection} reduces to
$\omega=\alpha\,\dd\theta$, and $K=0$. The operator $\Delta_s$ therefore
separates, and a mode $\psi_s=R_m(\rho)\e^{\ii m\theta}$ obeys
\begin{equation}
 \begin{gathered}
 R_m''+\frac{1}{\rho}R_m'
 +\left(\lambda^2-\frac{\nu_m^2}{\rho^2}\right)R_m=0,
 \\ \nu_m=\left|\frac{m}{\alpha}-s\right|.
 \end{gathered}
 \label{eq:conebessel}
\end{equation}
Here $m\in\mathbb Z+s+\delta$, and $K=0$ on the annulus. For $s=\pm\tfrac12$ and $\delta=\tfrac12$, the orders $|m/\alpha-s|$
coincide with the Bessel orders $|n/\alpha\pm\tfrac12|$, $n\in\mathbb Z$,
of the conical Dirac eigenfunctions in Ref.~\cite{DAncona2025}. For $s=\pm\tfrac12$ and $\delta=0$, the orders $|m/\alpha-s|$ with
$m\in\mathbb Z+\tfrac12$ coincide with the Bessel orders of the cosmic-string
eigenspinors in Ref.~\cite{BezerradeMello2008} [Eq.~(2.15) there, with
$q=1/\alpha$ and $j=m$]. For $\alpha=1$ and arbitrary $k$, the same equation holds with
\begin{equation}
 \nu_m=|m-sk|=|n+\delta|,\qquad n\in\mathbb Z.
 \label{eq:flatbessel}
\end{equation}
Recovering the flat Bessel orders for every $k$ provides a global check of the
spin-frame convention.

In the generic curved case, the indicial exponents of
Proposition~\ref{prop:indicial} do not depend on $\alpha$, but the finite-annulus eigenvalues do, because the
exponents describe only the scale-invariant problem. The term $\lambda^2\rho_0^2\e^{2u+2H(v)}$ in
Eq.~\eqref{eq:isothermalmaster} prevents the indicial factorization
from providing independent finite-$\lambda$ Bessel channels, and the
walls $\rho=a,b$ become angle-dependent curves in the $(u,v)$
cylinder. The flat
case $k=0$ admits Cartesian stretched coordinates, but its annular
boundary geometry still matters.

\section{Boundary domains and spectral consequences}
\label{sec:spectra}

The field equations of Sec.~\ref{sec:fields} do not by themselves
define a spectrum: boundary conditions must be specified. This section
fixes the physical boundary conditions on an annulus
(Sec.~\ref{sec:boundaries}), isolates the TEM sector (Sec.~\ref{sec:tem}), writes the Hermitian weak formulation
used later in the numerics (Sec.~\ref{sec:weak}), and derives two
analytic properties of the scalar spectrum: variational bounds
(Sec.~\ref{sec:bounds}) and a weak-anisotropy selection rule
(Sec.~\ref{sec:fourier}).

These analytic steps also determine how the numerical results of
Sec.~\ref{sec:numerics} can be read. On the annulus every problem
considered here has a real, discrete spectrum of isolated eigenvalues of
finite multiplicity, so each computed eigenvalue can be compared with a
definite level of the continuum problem. For the Dirac field this requires a self-adjoint Hamiltonian,
which provides real energies, a complete set of modes, and a vanishing
normal current at the walls, together with a compact resolvent, which
makes the levels isolated and of finite multiplicity. For the scalar and
Maxwell fields the variational characterization yields bounds and
monotonicity that hold for every eigenvalue and for all $k$ and
$\alpha$, whereas the computations sample finitely many cases. The Dirac
Hamiltonian is unbounded from below and from above, so no such bounds
are available, and the computations use its squared form instead. The
analytic results also supply exact relations against which the
implementation is tested.

\subsection{Physical boundary conditions on an annulus}
\label{sec:boundaries}

Fix the \textit{coordinate} domain
\begin{equation}
 \begin{gathered}
 \mathcal{A}=\mathcal A_{a,b}=\{a<\rho<b,\ 0\leq\theta<2\pi\},
 \\ 0<a<b<\infty.
 \end{gathered}
 \label{eq:annulus}
\end{equation}
A wall at $a>0$ is a confinement condition, not a regularity condition at the
axis, so the spectra computed on this domain belong to the chosen
boundary-value model, and removing a wall or taking a core limit defines a
different problem. For the ordinary cone without an inner wall, the Dirac operator is not
essentially self-adjoint: with spinor components periodic in the polar
frame (our $\delta=\tfrac12$), it admits a one-parameter family of
self-adjoint extensions~\cite{DAncona2025}, so the limit $a\to0$ requires
an additional choice.

For a scalar field, homogeneous Dirichlet or natural Neumann data
give standard self-adjoint choices. The outward unit normal to a
coordinate circle gives the conormal derivative
\begin{equation}
 \partial_n=\pm\frac{1}{\sqrt{\beta_1}}
       \left(\beta_1\partial_\rho
                   +\frac{\cM}{2\rho}\partial_\theta\right),
 \label{eq:conormal}
\end{equation}
with the plus sign at $b$ and the minus sign at $a$. Thus Neumann
data are not generally $\partial_\rho\phi=0$.

For the Dirac field, a concrete reflecting choice is the MIT bag
condition~\cite{Chodos1974},
\begin{equation}
 \begin{gathered}
 P_{\mathrm{bag}}\Psi\big|_{\partial\mathcal A}=0,
 \qquad P_{\mathrm{bag}}=\tfrac12(\Id+\ii n_\mu\gamma^\mu),
 \\ n_\mu n^\mu=-1,
 \end{gathered}
 \label{eq:bag}
\end{equation}
where $n$ is the outward spacelike unit normal and $\Id$ is the
$4\times4$ identity. Since $(\ii n_\mu\gamma^\mu)^2=\Id$,
$P_{\mathrm{bag}}$ is a projector of rank two, so the condition fixes
two of the four spinor components at each wall. In the representation
of Eq.~\eqref{eq:dirachamiltonian} it reads
$\psi_{\mathrm d}=-\ii\,n^a\sigma^a\,\psi_{\mathrm u}$ on
$\partial\mathcal A$, where $n=n^ae_a$ with $(n^1)^2+(n^2)^2=1$, so that
$n^a\sigma^a=n^1\sigma^1+n^2\sigma^2$. It implies $\bar\Psi\Psi=0$ and
$n_\mu\bar\Psi\gamma^\mu\Psi=0$ there, i.e.\ vanishing normal Dirac
current. Condition~\eqref{eq:bag} differs in sign from that of
Ref.~\cite{Chodos1974}, $\ii n_\mu\gamma^\mu\psi=\psi$ [Eq.~(5.3a) there].
That reference uses the inward normal and the metric signature
$(-,+,+,+)$. When both are converted to the conventions used here, the
matrix $\ii n_\mu\gamma^\mu$ is the same in the two papers, and the two
conditions select its opposite eigenvalues: Eq.~\eqref{eq:bag},
equivalently $\ii n_\mu\gamma^\mu\Psi=-\Psi$, selects $-1$, whereas
Ref.~\cite{Chodos1974} selects $+1$. As
noted in Ref.~\cite{Chodos1974} [Eq.~(5.11)], $\gamma^5\Psi$ satisfies the
condition with the opposite sign. In the representation of
Eq.~\eqref{eq:dirachamiltonian}, $\gamma^5\Psi=(\psi_{\mathrm d},\psi_{\mathrm u})$
commutes with the transition matrix of Eq.~\eqref{eq:twist} and gives
$\gamma^5H_D(M)\gamma^5=H_D(-M)$. For $M=0$, the case of all computations
below, the two conventions therefore give the same spectrum. For $M\neq0$,
condition~\eqref{eq:bag} at mass $M$ corresponds to the condition of
Ref.~\cite{Chodos1974} at mass $-M$.

We now specify the operator domain of $H_D$. In it, the
condition~\eqref{eq:bag} is imposed on the two walls $\rho=a$ and
$\rho=b$. No condition is imposed at $\theta=0$ or $\theta=2\pi$,
where the field obeys the transition rule~\eqref{eq:twist}, as for the
scalar spaces of Secs.~\ref{sec:weak} and~\ref{sec:bounds}. As in
Eq.~\eqref{eq:lichnerowicz}, each component of
$\psi_{\mathrm u}$ and $\psi_{\mathrm d}$ carries the spin weight
$s=\sigma/2$, where $\sigma=\pm1$ is its $\sigma^3$ eigenvalue, and
obeys Eq.~\eqref{eq:twist} with that $s$. Thus both two-spinors have
the transition matrix
$Q=\e^{2\pi\ii\delta}\e^{\ii\pi k\sigma^3}$.
Let $L^2_{\rm tw}(\mathcal A)^4$ be the space of measurable
$\Psi=(\psi_{\mathrm u},\psi_{\mathrm d})$ on the strip $a<\rho<b$,
$\theta\in\mathbb R$, that obey Eq.~\eqref{eq:twist} componentwise, with
inner product

\begin{equation}
 \langle\Phi,\Psi\rangle=\int_{\mathcal A_{a,b}}\dd A_h\,\Phi^\dagger\Psi .
 \label{eq:diracL2}
\end{equation}

The subscript ``tw'' (twisted) indicates that the components are not
periodic in $\theta$ but obey the transition rule~\eqref{eq:twist}, and
the exponent $4$ counts the components of $\Psi$, two in each of the
two-spinors $\psi_{\mathrm u}$ and $\psi_{\mathrm d}$.
Both fields obey Eq.~\eqref{eq:twist}, so
$\Phi(\theta+2\pi)^\dagger\Psi(\theta+2\pi)=\Phi(\theta)^\dagger
\operatorname{diag}(Q^\dagger Q,Q^\dagger Q)\Psi(\theta)$, and the integrand
is $2\pi$-periodic because $Q$ is unitary. Let
$H^1_{\rm tw}(\mathcal A)^4$ consist of those $\Psi$ whose components
have weak derivatives on the strip and finite norm
$\int_a^b\!\int_0^{2\pi}\rho\,(|\partial_\rho\psi_s|^2
+\rho^{-2}|\partial_\theta\psi_s|^2+|\psi_s|^2)\,\dd\theta\,\dd\rho$,
the norm of $H^1(\mathcal A)$ in Sec.~\ref{sec:bounds}. Equivalently,
$\e^{-\ii\chi_s\theta}\psi_s\in H^1(\mathcal A)$ for any representative
of $\chi_s$ modulo~1, as in the substitution of Sec.~\ref{sec:weak}.
Multiplication by $\e^{-\ii\chi_s\theta}$ changes only the angular
derivative, by $-\ii\chi_s$ times the field, and $\rho\geq a$, so the
triangle inequality gives
\begin{equation*}
 \frac{\|\psi_s\|_{H^1}}{1+|\chi_s|/a}
 \leq\|\e^{-\ii\chi_s\theta}\psi_s\|_{H^1}
 \leq\Bigl(1+\frac{|\chi_s|}{a}\Bigr)\|\psi_s\|_{H^1},
\end{equation*}
where we use the norm above.

For $j=1,2$, $\sigma^jQ=(-1)^{2k}Q\sigma^j$ and $\sigma^3Q=Q\sigma^3$.
Under $\theta\mapsto\theta+2\pi$, the coefficients of
Eqs.~\eqref{eq:inverseframe} and~\eqref{eq:dirac2d} acquire the factor
$(-1)^{2(k-1)}=(-1)^{2k}$. Hence $T$, and with it $H_D$ of
Eq.~\eqref{eq:dirachamiltonian}, maps fields obeying
Eq.~\eqref{eq:twist} to fields obeying it. Since $a>0$, these
coefficients are bounded, and $H_D$ maps $H^1_{\rm tw}(\mathcal A)^4$
boundedly into $L^2_{\rm tw}(\mathcal A)^4$.

For $f\in H^1(\mathcal A)$ with angular Fourier coefficients
$f_m(\rho)=(2\pi)^{-1}\int_0^{2\pi}f\,\e^{-\ii m\theta}\dd\theta$, each
$f_m$ lies in $H^1(a,b)$ and is therefore continuous on $[a,b]$. We
define $f|_{\rho=b}=\sum_mf_m(b)\e^{\ii m\theta}$, which is the restriction
of $f$ to the wall when $f$ is smooth, and likewise at $\rho=a$. By
Parseval's identity \cite{brezis2010functional} and the one-dimensional bound
$|f_m(b)|^2\leq2[(b-a)^{-1}\|f_m\|^2+(b-a)\|f_m'\|^2]$ in $L^2(a,b)$,
\begin{equation}
 \begin{split}
\|f|_{\rho=b}\|_{L^2(0,2\pi)}^2\leq\frac{2}{a}\max\{(b-a)^{-1},\,b-a\}\,\|f\|_{H^1(\mathcal A)}^2 .
 \end{split}
 \label{eq:diractrace}
\end{equation}
The trace of $\psi_s$ is
$\e^{\ii\chi_s\theta}$ times that of $\e^{-\ii\chi_s\theta}\psi_s$, and it
is independent of the representative of $\chi_s$. We take
\begin{equation}
 \begin{split}
 \operatorname{Dom}(H_D)=\bigl\{&\Psi\in H^1_{\rm tw}(\mathcal A)^4:\\
   &P_{\mathrm{bag}}\Psi=0\ \text{at}\ \rho=a\ \text{and}\ \rho=b\bigr\},
 \end{split}
 \label{eq:bagdomain}
\end{equation}
with the condition imposed on the traces for almost every $\theta$, that
is, except on a set of measure zero~\cite[Sec.~0.6]{Teschl2009}.
In Eq.~\eqref{eq:bag}, $n=N$ at $\rho=b$ and $n=-N$ at $\rho=a$, where
$N=\sqrt{\beta_1}\,\partial_\rho+\cM(2\rho\sqrt{\beta_1})^{-1}\partial_\theta$
is the unit vector of Eq.~\eqref{eq:conormal}, with frame components
$N^a=(\cos\xi,\,-\alpha^{-1}\sin\xi)/\sqrt{\beta_1}$. These components
also acquire $(-1)^{2k}$ under $\theta\mapsto\theta+2\pi$. Together
with $\sigma^jQ=(-1)^{2k}Q\sigma^j$ for $j=1,2$, this shows that the transition
rule maps $\ker P_{\mathrm{bag}}(\theta)$ onto
$\ker P_{\mathrm{bag}}(\theta+2\pi)$.

The space~\eqref{eq:bagdomain} is dense in
$L^2_{\rm tw}(\mathcal A)^4$. It contains every field whose components
are $\e^{\ii\chi_s\theta}$ times the periodic extension of a smooth
function with compact support in $a<\rho<b$, $0<\theta<2\pi$, since such a
field vanishes near the walls and satisfies the bag condition trivially.
Such functions are dense in $L^2$ of that rectangle with respect to
$\dd\rho\,\dd\theta$~\cite[Theorem~0.36]{Teschl2009}, and hence with
respect to $\dd A_h$, since $\alpha a\leq\alpha\rho\leq\alpha b$.
Multiplication by $\e^{\ii\chi_s\theta}$ preserves the norm and maps
$L^2$ of the rectangle onto the corresponding component space of
$L^2_{\rm tw}(\mathcal A)^4$, which proves the density. The
space~\eqref{eq:bagdomain} is therefore not closed in
$L^2_{\rm tw}(\mathcal A)^4$, because it is a proper dense subspace, but
it is closed in $H^1_{\rm tw}(\mathcal A)^4$. By Eq.~\eqref{eq:diractrace}
the traces are continuous, and $P_{\mathrm{bag}}$ is an orthogonal
projector that depends smoothly on $\theta$, so the condition defines the
kernel of a bounded map.

For real $M$ and $p_z$ and $\Phi,\Psi\in H^1_{\rm tw}(\mathcal A)^4$,
\begin{equation}
 \begin{gathered}
 \langle\Phi,H_D\Psi\rangle-\langle H_D\Phi,\Psi\rangle
 =-\ii\int_{\partial\mathcal A}\Phi^\dagger\mathcal B(n)\Psi\,\dd s,
 \\
 \mathcal B(n)=\begin{pmatrix}0&n^a\sigma^a\\ n^a\sigma^a&0\end{pmatrix},
 \end{gathered}
 \label{eq:diracgreen}
\end{equation}
where $n$ is the outward unit normal and $\dd s=\alpha\rho\sqrt{\beta_1}\,\dd\theta$
is the arclength of $h$ on each wall. For smooth fields this follows by
integrating by parts with the weight $\alpha\rho$. The zeroth-order term of
Eq.~\eqref{eq:dirac2d} equals $-\tfrac{\ii}{2}\sigma^a\operatorname{div}_h e_a$,
since $\operatorname{div}_h e_1=\omega(e_2)=k\cos\xi/\rho$ and
$\operatorname{div}_h e_2=-\omega(e_1)=-k\sin\xi/(\alpha\rho)$, so all bulk
terms cancel. The terms at $\theta=0$ and $2\pi$ cancel because
$e_a{}^\theta\,\Phi^\dagger\sigma^a\Psi$ is $2\pi$-periodic. Both sides are
continuous on $H^1_{\rm tw}(\mathcal A)^4$ by Eq.~\eqref{eq:diractrace}, and
smooth twisted fields are dense there, as follows by truncating the angular
Fourier series and approximating each $f_m$ in $H^1(a,b)$. On
$\operatorname{Dom}(H_D)$, $\psi_{\rm d}=-\ii n^a\sigma^a\psi_{\rm u}$ makes the
integrand $-\ii\varphi_{\rm u}^\dagger\psi_{\rm u}+\ii\varphi_{\rm u}^\dagger\psi_{\rm u}=0$,
so $H_D$ is symmetric on the domain~\eqref{eq:bagdomain}.

\begin{proposition}\label{ass:diracsa}
For real $M$ and $p_z$, the operator $H_D$ with
domain~\eqref{eq:bagdomain} is self-adjoint in
$L^2_{\rm tw}(\mathcal A)^4$.
\end{proposition}
\begin{proof}
We apply the theory of first-order elliptic boundary problems of
Ref.~\cite{BaerBallmann}. In this proof the Standard Setup, Example,
Lemma and Corollary numbers, and Eqs.~(29), (31), (32), (62) and (63),
refer to the published version of that reference.

The twisted fields are sections of a rank-four Hermitian vector bundle
over the closed annulus, with transition matrix
$\operatorname{diag}(Q,Q)$ from Eq.~\eqref{eq:twist}, and the measure is
$\dd A_h$. Since $a>0$, the coefficients of $H_D$ are smooth up to the
walls, and by Eq.~\eqref{eq:diracgreen} $H_D$ is formally self-adjoint.
Its principal symbol is $-\ii\mathcal S(\xi)$ with
$\mathcal S(\xi)=\begin{pmatrix}0&\xi_a\sigma^a\\ \xi_a\sigma^a&0\end{pmatrix}$
in frame components, so that $\mathcal S(n)=\mathcal B(n)$ of
Eq.~\eqref{eq:diracgreen}, and $\mathcal S(\xi)^2=|\xi|_h^2\Id$. The
Clifford relations (31) and (32) therefore hold, and $H_D$ is of Dirac
type in the sense of Example~4.3(a). The annulus is compact, so the
hypotheses of Standard Setup~1.5, including completeness, are satisfied.

The bag condition states that the trace of $\Psi$ lies in the $+1$
eigenbundle $\mathcal V_+$ of the Hermitian involution
$G=\begin{pmatrix}0&\ii n^a\sigma^a\\-\ii n^a\sigma^a&0\end{pmatrix}$.
The matrix $G$ anticommutes with $\mathcal S(n)$ and commutes with
$\mathcal S(\tau)$ for tangential $\tau$. It therefore anticommutes with
the symbol of the adapted boundary operator, which by Eq.~(29) is
proportional to $\mathcal S(n)\mathcal S(\tau)$, and that symbol exchanges
$\mathcal V_+$ and the $-1$ eigenbundle $\mathcal V_-$. By
Corollary~7.23 the bag condition is a local elliptic boundary condition,
and by Lemma~7.3 the corresponding closed extension of $H_D$ has exactly
the domain~\eqref{eq:bagdomain}. By Eqs.~(62) and (63), its adjoint is the
same differential operator with the boundary condition consisting of the
traces orthogonal to $\sigma_0\mathcal V_+$, where $\sigma_0$, the symbol at
the conormal, is proportional to $\mathcal S(n)$. Since $\mathcal S(n)$
anticommutes with $G$, it maps $\mathcal V_+$ onto $\mathcal V_-$. The
adjoint condition is therefore again the bag condition, and $H_D^*=H_D$.
\end{proof}
The self-adjointness can also be read off from a pointwise criterion.
In the setting of Ref.~\cite{Grosse2025}, $\mathcal S(\xi)$ is a
Hermitian Clifford multiplication on the rank-four bundle, and the bag
condition requires the boundary value to lie in the rank-two subbundle
$\mathcal V_+$. For $u,v\in\mathcal V_+$ one has $u^\dagger\mathcal S(n)v=0$,
which is the cancellation of the integrand in Eq.~\eqref{eq:diracgreen}.
By Theorem~1.2 of Ref.~\cite{Grosse2025}, a boundary subbundle of
half rank with this property defines a self-adjoint operator on the
graph-norm closure of the smooth fields obeying the condition. This
closure coincides with the domain~\eqref{eq:bagdomain} because the bag
condition is elliptic, by Corollary~7.23 and Lemma~7.3 of
Ref.~\cite{BaerBallmann}. The same reference notes that the MIT condition
of the geometric literature, which uses skew-Hermitian Clifford
multiplication, is skew-symmetric rather than self-adjoint.
By Proposition~\ref{ass:diracsa}, $H_D$ has compact resolvent. The domain
$\operatorname{Dom}(H_D)$ is complete in the $H^1$ norm, being closed in
$H^1_{\rm tw}(\mathcal A)^4$, and in the graph norm, since a self-adjoint
operator is closed. The $H^1$ norm dominates the graph norm, because $H_D$
maps $H^1_{\rm tw}(\mathcal A)^4$ boundedly into $L^2_{\rm tw}(\mathcal A)^4$,
so the two norms are equivalent by the open mapping theorem, in the form
of Corollary~2.8 of Ref.~\cite{brezis2010functional}. Hence
$(H_D-\ii)^{-1}$ maps $L^2_{\rm tw}(\mathcal A)^4$ boundedly into
$H^1_{\rm tw}(\mathcal A)^4$, whose embedding into
$L^2_{\rm tw}(\mathcal A)^4$ is compact by the Rellich argument of
Sec.~\ref{sec:bounds}, applied to $\e^{-\ii\chi_s\theta}\psi_s$. The spectrum
of $H_D$ therefore consists of isolated eigenvalues of finite
multiplicity~\cite[Theorem~III-6.29]{Kato}.

The squared operator has the inherited domain
\begin{equation}
 \begin{split}
 \operatorname{Dom}(H_D^2)
 =\{&\Psi\in\operatorname{Dom}(H_D):\\
                       &H_D\Psi\in\operatorname{Dom}(H_D)\}.
 \end{split}
 \label{eq:squareddomain}
\end{equation}
Every $\Psi$ in Eq.~\eqref{eq:squareddomain} satisfies the bag condition
for both $\Psi$ and $H_D\Psi$ in the trace sense, since
$\operatorname{Dom}(H_D)\subset H^1_{\rm tw}(\mathcal A)^4$. By
Proposition~\ref{ass:diracsa}, the form $q[\Psi]=\|H_D\Psi\|^2$ with
$\operatorname{Dom}(q)=\operatorname{Dom}(H_D)$ is
closed~\cite[Example~VI-1.13]{Kato}, and its associated operator is
$H_D^*H_D=H_D^2$ with the domain~\eqref{eq:squareddomain}
~\cite[Theorem~V-3.24 and Example~VI-2.13]{Kato}. Imposing Dirichlet data independently on every
squared component instead defines a different auxiliary problem.
It can test the local operator, but its eigenvalues cannot be
reported as the MIT bag Dirac spectrum.

For Maxwell fields, choose perfectly conducting walls parallel to
$z$, imposing vanishing tangential electric and normal magnetic
fields. For $E\ne0$, the residual gauge function with amplitude
$\ell=\varphi/(\ii E)$ sets $\varphi=0$ while preserving Lorenz
gauge. The remaining constraint is
\begin{equation}
 \operatorname{div}_h a^\sharp+\ii p_z a_z=0
 \quad\text{in }\mathcal A_{a,b}.
 \label{eq:temporalconstraint}
\end{equation}
In this gauge the electric field is $\bm E=\ii E\bm A$, where $E$
denotes the frequency, and the conductor conditions become
\begin{equation}
 a(\tau)|_{\partial\mathcal A}=0,\qquad
 a_z|_{\partial\mathcal A}=0,
 \label{eq:potentialpec}
\end{equation}
where $\tau$ is tangent to the transverse boundary.
The normal component of $a$ is not set to zero. A second-order
Hodge-Laplacian realization uses these tangential conditions
together with vanishing spatial codifferential at the boundary,
as inherited from Eq.~\eqref{eq:temporalconstraint}.
The static case $E=0$ requires a separate gauge and constraint
analysis.

\begin{figure*}[tbp]
 \centering
 \includegraphics[width=\linewidth]{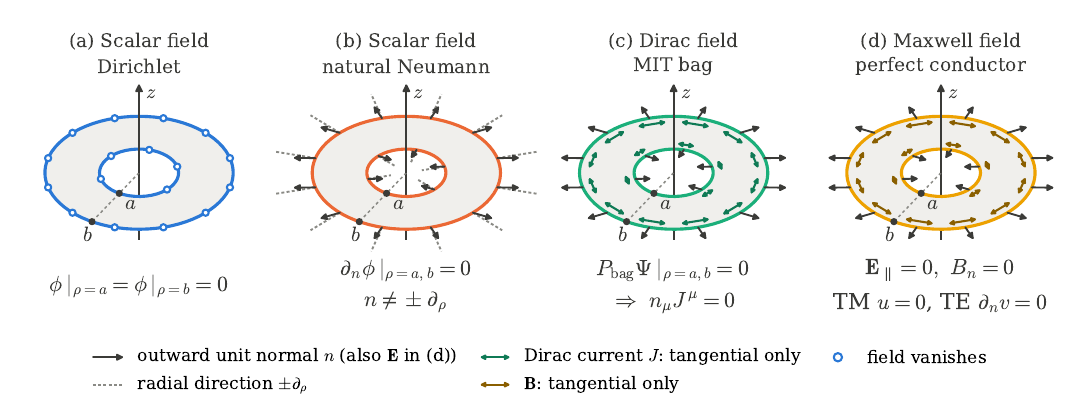}
 \caption{Boundary-value problems on the annulus $\mathcal A_{a,b}$ of
 Eq.~\eqref{eq:annulus}, shown as a transverse section for $k=2$ and $\alpha=0.7$ (for $k=1$ or $\alpha=1$ the normal $n$ would be radial), with $a/b$ enlarged and arrows not to scale. The walls are
 parallel to the $z$ axis. (b) The normal $n$ is computed from $h$ and is
 tilted from $\pm\partial_\rho$, so natural Neumann data, Eq.~\eqref{eq:conormal},
 are not $\partial_\rho\phi=0$. (c) The MIT condition, Eq.~\eqref{eq:bag},
 makes the normal Dirac current vanish. (d) For $\lambda^2>0$ the perfect-conductor
 conditions are Dirichlet data for the TM variable $u$ and Neumann data for the
 TE variable $v$, Eq.~\eqref{eq:pec}. Double-headed arrows indicate only
 tangential directions. The wall at $\rho=a>0$ is a confinement condition,
 not regularity at the axis.}
 \label{fig:boundaries}
\end{figure*}

Equivalently, for $\lambda^2>0$ the gauge-invariant scalar variables
in Eq.~\eqref{eq:maxwelllongitudinal} satisfy
\begin{equation}
 \text{TM: }u|_{\partial\mathcal A}=0,\qquad
 \text{TE: }\partial_n v|_{\partial\mathcal A}=0.
 \label{eq:pec}
\end{equation}
Together with Eq.~\eqref{eq:maxwellreconstruction}, these yield
the same physical fields. The constant Neumann scalar at
$\lambda=0$ is not a TE mode obtained by the reconstruction
with $\lambda^{-2}$. Figure~\ref{fig:boundaries} summarizes these four boundary-value problems and shows that the outward unit normal $n$, which defines the
conormal derivative $\partial_n$ of Eq.~\eqref{eq:conormal}, is
generally tilted away from the radial direction.

\subsection{The annular transverse-electromagnetic sector}
\label{sec:tem}

An annulus has an additional Maxwell sector at $\lambda=0$, which
must be treated separately from Eq.~\eqref{eq:maxwellreconstruction}.
Let $U$ solve
\begin{equation}
 \Delta_0 U=0,\qquad U|_{\rho=a}=0,\qquad U|_{\rho=b}=1.
 \label{eq:capacitary}
\end{equation}
For $E\ne0$, a temporal-gauge potential generating the TEM mode is
\begin{equation}
 \varphi=0,\qquad a_z=0,\qquad
 a=\frac{\dd_h U}{\ii E},\qquad p_z=\pm E.
 \label{eq:potentialtem}
\end{equation}
Here \(U\) is a scalar capacitary potential. Its differential \(d_hU\) is the transverse one-form that generates the physical TEM field and, in the circular basis of Eq.~\eqref{eq:potentialcircular}, supplies the \(s=\pm1\) components of the transverse potential. Since $\Delta_0U=0$,
the Lorenz constraint holds. For $E\ne0$ and $p_z=\pm E$, a TEM
field is, up to an overall amplitude,
\begin{equation}
 \begin{gathered}
 E_z=B_z=0,\qquad E_\perp^\flat=\dd U,\qquad B_\perp^\flat=\frac{p_z}{E}*_h\dd U.
 \end{gathered}
 \label{eq:tem}
\end{equation}
The constancy of $U$ on each boundary enforces the conductor conditions, and
its harmonicity yields the remaining Maxwell equations.
There is one independent such family per propagation direction on
the annulus: the difference between the two boundary constants is
the one free potential difference. Anisotropy changes the transverse
profile, but not the zero cutoff. This physical sector would be
missed by a calculation restricted to positive scalar eigenvalues.

\subsection{Hermitian weak formulation}
\label{sec:weak}

For a component problem with a self-adjoint boundary domain, the
natural inner product uses $\dd A_h$, not an unweighted coordinate
grid sum. The weak forms are
\begin{align}
 a_s(v,u)&=\int_{\mathcal A}\dd A_h\,
      \left[h^{ij}\overline{D_i^{(s)}v}D_j^{(s)}u
                         -c_sK\overline v u\right],\nonumber\\
 b(v,u)&=\int_{\mathcal A}\dd A_h\,\overline v u,
 \qquad a_s(v,u)=\lambda^2 b(v,u).
 \label{eq:weakform}
\end{align}
These forms follow from the master equation~\eqref{eq:master}, written as
$H_{s,c_s}u=\lambda^2u$. Let $v$ carry the same twist~\eqref{eq:twist} and
the same boundary condition as $u$. Multiplying by $\overline v$ and
integrating with $\dd A_h=\alpha\rho\,\dd\rho\,\dd\theta$ gives
$b(v,H_{s,c_s}u)=\lambda^2b(v,u)$, and the factor $\alpha\rho$ in
$\Delta_s$ [Eq.~\eqref{eq:masterdef}] cancels against $\dd A_h$. Because
$\omega$ is real, the connection terms of $\overline v$ and of
$W^i=\alpha\rho\,h^{ij}D^{(s)}_ju$ cancel in the product rule
\begin{equation}
 \partial_i\bigl(\overline v\,W^i\bigr)
   =\overline{D^{(s)}_iv}\,W^i+\overline v\,D^{(s)}_iW^i .
 \label{eq:productrule}
\end{equation}
Integration by parts therefore gives
\begin{equation}
 b(v,H_{s,c_s}u)=a_s(v,u)
   -\int_0^{2\pi}\!\!\int_a^b\partial_i\bigl(\overline v\,\alpha\rho\,
     h^{ij}D^{(s)}_ju\bigr)\,\dd\rho\,\dd\theta .
 \label{eq:greenweak}
\end{equation}
In Eq.~\eqref{eq:greenweak}, the boundary terms at
$\theta=0$ and $\theta=2\pi$ cancel, because $u$ and $v$ carry the
same twist of Eq.~\eqref{eq:twist} and $\overline v\,D^{(s)}_ju$ is
therefore $2\pi$-periodic. For real $c_s$ the form is Hermitian,
$a_s(u,v)=\overline{a_s(v,u)}$, because $h^{ij}$, $\omega$, and $K$
are real. Dirichlet data, or the covariant conormal condition
$h^{\rho j}D^{(s)}_ju=0$ at $\rho=a$ and $\rho=b$, which is the natural
boundary condition of $a_s$, remove the remaining boundary terms, so that
$a_s$ represents $H_{s,c_s}$ with that boundary condition. Because
$\omega^\rho=0$ in the coframe~\eqref{eq:coframe}
[Eq.~\eqref{eq:operatorchecks}], this condition is $\partial_nu=0$ with
$\partial_n$ of Eq.~\eqref{eq:conormal}, for every $s$.
For the self-adjoint realization defined by $a_s$ below, a nonzero $u$
in the form domain is an eigenfunction with eigenvalue $\lambda^2$
exactly when $a_s(v,u)=\lambda^2b(v,u)$ for every $v$ in the form
domain. The formulation requires only first derivatives of $u$ and $v$,
which is the setting of the finite-element computations of
Sec.~\ref{sec:numerics}, and the natural boundary condition is not
imposed on the trial functions but follows from the weak form. Here $u$ and $v$ denote functions on
$\mathcal A$, unrelated to the isothermal coordinates of
Sec.~\ref{sec:indicial}.

For these two boundary conditions each component realization has a
discrete, real spectrum bounded below. The substitution
$u=\e^{\ii\chi_s\theta}\tilde u$ maps functions with the twist of
Eq.~\eqref{eq:twist} one-to-one onto $2\pi$-periodic functions $\tilde u$.
It preserves $b$ and the vanishing of traces, and it replaces
$D^{(s)}u$ by $\e^{\ii\chi_s\theta}(\partial+\ii\kappa)\tilde u$ with the real
one-form $\kappa=\chi_s\dd\theta-s\omega$. On a fixed annulus with
$a>0$, the quantities $h^{ij}\kappa_i\kappa_j$ and $K$ are bounded. The
Cauchy--Schwarz inequality then gives, for every $\varepsilon\in(0,1)$,
a constant $C_\varepsilon$ such that
$(1-\varepsilon)a_0(\tilde u,\tilde u)-C_\varepsilon b(\tilde u,\tilde u)\leq a_s(u,u)
\leq(1+\varepsilon)a_0(\tilde u,\tilde u)+C_\varepsilon b(\tilde u,\tilde u)$. The form $a_s$ is
therefore bounded below and closed on the domain of $a_0$ in
Sec.~\ref{sec:bounds}, $H^1_0(\mathcal A)$ for Dirichlet data and
$H^1(\mathcal A)$ for the conormal condition, transported by the substitution, and the compactness argument given there
carries over without modification. By the first representation theorem~\cite[Theorems~VI-2.1 and VI-2.6]{Kato},
as in Sec.~\ref{sec:bounds}, $a_s$ therefore defines a self-adjoint
operator that is bounded below and has compact resolvent, so its spectrum
consists of real isolated eigenvalues of finite multiplicity that
accumulate only at $+\infty$.
An arbitrary curvature coupling need not give a nonnegative operator.
Nonnegativity of the squared Dirac problem and of the Maxwell problems is
not inferred from this estimate. It holds by construction for the squared MIT form of
Eq.~\eqref{eq:vi_bagform}, and for the TE and TM scalar problems by
Sec.~\ref{sec:bounds}.

Complex conjugation gives the local identity
\begin{equation}
 H_{-s,c}\,\overline{\psi_s}
       =\overline{H_{s,c}\psi_s}.
 \label{eq:conjugation}
\end{equation}
For conjugate twists and matching component boundary domains, the
two signed-weight spectra coincide. This gives a further numerical
check, but it does not prove an independent physical Dirac
polarization degeneracy with arbitrary boundary conditions.

\subsection{Scalar eigenvalue bounds and monotonicity}
\label{sec:bounds}

We now consider the scalar problem of Eq.~\eqref{eq:scalar}, the case
$s=0$, $c_0=0$ of Eq.~\eqref{eq:master}, on the
annulus~\eqref{eq:annulus} with homogeneous Dirichlet or natural Neumann
data on both walls. Its functions are $2\pi$-periodic in $\theta$, as
Eq.~\eqref{eq:twist} requires for $s=0$ and $\delta=0$. We write
$H^1(\mathcal A)$ for the functions $f(\rho,\theta)$ on the strip
$a<\rho<b$, $\theta\in\mathbb R$, that are $2\pi$-periodic in $\theta$,
have weak derivatives $\partial_\rho f$ and $\partial_\theta f$ on this
strip, and have finite norm
\begin{equation*}
 \|f\|_{H^1}^2=\int_a^b\!\int_0^{2\pi}\rho\,\bigl(|\partial_\rho f|^2
 +\rho^{-2}|\partial_\theta f|^2+|f|^2\bigr)\,\dd\theta\,\dd\rho ,
\end{equation*}
and $H^1_0(\mathcal A)$ for the elements of $H^1(\mathcal A)$ whose
traces vanish at $\rho=a$ and $\rho=b$. For $s=0$ one has
$D^{(0)}_i=\partial_i$, and for $c_0=0$ the curvature term is absent, so
the forms of Eq.~\eqref{eq:weakform} reduce to
\begin{equation*}
 a_0(f,f)=\int_{\mathcal A}\dd A_h\,h^{ij}\,\overline{\partial_if}\,\partial_jf,
 \qquad
 b(f,f)=\int_{\mathcal A}\dd A_h\,|f|^2 .
\end{equation*}
The form $b$ is the inner product of $L^2(\mathcal A,\dd A_h)$, so
$b(f,f)>0$ for every nonzero $f$ in this space. The form domain
$\operatorname{Dom}(a_0)$ is $H^1_0(\mathcal A)$ for homogeneous
Dirichlet data and $H^1(\mathcal A)$ for natural Neumann data. Neumann condition, Eq.~\eqref{eq:conormal}, is not imposed on the form
domain but arises as the natural boundary condition of $a_0$, satisfied
by every element of the operator domain (Sec.~\ref{sec:weak}).

To write $a_0$ in coordinates, define the two $\alpha$-independent
vector fields
\begin{equation}
 \begin{gathered}
 X=\cos\xi\,\partial_\rho+\frac{\sin\xi}{\rho}\partial_\theta,
 \\
 Y=-\sin\xi\,\partial_\rho+\frac{\cos\xi}{\rho}\partial_\theta.
 \end{gathered}
 \label{eq:XY}
\end{equation}
By Eq.~\eqref{eq:inverseframe}, the orthonormal frame of $h$ is
$e_1=X$ and $e_2=\alpha^{-1}Y$. Since $h=(e^1)^2+(e^2)^2$
[Eq.~\eqref{eq:metric}], the inverse metric is
$h^{ij}=e_1{}^ie_1{}^j+e_2{}^ie_2{}^j$, and therefore
$h^{ij}\,\overline{\partial_if}\,\partial_jf=|Xf|^2+\alpha^{-2}|Yf|^2$.
The anisotropy enters this integrand only through the factor
$\alpha^{-2}$ from $e_2$. The fields $X$ and $Y$ are obtained from
$\partial_\rho$ and $\rho^{-1}\partial_\theta$ by a rotation through the
angle $\xi$, so that, almost everywhere for $f\in H^1(\mathcal A)$,
\begin{equation}
 |Xf|^2+|Yf|^2=|\partial_\rho f|^2+\rho^{-2}|\partial_\theta f|^2 .
 \label{eq:XYnorm}
\end{equation}
The area element $\dd A_h=\alpha\rho\,\dd\rho\,\dd\theta$
[Eq.~\eqref{eq:metriccomponents}] contributes the same constant factor
$\alpha$ and the same weight $\rho$ to $a_0$ and $b$. The Rayleigh
quotient \cite{brezis2010functional} of $a_0$ relative to $b$ is therefore
\begin{equation}
 \begin{gathered}
 Q_\alpha[f]=
 \frac{\displaystyle\int_a^b\!\int_0^{2\pi}
       \rho\left(|Xf|^2+\alpha^{-2}|Yf|^2\right)
           \dd\theta\dd\rho}
      {\displaystyle\int_a^b\!\int_0^{2\pi}
       \rho|f|^2\dd\theta\dd\rho}
 \\
 =\frac{a_0(f,f)}{b(f,f)},
 \qquad f\in\operatorname{Dom}(a_0)\setminus\{0\}.
 \end{gathered}
 \label{eq:rayleigh}
\end{equation}
For $0<\alpha\leq1$, Eq.~\eqref{eq:XYnorm} gives
$|Xf|^2+|Yf|^2\leq|Xf|^2+\alpha^{-2}|Yf|^2
\leq\alpha^{-2}(|Xf|^2+|Yf|^2)$. Hence $Q_\alpha[f]$ is finite, and the
norm $\|f\|_{a_0}^2=a_0(f,f)+b(f,f)$ of $\operatorname{Dom}(a_0)$
satisfies
\begin{equation*}
\alpha\|f\|_{H^1}^2\leq\|f\|_{a_0}^2\leq\alpha^{-1}\|f\|_{H^1}^2.
\end{equation*}
The term $b(f,f)$ in $\|f\|_{a_0}$ is needed because, for Neumann data,
$a_0$ vanishes on constants. The two norms are equivalent for each fixed
$\alpha$ (inequality above), though not uniformly as $\alpha\to0$. The form domains do not
depend on $\alpha$, whereas the strong Neumann condition,
Eq.~\eqref{eq:conormal}, does. The numerator of $Q_\alpha$ is a
nonnegative Dirichlet integral, not the physical energy of the field,
and $Q_\alpha$ has the dimension of $\lambda^2$ in
Eq.~\eqref{eq:scalar}. The quotient is built from the scalar form alone, whereas the Dirac form
$\langle\Psi,H_D\Psi\rangle$ is indefinite (Sec.~\ref{sec:implementation})
and plays no role here.

The quotient is related to the eigenvalue problem through the operator
that $a_0$ defines. For $a>0$, $a_0$ is a densely defined, closed,
nonnegative form on $\operatorname{Dom}(a_0)$. It is densely defined,
because the periodic extensions of smooth functions with compact support
in the rectangle $a<\rho<b$, $0<\theta<2\pi$ lie in $H^1_0(\mathcal A)$
and are dense in $L^2$~\cite[Theorem~0.36]{Teschl2009}. It is closed,
because $H^1(\mathcal A)$ is complete, $H^1_0(\mathcal A)$ is a closed
subspace of it by the continuity of the traces
[Eq.~\eqref{eq:diractrace}], and the norm equivalence above transfers
completeness to $\|\cdot\|_{a_0}$. By the first representation
theorem~\cite[Theorems~VI-2.1 and VI-2.6]{Kato}, it defines a
self-adjoint operator $T_\alpha\geq0$ in $L^2(\mathcal A,\dd A_h)$, with
$\operatorname{Dom}(T_\alpha)\subset\operatorname{Dom}(a_0)$ and
$a_0(v,u)=b(v,T_\alpha u)$ for $u\in\operatorname{Dom}(T_\alpha)$ and
$v\in\operatorname{Dom}(a_0)$.
In particular, $Q_\alpha[u]=b(u,T_\alpha u)/b(u,u)$ for nonzero
$u\in\operatorname{Dom}(T_\alpha)$, so $Q_\alpha$ is the Rayleigh
quotient of $T_\alpha$ on its operator domain and extends it to the form
domain. For $u\in\operatorname{Dom}(T_\alpha)$, choosing $v$ smooth,
$2\pi$-periodic in $\theta$, and vanishing near $\rho=a$ and $\rho=b$
gives $T_\alpha u=-\Delta_0u$ in the sense of distributions \cite[Sec.~5.2.1]{evans1998partial}, independently
of the boundary condition. Because $0<a<b<\infty$, polar
coordinates identify $H^1(\mathcal A)$, with the norm
$\|\cdot\|_{H^1}$, isometrically with the Sobolev space $H^1$ of the
planar region $a<|x|<b$. Rellich's theorem \cite{brezis2010functional, evans1998partial} for this planar region
therefore makes the embedding of $H^1(\mathcal A)$ into
$L^2(\mathcal A,\dd A_h)$ compact. A direct proof uses the angular
Fourier coefficients $f_m$ defined above Eq.~\eqref{eq:diractrace}. Let
$\|f^{(n)}\|_{H^1}\leq C_0$. Since $\rho^{-1}\geq\rho/b^2$, Parseval's
identity gives
$2\pi\sum_{|m|>J}\int_a^b\rho|f^{(n)}_m|^2\dd\rho\leq b^2J^{-2}C_0^2$ for
every $n$ and every integer $J\geq1$. For each fixed $m$, $(f^{(n)}_m)_n$ is bounded in $H^1(a,b)$,
hence uniformly bounded and equicontinuous on $[a,b]$, and the
Arzel\`a--Ascoli theorem \cite[App.~C.7]{evans1998partial} gives a uniformly convergent subsequence. A
diagonal subsequence converges for every $m$, and the uniform tail bound
makes it a Cauchy sequence in $L^2(\mathcal A,\dd A_h)$.
By the norm equivalence above, the embedding of
$(\operatorname{Dom}(a_0),\|\cdot\|_{a_0})$ is compact too. For
$g\in L^2(\mathcal A,\dd A_h)$ and $u=(T_\alpha+1)^{-1}g$,
$\|u\|_{a_0}^2=b(u,(T_\alpha+1)u)=b(u,g)\leq\|u\|_{a_0}\|g\|$, so
$(T_\alpha+1)^{-1}$ maps $L^2(\mathcal A,\dd A_h)$ boundedly into
$(\operatorname{Dom}(a_0),\|\cdot\|_{a_0})$. Composed with the compact
embedding it is compact, so $T_\alpha$ has compact
resolvent. By Theorem~III-6.29 of Ref.~\cite{Kato}, the spectrum of
$T_\alpha$ consists of isolated eigenvalues of finite multiplicity. By
Sec.~V-3.8 of Ref.~\cite{Kato}, the eigenvectors can be chosen to form
an orthonormal basis $\{\varphi_j\}$ of $L^2(\mathcal A,\dd A_h)$, with
$T_\alpha\varphi_j=\lambda_j^2\varphi_j$. The eigenvalues
$\lambda_j^2(\alpha)$ are taken in increasing order and counted with
multiplicity, including the Neumann zero eigenvalue. They satisfy the
min--max principle on the form domain,
\begin{equation*}
 \lambda_j^2(\alpha)=\min_{\substack{L\subset\operatorname{Dom}(a_0)\\ \dim L=j}}
 \ \max_{f\in L\setminus\{0\}}Q_\alpha[f].
\end{equation*}
Theorem~4.10 of Ref.~\cite{Teschl2009} states the corresponding maximin
principle over the operator domain, and the remark following it notes
that the operator domain may be replaced by the form domain. The minimax
form used here also follows directly from the eigenbasis. For $f\in\operatorname{Dom}(a_0)$ and
$c_j=b(\varphi_j,f)$, the identity $a_0(\varphi_j,f)=\lambda_j^2c_j$ and
$a_0\geq0$ give $a_0(f,f)\geq\sum_j\lambda_j^2|c_j|^2$, with equality
when $f$ is a finite linear combination of the $\varphi_j$, while
$\sum_j|c_j|^2=b(f,f)$. Hence $Q_\alpha[f]\leq\lambda_j^2$ for every
nonzero $f$ in the span of $\varphi_1,\dots,\varphi_j$, with equality at
$f=\varphi_j$. Conversely, every $j$-dimensional
$L\subset\operatorname{Dom}(a_0)$ contains a nonzero $f$ orthogonal to
$\varphi_1,\dots,\varphi_{j-1}$, for which $Q_\alpha[f]\geq\lambda_j^2$.
The minimum is therefore attained by the span of
$\varphi_1,\dots,\varphi_j$.
In particular,
$\lambda_1^2(\alpha)=\min_{f\in\operatorname{Dom}(a_0)\setminus\{0\}}
Q_\alpha[f]$, attained by a first eigenfunction. For Dirichlet data
$\lambda_1^2(\alpha)>0$. Since the minimum is attained, it suffices to
show that $Q_\alpha[f]=0$ is impossible for nonzero
$f\in H^1_0(\mathcal A)$. Suppose $Q_\alpha[f]=0$. The integrand in the
numerator of Eq.~\eqref{eq:rayleigh} is nonnegative, and $\rho>0$ and
$\alpha>0$, so $Xf=Yf=0$ almost
everywhere~\cite[Lemma~0.27]{Teschl2009}. Equation~\eqref{eq:XYnorm}
then gives $\partial_\rho f=\partial_\theta f=0$ almost everywhere. In
terms of the angular Fourier coefficients $f_m\in H^1(a,b)$ defined above
Eq.~\eqref{eq:diractrace}, $\partial_\theta f=0$ gives $\ii m f_m=0$, so
$f_m=0$ for $m\neq0$, and $\partial_\rho f=0$ gives $f_0'=0$, so $f_0$ is
constant~\cite[Problem~2.30]{Teschl2009}. Hence $f=c$ for some constant
$c$. The trace of a
constant function is that constant, so the single condition that the
trace vanishes at $\rho=a$ already gives $c=0$, and the condition at
$\rho=b$ gives the same. Hence $f=0$ in $L^2$, which contradicts the
assumption $f\neq0$. For natural Neumann data $\lambda_1^2(\alpha)$ is
zero and is attained by constants.
For Dirichlet data the first eigenvalue is moreover simple, and its
eigenfunction can be chosen positive in $\mathcal A$. In Cartesian
coordinates of the plane, $\dd A_h=\alpha\,\dd x\,\dd y$, so
$-\Delta_0=-\partial_i(h^{ij}\partial_j)$, where $h^{ij}$ is symmetric,
has eigenvalues between $1$ and $\alpha^{-2}$, and is smooth on the
closed annulus because $2(k-1)\in\mathbb Z$. The annulus is bounded and
connected, so Theorem~2 of Sec.~6.5.1 of Ref.~\cite{evans1998partial}
applies. Although the theorem is stated for
real functions, simplicity also holds in $L^2(\mathcal A,\dd A_h)$, because the
real and imaginary parts of a complex eigenfunction are themselves real
eigenfunctions.

\begin{proposition}[Scalar spectral comparison]
\label{prop:bounds}
Fix $k,a,b$ and either homogeneous Dirichlet or natural Neumann data.
Let $\lambda_j^2(\alpha)$ denote the scalar eigenvalues in increasing
order, with multiplicity. For $0<\alpha\leq1$,
\begin{equation}
 \lambda_j^2(1)\leq\lambda_j^2(\alpha)
                       \leq\alpha^{-2}\lambda_j^2(1).
 \label{eq:bounds}
\end{equation}
Each ordered eigenvalue is nonincreasing as $\alpha$ increases.
For a simple differentiable eigenvalue branch with eigenfunction
$f_j$ normalized by $\int\rho|f_j|^2\dd\rho\dd\theta=1$, so that
$f_j=\sqrt\alpha\,\varphi_j$ up to a phase because
$\dd A_h=\alpha\rho\,\dd\rho\dd\theta$,
\begin{equation}
 \frac{\dd\lambda_j^2}{\dd\alpha}
       =-2\alpha^{-3}\int\rho|Yf_j|^2\dd\rho\dd\theta\leq0.
 \label{eq:hellmannfeynman}
\end{equation}
\end{proposition}

\begin{proof}
Since $1\leq\alpha^{-2}$, Eq.~\eqref{eq:rayleigh} gives
$Q_1\leq Q_\alpha\leq\alpha^{-2}Q_1$ on the $\alpha$-independent form
domain. Applying the min--max
principle proves Eq.~\eqref{eq:bounds}, and form ordering proves
monotonicity. For the derivative, let $\eta=\alpha^{-2}-1$ and
$q_\alpha(v,u)=\int\rho\,(\overline{Xv}\,Xu+\alpha^{-2}\,\overline{Yv}\,Yu)
\dd\rho\dd\theta$ on the $\alpha$-independent domain
$\operatorname{Dom}(a_0)$. Since $\dd A_h=\alpha\rho\,\dd\rho\dd\theta$,
$a_0(v,u)=\alpha\,q_\alpha(v,u)$ and
$b(v,u)=\alpha\int\rho\,\overline vu\,\dd\rho\dd\theta$, so $T_\alpha$ is
also the selfadjoint operator associated with $q_\alpha$ in the
$\alpha$-independent space $L^2(\rho\,\dd\rho\dd\theta)$, in which $f_j$
is normalized. For a fixed $\alpha_0$ with $\eta_0=\alpha_0^{-2}-1$,
$q_\alpha=q_{\alpha_0}+(\eta-\eta_0)\,q_Y$ with
$q_Y[f]=\int\rho|Yf|^2\dd\rho\dd\theta$, and
$q_Y[f]\leq\alpha_0^2\,q_{\alpha_0}[f]$ because
$q_{\alpha_0}[f]\geq\alpha_0^{-2}q_Y[f]$. The closed nonnegative form
$q_{\alpha_0}$, the first-order term $q_Y$ with relative bound
$\alpha_0^2$, and no higher-order terms satisfy the hypotheses of
Theorem~VII-4.8 of Ref.~\cite{Kato}. The operators $T_\alpha$ therefore
form a holomorphic family of type~(B) in $\eta$ on a complex disc about
each $\eta_0>-1$, and these discs cover a neighborhood of the half-line
$\eta>-1$. The local families agree on overlapping discs, since each
consists of the operators associated with the same quadratic form. Because $q_\alpha$ is real for real $\eta$, the family is
selfadjoint by Remark~VII-4.7 of Ref.~\cite{Kato}, and each $T_\alpha$
has compact resolvent, as shown above. On every interval
$[\eta_1,\eta_2]\subset(-1,\infty)$ the family satisfies condition
VII-(4.54) of Ref.~\cite{Kato} with $a'=0$ and $b'=(1+\eta_1)^{-1}$,
since $q_\alpha[f]\geq\alpha^{-2}q_Y[f]=(1+\eta)\,q_Y[f]$ gives
$0\leq q_Y\leq(1+\eta_1)^{-1}q_\alpha$. By Remark~VII-4.22 of
Ref.~\cite{Kato}, which extends Theorem~VII-3.9 to such families, the
eigenvalues are represented by functions holomorphic in real $\eta>-1$.
Near a simple eigenvalue the ordered eigenvalue $\lambda_j^2$ coincides
with one of these functions, and Eq.~VII-(4.44) of Ref.~\cite{Kato} for
a one-dimensional eigenspace, equivalently Eq.~VII-(4.56), gives
$\dd\lambda_j^2/\dd\eta=q_Y[f_j]$. Since $\dd\eta/\dd\alpha=-2\alpha^{-3}$,
this is Eq.~\eqref{eq:hellmannfeynman}.
\end{proof}

The Neumann zero eigenvalue is included in the proposition. The
positive TE and TM cutoff eigenvalues obey the same bounds by
Eq.~\eqref{eq:pec}. These bounds are not asserted for a general
spin-weighted curvature-coupled component operator or for the bag
Dirac spectrum.

Two scaling properties provide further normalization checks. Under
$(a,b)\mapsto(La,Lb)$, a transverse eigenvalue scales as
$\lambda_j^2\mapsto L^{-2}\lambda_j^2$. The scalar Weyl law \cite{Weyl1912,Ivrii2016} has
leading term
\begin{equation}
 N(\Lambda)\sim\frac{\operatorname{Area}(\mathcal A)}{4\pi}\Lambda,
 \qquad \operatorname{Area}(\mathcal A)=\pi\alpha(b^2-a^2),
 \label{eq:weyl}
\end{equation}
where $N(\Lambda)$ counts eigenvalues of $-\Delta_0$ below $\Lambda$.
Reporting both $b^2\lambda_j^2$ and
$\operatorname{Area}(\mathcal A)\lambda_j^2$ helps distinguish the
overall scale from the remaining geometry dependence.

\subsection{Fourier blocks and a perturbative selection rule}
\label{sec:fourier}

For $k\ne1$, set
\begin{equation}
 \begin{gathered}
 N=2(k-1)\in\mathbb Z\setminus\{0\},\qquad
 b_0=\tfrac12(1+\alpha^{-2}),\\
 d_0=\tfrac12(1-\alpha^{-2}).
 \end{gathered}
 \label{eq:fourierparameters}
\end{equation}
Since $\xi=N\theta/2$ enters the coefficients of
Eq.~\eqref{eq:masterexpanded} only through $\cos2\xi$ and $\sin2\xi$,
and $\rho^2K=-2k(k-1)d_0\cos N\theta$ by
Eq.~\eqref{eq:operatorchecks}, each coefficient is a constant plus
terms proportional to $\e^{\pm\ii N\theta}$.
Consequently, the expansion
\begin{equation}
 \psi_s(\rho,\theta)=\sum_{m\in\mathbb Z+\chi_s}
                R_m(\rho)\e^{\ii m\theta}
 \label{eq:fourierexpansion}
\end{equation}
couples $m$ only to $m\pm N$. The component operator decomposes into
$|N|$ invariant blocks $m=m_0+jN$, provided the chosen boundary
domain respects the same symmetry. For the scalar operator, Dirichlet
data do, and so does the conormal condition, because $\beta_1$ and
$\cM$ in Eq.~\eqref{eq:conormal} contain the same harmonics and the
nonvanishing prefactor $\beta_1^{-1/2}$ does not affect a homogeneous
condition. The exact radial matrices are
given in Appendix~\ref{app:radial}.

An eigenfunction confined to one block has a density with angular
period $2\pi/|N|$. A linear combination from degenerate different
blocks need not display that pattern. For the real scalar operator,
complex conjugation exchanges the blocks containing $m$ and $-m$.
If these blocks are distinct, their spectra agree. These facts should
be considered before interpreting lobes or apparent splittings in
numerical eigenvectors.

There is also a weak-anisotropy test for the scalar spectrum.
Let $\eta=\alpha^{-2}-1$, and let $f_m=R(\rho)\e^{\ii m\theta}$
be flat-annulus scalar modes normalized by
$\int\rho|f_m|^2\dd\rho\dd\theta=1$. The numerator of
Eq.~\eqref{eq:rayleigh} equals
$\int\rho(|Xf|^2+|Yf|^2)\dd\rho\dd\theta
+\eta\int\rho|Yf|^2\dd\rho\dd\theta$, the first term being the flat
Dirichlet integral, and the denominator does not depend on $\alpha$.
Let $\lambda_0^2$ be a flat eigenvalue whose eigenspace is spanned by the
modes $f_m$. First-order perturbation theory in $\eta$ within this
eigenspace involves the matrix $V$ with entries
\begin{equation}
 \begin{gathered}
 V_{m'm}=\int\rho\,\overline{Yf_{m'}}Yf_m\,\dd\rho\dd\theta,
 \\ \lambda^2=\lambda_0^2+\eta\,\mu+O(\eta^2),
 \end{gathered}
 \label{eq:perturbationmatrix}
\end{equation}
where $\mu$ runs over the eigenvalues of $V$.
For an isolated $\{m,-m\}$ doublet with $m>0$, its off-diagonal
entry can be nonzero at first order only if
\begin{equation}
 2m=|N|=2|k-1|.
 \label{eq:firstordersplitting}
\end{equation}
Indeed, writing $\sin\xi$ and $\cos\xi$ in exponentials shows that
$Yf_m$ contains only the harmonics $m\pm N/2$, so $Yf_{-m}$ and
$Yf_m$ share a harmonic only if $2m=|N|$.
The selection rule is necessary but not sufficient, and it does not imply that
every allowed matrix element is nonzero. For $N\ne0$ the cross terms between the two
harmonics of $Yf_m$ integrate to zero over $\theta$. This gives
$V_{mm}=\pi\int\rho(|R'|^2+m^2\rho^{-2}|R|^2)\dd\rho$, which equals
$\lambda_0^2/2$ after integration by parts with the flat boundary
condition.
When the off-diagonal entry vanishes, both members of the doublet
therefore have $\lambda^2=\lambda_0^2(1+\eta/2)+O(\eta^2)$, and since
$\dd\eta/\dd\alpha=-2$ at $\alpha=1$,
\begin{equation}
 \left.\frac{\dd\lambda^2}{\dd\alpha}\right|_{\alpha=1}
     =-\lambda_0^2,
 \qquad
 \left.\frac{\dd(\alpha\lambda^2)}{\dd\alpha}\right|_{\alpha=1}=0.
 \label{eq:unsplitslope}
\end{equation}
If an accidental larger degeneracy is present, the full degenerate
subspace must be used in Eq.~\eqref{eq:perturbationmatrix}.

\section{Numerical spectra and validation}
\label{sec:numerics}

A finite-element discretization always produces eigenvalues, since a
generalized problem such as Eq.~\eqref{eq:discreteeigenproblem}, with
Hermitian $\mathsf K$ and positive definite $\mathsf M$, has as many real
eigenvalues as unknowns, and their stability under mesh refinement does
not by itself show that they belong to the continuum problem. For operators that
are not self-adjoint, whose spectrum is not discrete, or that are
unbounded from below, discrete values can lie outside the spectrum, and
some of them remain stable under refinement. The analysis of
Sec.~\ref{sec:spectra} removes these possibilities for the problems
computed here. Each scalar and Maxwell problem is a self-adjoint operator
that is bounded below and has compact resolvent
(Sec.~\ref{sec:bounds}), and the same holds for the squared Dirac form
of Eq.~\eqref{eq:vi_bagform} by Proposition~\ref{ass:diracsa}. The trial spaces are contained in the corresponding form domains. With exactly integrated forms the min--max principle would make the $j$th discrete eigenvalue an upper bound for the $j$th exact one, but with the Gaussian quadrature used here this one-sided bound is not guaranteed. For conforming finite elements with quadrature of sufficient precision, the discrete eigenvalues still converge to the exact ones, with the same order as for exact integration when the precision is high enough~\cite{Banerjee1992}. That analysis is carried out for simplicial elements with Dirichlet data, and we use it here only as a guide for the present tensor-product discretization. In all refinement sequences reported below the computed eigenvalues decrease toward their limits. The comparisons with analytic
benchmarks and exact relations below therefore test the implementation
and its convergence, not the existence of the spectrum.

This section solves the boundary-value problems of
Sec.~\ref{sec:spectra} numerically. We first describe the
discretization (Sec.~\ref{sec:implementation}), then validate it
against analytic benchmarks (Sec.~\ref{sec:benchmarks}), and then
report the physical cutoffs, the weak-anisotropy test, the
reconstructed fields, the TEM sector, and a
strong-anisotropy resolution test.

\subsection{Physical problems and numerical implementation}
\label{sec:implementation}

We consider the annulus $a<\rho<b$ with $a/b=0.2$, and use $b$ as the
unit of length. The sampled geometries have
$k=0,\tfrac12,1,\tfrac32,2$ and $\alpha=0.6,0.7,0.8,0.9,1$.
The minimally coupled scalar has homogeneous Dirichlet data.
For conducting Maxwell walls, the positive Dirichlet and natural
conormal Neumann eigenvalues give the TM and TE cutoffs, respectively.
The Neumann constant is excluded from the propagating TE spectrum,
and the TEM sector is obtained from a separate capacitary solve.
Potentials and fields are reconstructed as in Appendix~\ref{app:maxwell}, with
$p_zb=1$ and $E^2=p_z^2+\lambda^2$ for the positive-cutoff modes.
The Dirac calculation uses the MIT bag domain, $M=p_z=0$, and
$\delta=0$ unless stated otherwise. These are different physical
boundary-value problems, and comparing their cutoffs does not identify
independent circular components with polarizations.

The scalar and auxiliary component operators are assembled with
conforming bilinear finite elements in $(\rho,\theta)$, a consistent
mass matrix for $\alpha\rho\,d\rho\,d\theta$, and three-point Gaussian
quadrature in each cell direction. Dirichlet degrees of freedom are
eliminated, while Neumann conditions enter naturally through the weak form, and the angular
transition rule is built into the trial space. The
resulting generalized problem is
\begin{equation}
 \mathsf K\mathbf u=\lambda^2\mathsf M\mathbf u,
 \qquad \mathsf K=\mathsf K^\dagger,
 \qquad \mathsf M=\mathsf M^\dagger>0.
 \label{eq:discreteeigenproblem}
\end{equation}
Shift-invert iteration uses tolerance $10^{-11}$. The eigenvectors are
mass-orthonormalized, and their residuals are checked independently.
The parameter sweeps use $49\times64$ nodes, including
both radial walls and excluding a duplicate angular endpoint.
The inversion test of Eq.~\eqref{eq:vi_tem_inversion} uses radial
nodes uniformly spaced in $\rho$.
The plotted sweep values are finite-mesh results and are not extrapolated.
The strong-anisotropy point $(k,\alpha)=(3,0.25)$, which lies outside
this sweep, is examined for the scalar/TM mode in
Sec.~\ref{sec:vi_strong} on meshes up to $769\times1024$. Field maps
at further points outside the sweep, computed on $193\times256$, are
collected in Appendix~\ref{app:visual} as visual robustness checks.

For the Dirac field, the positive quadratic form is assembled directly
from the differential Hamiltonian,
\begin{equation}
 \begin{gathered}
 \mathsf S_{ij}=\langle H_D\varphi_i,H_D\varphi_j\rangle,
 \qquad P_{\rm bag}\varphi_i|_{\partial\mathcal A}=0,
 \\ \mathsf S\mathbf v=\mu\mathsf M\mathbf v.
 \end{gathered}
 \label{eq:vi_bagform}
\end{equation}

As noted earlier, the operator $H_D$ is neither bounded below nor bounded above. Let
$\Gamma(\psi_{\mathrm u},\psi_{\mathrm d})=(-\psi_{\mathrm d},\psi_{\mathrm u})$.
This map is unitary on $L^2_{\rm tw}(\mathcal A)^4$, commutes with the
transition matrix of Eq.~\eqref{eq:twist}, and preserves the condition
$\psi_{\mathrm d}=-\ii n^a\sigma^a\psi_{\mathrm u}$ because
$(n^a\sigma^a)^2$ is the identity, so it maps $\operatorname{Dom}(H_D)$ onto itself. By
Eq.~\eqref{eq:dirachamiltonian}, $\Gamma H_D=-H_D\Gamma$ for all real $M$ and
$p_z$, so $\langle\Gamma\Psi,H_D\Gamma\Psi\rangle=-\langle\Psi,H_D\Psi\rangle$.
Since $H_D$ is symmetric [Eq.~\eqref{eq:diracgreen}], a bound
$\langle\Psi,H_D\Psi\rangle\ge E_0\|\Psi\|^2$ on $\operatorname{Dom}(H_D)$
would give $|\langle\Psi,H_D\Psi\rangle|\le|E_0|\,\|\Psi\|^2$, and
polarization and the density of $\operatorname{Dom}(H_D)$ would give
$\|H_D\Psi\|\le|E_0|\,\|\Psi\|$ for every $\Psi$ in the domain,
which the following family rules out. For $0\neq\Psi_0$ in the
dense class described after Eq.~\eqref{eq:bagdomain} and $\ell\in\mathbb Z$,
$\e^{\ii\ell\theta}\Psi_0\in\operatorname{Dom}(H_D)$ and
$\|\e^{\ii\ell\theta}\Psi_0\|=\|\Psi_0\|$, while the coefficient of
$\partial_\theta$ in $H_D$ squares to $-h^{\theta\theta}\Id$, where
$h^{\theta\theta}=\beta_2/\rho^2\geq b^{-2}>0$ on $\mathcal A$, so
$\|H_D(\e^{\ii\ell\theta}\Psi_0)\|\ge|\ell|\,\|(h^{\theta\theta})^{1/2}\Psi_0\|
-\|H_D\Psi_0\|$, which grows without bound. The same argument applied to
$-H_D$ excludes an upper bound. The Friedrichs extension is defined for
densely defined sectorial operators, which for symmetric operators means
bounded below~\cite[Sec.~VI-\S2.3]{Kato},~\cite[Theorem~2.12]{Teschl2009},
so it is not available for $H_D$~\cite{DAncona2025}, and the indefinite form
$\langle\varphi,H_D\varphi\rangle$ does not give the Rayleigh--Ritz upper
bounds of Sec.~\ref{sec:bounds}, whereas the positive squared form above
does.
In Eq.~\eqref{eq:vi_bagform} the form domain imposes the bag condition, and,
by Proposition~\ref{ass:diracsa}, the associated squared operator $H_D^2$
inherits the condition $P_{\rm bag}H_D\Psi|_{\partial\mathcal A}=0$.
The matrix $\mathsf S$ is thus assembled by integrating products of first
derivatives of the basis functions, rather than by squaring a discrete
centered Dirac matrix.
An adapted orthonormal frame makes the first transverse leg normal
to every coordinate circle:
\begin{equation}
 \widehat e_1=\sqrt{\beta_1}\,\partial_\rho
      +\frac{c_{\rm mix}}{2\rho\sqrt{\beta_1}}\partial_\theta,
 \qquad
 \widehat e_2=\frac{1}{\alpha\rho\sqrt{\beta_1}}\partial_\theta.
 \label{eq:vi_normalframe}
\end{equation}
Its connection obeys $\widehat\omega_\rho=\omega_\rho$ and
$\widehat\omega_\theta=\omega_\theta-(k-1)/(\alpha\beta_1)$.
In this frame, $\psi_{\rm d}=-i\epsilon\sigma^1\psi_{\rm u}$ at a wall,
where $\epsilon=-1$ at $a$ and $+1$ at $b$. With
$\chi=-\arg(\cos\xi+\ii\alpha^{-1}\sin\xi)$ continued continuously from
$\chi(0)=0$, so that $\chi(\theta+2\pi)=\chi(\theta)-2\pi(k-1)$, the
components $\widehat\psi=\e^{\ii\chi\sigma^3/2}\psi$ acquire the common
factor $-\e^{2\pi\ii\delta}$ under $\theta\mapsto\theta+2\pi$, so the full
spinor is antiperiodic for $\delta=0$ and periodic for $\delta=\tfrac12$ in
this frame. In the frame~\eqref{eq:inverseframe} the components obey
Eq.~\eqref{eq:twist}.
The first-order Hamiltonian is diagonalized within the computed
low-lying squared-operator subspace to recover signed energies.
Candidates from an incomplete subspace are screened using the actual
first-order residual, rather than interpreted as zero modes from a
small projected energy alone.
The MIT energies reported below are the energies obtained in this
projected first-order step, not the eigenvalues $\mu$ of the squared
form~\eqref{eq:vi_bagform}. This applies to the cone benchmark in
Fig.~\ref{fig:vi_validation}(a), the MIT curves in
Fig.~\ref{fig:vi_cutoffs}, the MIT entry of Table~\ref{tab:vi_cutoffs},
and the spin-structure comparison. These recovered energies are not
variational upper bounds, but on the cone benchmark of
Fig.~\ref{fig:vi_validation}(a) they are observed to converge faster than
second order.
The squared form evaluated on the recovered modes enters only
Fig.~\ref{fig:vi_validation}(b)
and the Dirac convergence order quoted below.

\subsection{Analytic benchmarks and measured accuracy}
\label{sec:benchmarks}

The scalar flat and conical spectra are compared with the Dirichlet
Bessel determinant
\begin{equation}
 \begin{gathered}
 J_\nu(\lambda a)Y_\nu(\lambda b)
 -J_\nu(\lambda b)Y_\nu(\lambda a)=0,
 \\ \nu=|m|/\alpha,
 \end{gathered}
 \label{eq:besseldet}
\end{equation}
and its derivative counterpart for Neumann data. The MIT comparison
uses its own first-order radial determinant. For example, in one
massless conical block with $j>0$, $\nu=j/\alpha-\tfrac12$ and
$(f,g)=(J_\nu(E\rho),iJ_{\nu+1}(E\rho))$. The inner and outer
bag conditions require the combinations $J_{\nu+1}-J_\nu$ and
$J_{\nu+1}+J_\nu$, respectively, with the corresponding $Y$ solutions.
The determinant built from these two combinations is independent
of the finite-element calculation.

\begin{figure*}[tbp]
 \centering
 \includegraphics[width=\linewidth]{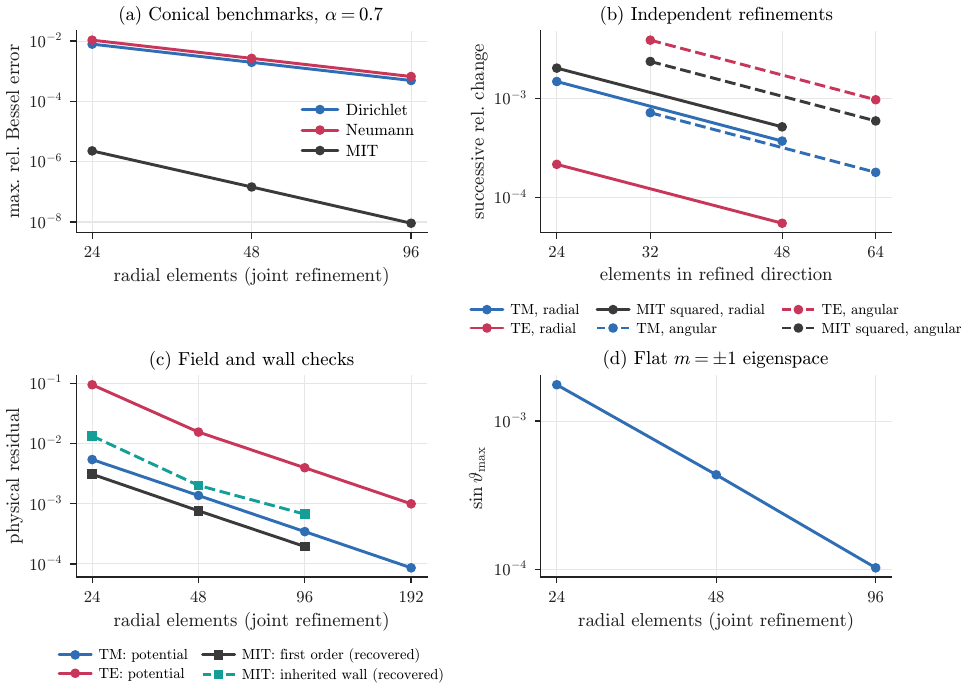}
 \caption{Validation at fixed $a/b=0.2$. (a) Largest relative Bessel
 errors on the $\alpha=0.7$ cone for the first six scalar eigenvalues
 (the five positive ones for Neumann data) and the first four positive
 MIT energies. The MIT energies are recovered first-order
 energies (Sec.~\ref{sec:numerics}). (b) Successive changes under separate radial
 and angular refinement at $(k,\alpha)=(2,0.7)$. The Dirac curve uses
 the squared form evaluated on the recovered mode. Each curve shows the two
 successive changes from three resolutions, so a decrease by a factor of
 four per doubling corresponds to second order. (c) Independent field and inherited-wall
 residuals, with the reconstruction conventions specified in the text.
 (d) Basis-independent convergence of a degenerate flat eigenspace.}
 \label{fig:vi_validation}
\end{figure*}

Figure~\ref{fig:vi_validation} summarizes the verification, for which the
finest benchmark mesh is $97\times128$.
Across the first six scalar eigenvalues, including the separate
Neumann zero mode, the largest positive-eigenvalue errors are
$1.07\times10^{-3}$ for Dirichlet data and $1.48\times10^{-3}$ for Neumann data,
over the flat and $\alpha=0.7$ cone tests.
The first four positive MIT energies, which are recovered first-order
energies, agree with their Bessel values
to $9.11\times10^{-9}$ relative error. This particularly accurate energy
comparison does not imply an equally small field-reconstruction error.
All Bessel-determinant benchmarks are at $k=1$. Repeating the flat benchmarks
at $(k,\alpha)=(2,1)$, where the frame winds, gives the same scalar errors and
reproduces the flat MIT Bessel energies to $3.7\times10^{-9}$ for $\delta=0$ and
$3.9\times10^{-8}$ for $\delta=\tfrac12$, identically to $k=1$, which checks that the frame transition cancels the
$k$-dependence of the Bessel orders in Eq.~\eqref{eq:flatbessel}. For $k\neq1$ and $\alpha\neq1$
the checks include the weak-anisotropy splitting of Fig.~\ref{fig:vi_weak}, the
TEM inversion identity, and mesh refinement.

At $(k,\alpha)=(2,0.7)$, independent radial refinement gives observed
orders $2.00$ and $1.98$ for the first TM and TE cutoffs,
and $1.97$ for the squared Dirac form, i.e.\ for the squared form
evaluated on the recovered mode rather than the recovered energy. The
corresponding angular orders are $2.00$ in all three cases
[Fig.~\ref{fig:vi_validation}(b)]. Each order is inferred from three
successive resolutions, with the other direction held at its finest level.
Comparison of the flat $m=\pm1$ eigenspace uses the largest principal
angle after interpolation and mass orthonormalization. The sine of
this angle decreases from $1.76\times10^{-3}$ to $1.02\times10^{-4}$.
Individual eigenvectors inside a degenerate eigenspace are not
used as convergence invariants.

For positive eigenvalues we monitor
\begin{equation}
 r_j=\frac{\|\mathsf K\mathbf u_j-\lambda_j^2\mathsf M\mathbf u_j\|_2}
 {\|\mathsf K\mathbf u_j\|_2+|\lambda_j^2|\|\mathsf M\mathbf u_j\|_2},
 \label{eq:vi_matrix_residual}
\end{equation}
using an absolute residual for the known Neumann zero mode.
In the benchmark and Dirac refinement runs, the largest measured
matrix residual is $1.4\times10^{-10}$ and the largest mass-orthogonality
defect is $1.2\times10^{-13}$, and the
Hermitian defect is at the level of roundoff. Flat component and bag matrices
and spectra are independent of $k$ with consistently transformed domains, and
the component conjugation identity holds for both spin structures. Uniform radius
rescaling reproduces the $L^{-2}$ eigenvalue law of Sec.~\ref{sec:bounds}
exactly at the discrete level, which checks the implementation rather
than the discretization accuracy.

\subsection{Physical cutoffs and the role of the boundary domain}

\begin{figure*}[htbp]
 \centering
 \includegraphics[width=\linewidth]{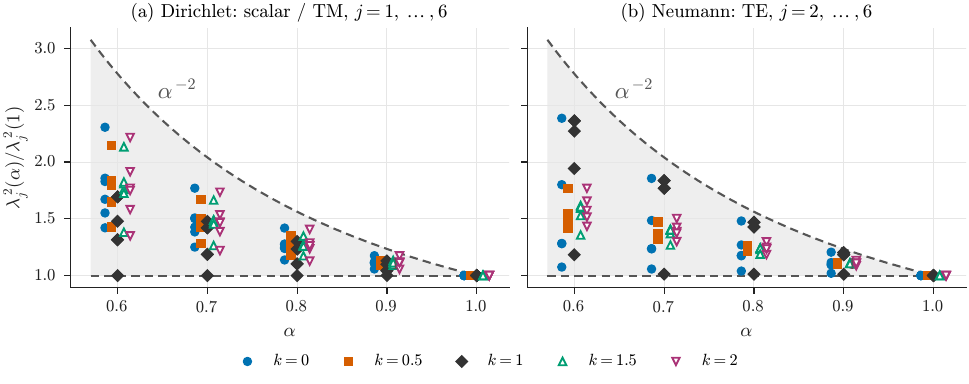}
 \caption{Numerical check of the bounds of
 Proposition~\ref{prop:bounds}. Each marker is the ratio
 $\lambda_j^2(\alpha)/\lambda_j^2(1)$ for one eigenvalue at one sampled
 geometry of the $49\times64$ sweep of Fig.~\ref{fig:vi_cutoffs}, with
 $a/b=0.2$, $k=0,\tfrac12,1,\tfrac32,2$ and $\alpha=0.6,\dots,1$.
 Markers are offset slightly in $\alpha$ for legibility. (a) The first six
 Dirichlet eigenvalues (scalar and Maxwell TM). (b) The positive
 Neumann eigenvalues $j=2,\dots,6$ (Maxwell TE); the zero eigenvalue
 $j=1$ is omitted. The shaded band is $1\leq\lambda_j^2(\alpha)/\lambda_j^2(1)\leq\alpha^{-2}$
 [Eq.~\eqref{eq:bounds}], with the flat reference $\lambda_j^2(1)$, which is
 the same for all $k$. All $220$ ratios lie in the band. The lower
 bound is attained only on the cone, $k=1$, whose lowest TM mode has
 $m=0$ and does not depend on $\alpha$. Because the discrete forms inherit the structure of Eq.~\eqref{eq:rayleigh}, these discrete bounds hold by construction, so the figure
 checks the implementation, not the proposition itself.}
 \label{fig:vi_bounds}
\end{figure*}

\begin{figure*}[tbp]
 \centering
 \includegraphics[width=\linewidth]{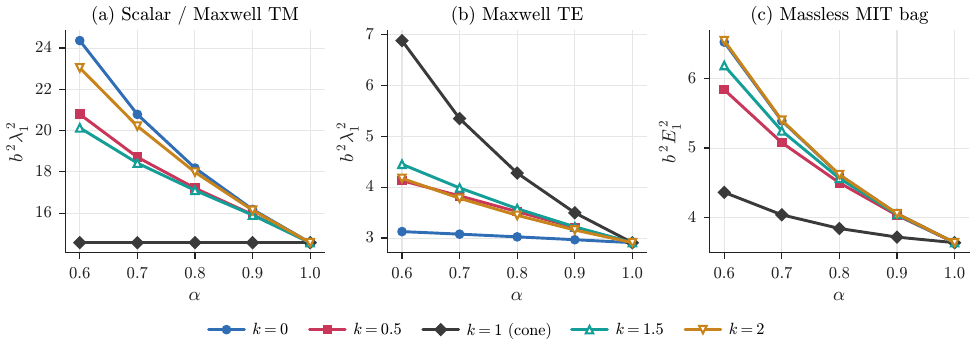}
 \caption{First positive cutoffs at fixed $a/b=0.2$, from the
 $49\times64$ parameter sweep over $k=0,\tfrac12,1,\tfrac32,2$ and
 $\alpha=0.6,\dots,1$. The panels show the Dirichlet scalar/Maxwell TM
 cutoff $b^2\lambda_1^2$, the Maxwell TE cutoff $b^2\lambda_1^2$
 (indexed as in Table~\ref{tab:vi_cutoffs}), and
 the massless MIT energy $b^2E_1^2$. The massless MIT
 curves show recovered first-order energies (Sec.~\ref{sec:numerics})
 and use $p_z=0$ and $\delta=0$. All winding cases meet in the
 flat limit within each physical boundary problem. Colors have the
 same meaning in all panels.}
 \label{fig:vi_cutoffs}
\end{figure*}

Figure~\ref{fig:vi_cutoffs} compares the first positive physical
cutoffs. The Dirichlet scalar and Maxwell TM spectra coincide,
whereas the Maxwell TE spectrum is the positive Neumann spectrum. Under $(a,b)\mapsto(La,Lb)$ every transverse eigenvalue scales as
$\lambda_j^2\mapsto L^{-2}\lambda_j^2$ (Sec.~\ref{sec:bounds}), and so
does $E^2$ for the massless MIT problem with $p_z=0$. We therefore
report the dimensionless combinations $b^2\lambda_1^2$ and $b^2E_1^2$,
where $b$ is the outer radius. They depend only on $k$, $\alpha$ and
$a/b$, and the computations use $b=1$. We normalize by $b^2$ rather than
by $\operatorname{Area}(\mathcal A)=\pi\alpha(b^2-a^2)$
[Eq.~\eqref{eq:weyl}], whose factor $\alpha$ would absorb part of the
anisotropy dependence under study. The first positive eigenvalue is the
propagation threshold. A guided scalar or Maxwell mode has
$E^2=p_z^2+\lambda^2$ [Eqs.~\eqref{eq:scalar}
and~\eqref{eq:potentialansatz}, with $M_0=0$], so no such mode
propagates with energy below $\lambda_1$.
The first eigenvalue is also the one fixed most directly by the
variational characterization, since $\lambda_1^2$ is the minimum of the
Rayleigh quotient~\eqref{eq:rayleigh}. For Dirichlet data (scalar and
Maxwell TM) it is simple for every $(k,\alpha)$ (Sec.~\ref{sec:bounds}),
so the lowest cutoff can be followed without ambiguity as $\alpha$
varies. For the MIT problem the spectrum is symmetric about zero,
because $\Gamma$ anticommutes with $H_D$ (Sec.~\ref{sec:implementation}),
so $E_1$ is the smallest $|E|$ at $M=p_z=0$ and $E_1^2$ is the lowest
eigenvalue of the form~\eqref{eq:vi_bagform}, and no dispersion
relation is assumed for the bag spectrum at $p_z\neq0$. Higher eigenvalues are
computed and checked, against the Bessel benchmarks and the bounds
below, but only the first is reported.
Their first six ordered eigenvalues satisfy
$\lambda_j^2(1)\leq\lambda_j^2(\alpha)\leq
\alpha^{-2}\lambda_j^2(1)$ at every sampled geometry
(Fig.~\ref{fig:vi_bounds}), and the first
cutoffs are nonincreasing with increasing $\alpha$. Because the
discretization is conforming, these discrete bounds and the
monotonicity hold by construction on every mesh, so the comparison verifies the
implementation rather than
Proposition~\ref{prop:bounds}, which is proved analytically.
On the ordinary cone, the lowest TM mode has $m=0$ and its cutoff
is independent of $\alpha$, while the first TE cutoff changes.
The $k=0$ results also change with $\alpha$, although the local
curvature vanishes and the stretched Cartesian annulus has different
boundary geometry. Thus a change in cutoff cannot be attributed
to curvature alone.

The MIT spectrum is obtained with the coupled spinor wall condition,
to which the scalar comparison theorem does not apply. At
$(k,\alpha)=(2,0.7)$ the refined values are listed in
Table~\ref{tab:vi_cutoffs}. The quoted mesh changes measure
discretization sensitivity and are not rigorous continuum error bounds.
Changing the spin structure to $\delta=\tfrac12$ changes the lowest
$b^2E^2$ from $5.400051$ to $5.728845$ on the same fine mesh, a shift that reflects the
different global field domain rather than a different local spin-frame
convention.

\begin{table}[tbp]
 \centering
 \caption{First positive cutoffs at $k=2$, $\alpha=0.7$, $a/b=0.2$. Here $b$ is the outer radius, so $b^2\lambda_1^2$ and $b^2E_1^2$ do
 not depend on the overall size of the annulus (Sec.~\ref{sec:numerics}).
 The last column compares the two finest joint refinements for each
 problem. Maxwell uses $193\times256$ nodes and
 MIT uses $97\times128$ nodes. The MIT entry is the recovered
 first-order energy (Sec.~\ref{sec:numerics}), so its last relative
 change refers to a different kind of quantity from the TM and TE
 entries, which are eigenvalues of the discrete quadratic forms.
 For Maxwell TE, $\lambda_1^2$ is the first positive TE eigenvalue,
 that is, the second scalar Neumann eigenvalue $\lambda_2^2$ in the
 ordering of Proposition~\ref{prop:bounds}, which includes the zero
 eigenvalue. The TM and TE entries are finite-mesh values, not
 extrapolated, rounded to two decimal places.}
 \label{tab:vi_cutoffs}
 \begin{tabular}{lcc}
 \toprule
 Problem & $b^2\lambda_1^2$ or $b^2E_1^2$ & Last relative change\\
 \midrule
 Scalar / Maxwell TM & $20.20$ & $1.38\times10^{-4}$\\
 Maxwell TE & $3.78$ & $2.57\times10^{-4}$\\
 Massless MIT bag & $5.400051$ & $9.16\times10^{-7}$\\
 \bottomrule
 \end{tabular}
\end{table}

The transverse eigenvalues have a simple interpretation in the remaining $(t,z)$ dynamics. Expanding the scalar field as $\Phi=\sum_j q_j(t,z)\phi_j(\rho,\theta)$ gives
\begin{equation*}
 \left(\partial_t^2-\partial_z^2+M_0^2+\lambda_j^2\right)q_j=0.
\end{equation*}
Each transverse mode therefore contributes an effective mass squared $M_0^2+\lambda_j^2$ to the longitudinal equation. For Maxwell TE and TM modes, $\lambda_j$ similarly gives the confinement gap: $p_{z,j}^2=E^2-\lambda_j^2$. Above threshold the mode propagates, whereas below threshold it is evanescent, with decay length $(\lambda_j^2-E^2)^{-1/2}$.

At the parameters of Table~\ref{tab:vi_cutoffs}, the first thresholds are $b\lambda_{\mathrm{TE},1}\simeq1.94$ and $b\lambda_{\mathrm{TM},1}\simeq4.49$. Energies between these thresholds admit TE propagation while every TM branch remains evanescent. The zero-cutoff TEM channel remains available throughout. Thus the computed eigenvalues specify which electromagnetic sectors can transport energy along the axis.

\subsection{Fourier blocks and weak-anisotropy splitting}

For $k=2$, the angular coupling changes $m$ by multiples of two.
As an auxiliary check of the discretized component operator, the
component problems with $(s,c_s)=(0,0)$, $(\pm\tfrac12,-\tfrac12)$, and
$(\pm1,-1)$ were solved at $\alpha=0.7$, $a/b=0.2$, and $\delta=0$ on
the $49\times64$ mesh, with homogeneous Dirichlet data on both walls.
Each problem was restricted to the even and to the odd angular block of
the reference-frame component $\widetilde\psi_s$ of
Eq.~\eqref{eq:framerelation}, which is $2\pi$-periodic here. In each
block the lowest eigenvector, extended to the full mesh, also solves
the unrestricted discrete Dirichlet problem to rounding accuracy, which confirms the half-turn symmetry of the discretized
operator. The test does not involve the conormal, MIT bag, or conductor
conditions and is distinct from the physical spinor and Maxwell calculations
above.

Let $\eta=\alpha^{-2}-1$. For the isolated flat Dirichlet
$m=\pm1$ doublet at $k=2$, the perturbation matrix has diagonal
entries $\lambda_0^2/2$. Integration by parts, with
$R(a)=R(b)=0$, gives
$|V_{1,-1}|=\lambda_0^2/4$. Therefore
\begin{equation}
 \begin{gathered}
 \lambda_-^2=\lambda_0^2(1+\eta/4)+O(\eta^2),\\
 \lambda_+^2=\lambda_0^2(1+3\eta/4)+O(\eta^2).
 \end{gathered}
 \label{eq:vi_doublet_slopes}
\end{equation}
As $\eta\to0$, the two split branches of the Dirichlet $m=\pm1$ doublet at $k=2$ approach real standing-wave combinations with angular dependence $\cos\theta$ (lower branch, $\lambda_-$) and $\sin\theta$ (upper branch, $\lambda_+$), with $\theta$ measured from the seam. Anisotropy distinguishes these orthogonal spatial orientations and gives them different excitation thresholds. Equation~\eqref{eq:vi_doublet_slopes} implies
\begin{equation*}
 \frac{\lambda_+-\lambda_-}{\lambda_0} =\frac{\eta}{4}+O(\eta^2).
\end{equation*}
The relative splitting therefore measures directional anisotropy independently of an overall rescaling of the annulus, which shifts both thresholds by the same factor. The absence of a linear splitting for the $m=\pm2$ doublet expresses the angular selection rule: the first-order geometric perturbation cannot directly couple those two harmonics.

\begin{figure*}[tbp]
 \centering
 \includegraphics[width=\linewidth]{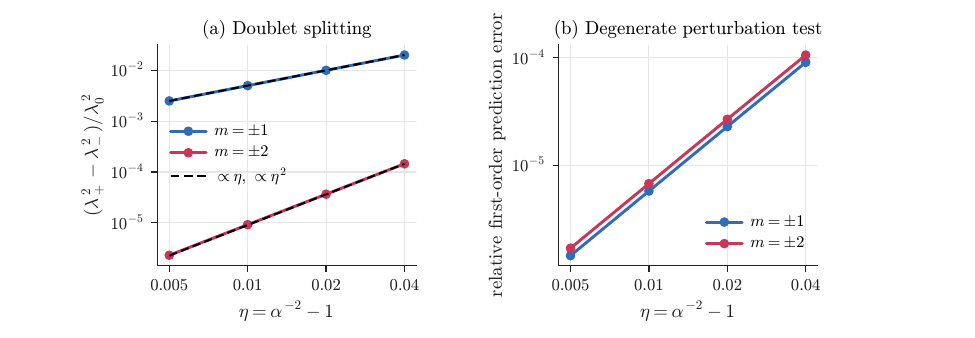}
 \caption{Weak anisotropy at $k=2$ for the first radial Dirichlet
 doublets. (a) The permitted $m=\pm1$ splitting is linear in
 $\eta$, while the $m=\pm2$ splitting starts at second order.
 Dashed lines indicate the corresponding powers, normalized at
 the largest sampled $\eta$. (b) Errors of the first-order
 perturbation matrix, assembled separately from the direct solves with
 the same discretization. The common finite-grid
 flat eigenvalue is used in the comparison.}
 \label{fig:vi_weak}
\end{figure*}

The $m=\pm2$ pair has no off-diagonal matrix element at first order.
Figure~\ref{fig:vi_weak} shows a linear splitting for $m=\pm1$
and an $m=\pm2$ splitting without a first-order term, with fitted powers
$1.000$ and $1.992$ over
$0.005\leq\eta\leq0.04$ on the $97\times128$ mesh.
The first-order perturbative eigenvalues differ from the direct
solutions by terms of order $\eta^2$. The mean shift also verifies
the stationarity, at first order, of the area-rescaled unsplit
spectrum. Fourier labels here describe a coupled expansion and
do not provide separated finite-frequency radial channels
in the generic geometry.

\subsection{Reconstruction of the physical fields}

\begin{figure*}[tbp]
 \centering
 \includegraphics[width=\linewidth]{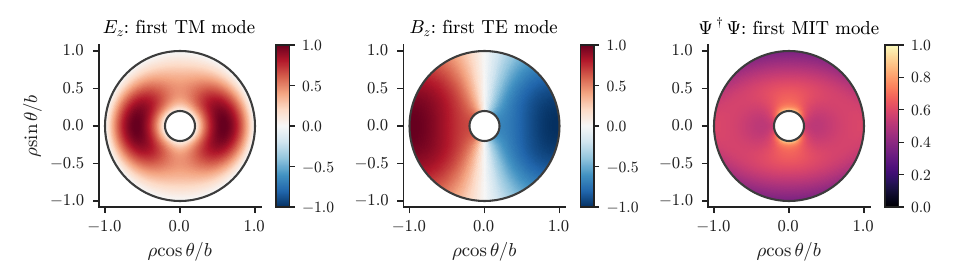}
 \caption{Representative physical modes at $k=2$, $\alpha=0.7$,
 $a/b=0.2$, on the $97\times128$ mesh.
 The real longitudinal Maxwell amplitudes have a fixed global phase
 and unit maximum modulus. The MIT panel shows the full four-spinor
 density, scaled to unit maximum. The coordinates displayed are
 $(\rho\cos\theta,\rho\sin\theta)$, so the plots are coordinate maps,
 not isometric embeddings of the transverse geometry.}
 \label{fig:vi_fields}
\end{figure*}

Figure~\ref{fig:vi_fields} displays the first TM and TE longitudinal
fields and the full positive-energy MIT spinor density at
$(k,\alpha)=(2,0.7)$. The electromagnetic potentials are reconstructed
in temporal gauge, and $F_{\mu\nu}$ is obtained by differentiating them.
We then check both curl equations, both divergence constraints,
the potential wave equation, and the conductor traces. Not all of
these checks are independent: the Faraday and magnetic-divergence
equations hold by construction, and further relations are noted below.
Pure-gauge modes are not counted: at nonzero frequency temporal
gauge removes the residual scalar gauge amplitude, and the retained
TE/TM reconstructions have nonzero field strength.

The Maxwell bulk diagnostics, which are independent of the
finite-element derivatives but use the same metric coefficients as the
solver, use five-point radial
derivatives and Fourier angular derivatives on
$0.28\leq\rho/b\leq0.92$, away from the walls,
whose conditions are checked separately.
For the finest TM and TE calculations, the relative potential-wave
defects (bulk $L^2$ residual normalized by $(\lambda^2+b^{-2})\|A\|$) are $8.63\times10^{-5}$ and $9.99\times10^{-4}$, and the Amp\`ere defects are
$3.91\times10^{-5}$ and $4.99\times10^{-4}$.
On the finest meshes the normalized tangential-potential wall defect is
$2.4\times10^{-4}$ for TE. For TM it vanishes identically, because the
longitudinal field is zero at the wall nodes.
The Faraday and magnetic-divergence checks are near numerical
precision, as expected from constructing the fields from a potential.
For TM the Gauss-law defect equals the Lorenz defect, since
$\mathbf E=\ii E\mathbf A$ in temporal gauge. The TE potential is
divergence-free by construction, so both vanish to rounding, and with
$p_zb=1$ the TE Amp\`ere defect is one half of the TE potential-wave
defect, so it is not a separate check. For Neumann data the algebraic
residual~\eqref{eq:vi_matrix_residual} includes the boundary rows, in
which the natural condition is imposed weakly, so it measures only the
accuracy of the discrete eigensolution. The continuum Neumann condition
is tested by the TE tangential-potential wall defect.

For the Dirac field, the finite-element first-order residual, evaluated
from the assembled forms through
$\|H_D\Psi-E\Psi\|^2=\|H_D\Psi\|^2-E^2$ for the normalized discrete
mode, is
\begin{equation}
 \epsilon_D=
 \frac{\|H_D\Psi-E\Psi\|_{L^2}}
 {\|H_D\Psi\|_{L^2}+|E|\|\Psi\|_{L^2}}.
 \label{eq:vi_dirac_residual}
\end{equation}
It is $9.65\times10^{-3}$ on the finest mesh, reflecting the first-order
accuracy of the piecewise-linear derivatives in this norm.
An independent smooth reconstruction from the same nodal spinor,
using five-point radial derivatives and a twisted angular Fourier
derivative, gives $\epsilon_D=1.95\times10^{-4}$.
Its inherited-wall defect
$\sqrt b\,\|P_{\rm bag}H_D\Psi\|_{L^2(\partial\mathcal A)}
/\|H_D\Psi\|_{L^2(\mathcal A)}$ is $6.71\times10^{-4}$.
For the unrecovered piecewise-linear derivative this wall diagnostic
is $7.81\times10^{-2}$ and also decreases with refinement.
The bag condition is imposed exactly on the discrete boundary trace:
in the normal frame the projector is constant along each wall, and the
bilinear trace interpolates nodal values that satisfy it. The normal
Dirac current, which vanishes identically under this condition, is zero
to rounding. The inherited condition $P_{\rm bag}H_D\Psi=0$ is natural
and holds only approximately, as measured above. Both reconstruction measures
decrease under refinement, and the small algebraic eigenvalue residual is not
used in their place.

\subsection{The TEM sector and an inversion identity}

The capacitary potential $U$ solves the scalar Laplace equation with
$U(a)=0$ and $U(b)=1$. Its reconstructed TEM field has exactly zero
cutoff and $p_z=\pm E$, while its profile changes with the geometry.
Define the dimensionless capacity
\begin{equation}
 C(k,\alpha;a/b)=\int_{\mathcal A}|dU|_h^2\,dA_h.
 \label{eq:vi_capacity}
\end{equation}
For unit complex electric amplitude and the usual time-average
convention, the total electromagnetic energy per unit length is
$C/2$. On the cone, $U=\log(\rho/a)/\log(b/a)$ and
$C=2\pi\alpha/\log(b/a)$.
The numerical cone capacity agrees with this expression to
$4.31\times10^{-5}$ relative error on the
$97\times128$ mesh.
At $k=2$, $\alpha=0.7$, the finest ($193\times256$) capacitary solve gives
$C=4.046$, with normalized weak residual of order $10^{-12}$.
The TEM Lorenz and Amp\`ere defects, $1.91\times10^{-4}$ and
$9.55\times10^{-5}$, share the same numerator, the bulk residual of the
capacitary equation, and differ only by their normalization. The
Amp\`ere defect is therefore one half of the Lorenz defect and is not an
independent check.
The derivative-sensitive potential-wave defect is $3.14\times10^{-3}$.

\begin{figure*}[tbp]
 \centering
 \includegraphics[width=\linewidth]{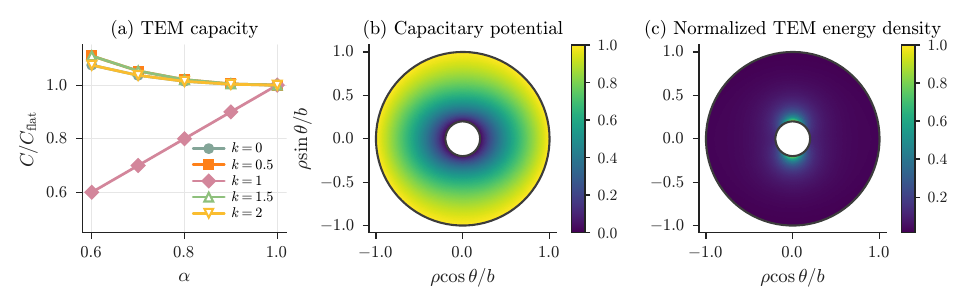}
 \caption{TEM sector at $a/b=0.2$. (a) Capacity normalized by
 $C_{\rm flat}=2\pi/\log5$ on the $49\times64$ sweep mesh. The $k=0,2$ and $k=\tfrac12,\tfrac32$
 curves coincide in the continuum by the inversion identity.
 (b,c) Capacitary potential and normalized local TEM energy density
 for $k=2$, $\alpha=0.7$ on the $97\times128$ mesh. The spectral cutoff remains zero despite
 the redistribution of the transverse field.}
 \label{fig:vi_tem}
\end{figure*}

The coincident capacity curves for $k$ and $2-k$ in
Fig.~\ref{fig:vi_tem} have an exact explanation. With
$x=\log(\rho/a)$ and $\ell=\log(b/a)$, the Dirichlet energy is
\begin{equation}
 \mathcal E_k[U]=\alpha\int_0^\ell\!\int_0^{2\pi} \left[\beta_1U_x^2+c_{\rm mix}U_xU_\theta +\beta_2U_\theta^2\right]d\theta\,dx.
\label{eq:vi_logenergy}
\end{equation}
Under $k\mapsto2-k$, $\beta_1$ and $\beta_2$ are unchanged and
$c_{\rm mix}$ reverses sign. The admissible transformation
$U(x,\theta)\mapsto1-U(\ell-x,\theta)$ preserves the energy.
Uniqueness of the capacitary minimizer therefore yields
\begin{equation}
 \begin{gathered}
 U_{2-k}(\rho,\theta)=1-U_k(ab/\rho,\theta),\\
 C(2-k,\alpha;a/b)=C(k,\alpha;a/b).
 \end{gathered}
 \label{eq:vi_tem_inversion}
\end{equation}
The capacity also determines the power carried by a traveling TEM mode. Let $\mathcal V$ be the complex voltage amplitude between the conducting walls, so that $E_\perp^\flat=\mathcal V\,\dd_hU$. Using Eq.~\eqref{eq:tem}, the time-averaged energy per unit length and longitudinal power, in Heaviside--Lorentz units, are
\begin{equation*}
 \overline{\mathcal U}=\frac{|\mathcal V|^2}{2}\,C, \qquad \overline P_z=\frac{p_z}{E}\,\overline{\mathcal U}             =\pm\frac{|\mathcal V|^2}{2}\,C.
\end{equation*}
Consequently, the inversion identity equates the integrated TEM energy and transmitted power at fixed voltage amplitude for the pair $k$ and $2-k$. In particular, the flat member $k=0$ and the curved member $k=2$ have the same integrated TEM response at the same $\alpha<1$ and $a/b$, despite their different local geometries. This gives an explicit example in which an integrated transport observable does not uniquely determine the exterior curvature.

For the $k=0,2$ pair at $\alpha=0.7$, the finest uniform-radial-grid
test gives relative capacity discrepancy $4.62\times10^{-7}$
and maximum interpolated profile discrepancy $3.47\times10^{-4}$.
These discrepancies are nonzero because a uniform $\rho$ grid
does not preserve the inversion exactly, and they decrease under
refinement. This identity concerns
the two-dimensional Dirichlet energy. It does not equate the
finite-annulus TE or TM spectra, whose eigenvalue norm contains
the radial area weight. Because the two-dimensional Dirichlet energy
is conformally invariant, $C$ depends only on the conformal type of
the annular domain. For $k\ne1$, its variation with $\alpha$ in
Fig.~\ref{fig:vi_tem}(a) therefore reflects the shape of the
coordinate walls rather than the bulk curvature. The inversion
identity is an instance of this invariance.

\subsection{Strong anisotropy and parameter-dependent resolution}
\label{sec:vi_strong}

By Eqs.~\eqref{eq:pec} and \eqref{eq:maxwellscalars}, the TM
longitudinal field $u=E_z$ satisfies the scalar eigenproblem
$(\Delta_0+\lambda^2)u=0$ with homogeneous Dirichlet data, which is the
minimally coupled scalar problem of Sec.~\ref{sec:implementation}. The
first TM mode and the first Dirichlet scalar mode $\psi_1$ therefore
coincide, with the phase and normalization convention of
Fig.~\ref{fig:vi_fields}. The TM panel of Fig.~\ref{fig:vi_fields}
thus also shows $\psi_1$ at the reference point $(k,\alpha)=(2,0.7)$.
The scalar/TM benchmarks validate the same solver, and the
refinements and TM diagnostics reported above refer to this mode. On the $97\times128$ mesh of that figure, its
relative TM potential-wave defect is $3.45\times10^{-4}$.

To test whether a resolution adequate at the reference point remains adequate
under much stronger anisotropy, we evaluate the same first Dirichlet scalar/TM
mode, the only field considered at this point, at $(k,\alpha)=(3,0.25)$. The
discretization, solver settings, mode selection, and normalization are those
of Fig.~\ref{fig:vi_fields}, and only the mesh is varied, from $49\times64$ to
$769\times1024$.

To compare the angular structure of the two modes, we expand a mode on
the annulus as
\begin{equation}
 \begin{gathered}
 \psi(\rho,\theta)=\sum_n\sigma_n f_n(\rho)g_n(\theta),\\
 \mathrm{NS}=1-\frac{\sigma_1^2}{\sum_n\sigma_n^2},
 \end{gathered}
 \label{eq:vi_ns}
\end{equation}
where $\sigma_1\geq\sigma_2\geq\dots\geq0$, the $f_n$ are orthonormal
in $L^2((a,b);\alpha\rho\,\dd\rho)$, and the $g_n$ are orthonormal in
$L^2((0,2\pi);\dd\theta)$, so that
$\sum_n\sigma_n^2=\int_{\mathcal A}|\psi|^2\dd A_h$. NS vanishes
exactly when $\psi=f(\rho)g(\theta)$. Otherwise it is the fraction of
the norm not captured by the best single such product. We evaluate it
from the nodal values with the trapezoidal rule in $\rho$ and the
uniform rule in $\theta$. NS refers to the polar coordinates, in which
the walls are coordinate circles, and is not a coordinate-invariant
property of the geometry, and on the cone it
vanishes to rounding error.

At the reference point $\mathrm{NS}=5.92\times10^{-3}$ on the
$97\times128$ mesh of Fig.~\ref{fig:vi_fields}. At
$(3,0.25)$, $\mathrm{NS}=6.11\times10^{-2}$ on the finest mesh, about
ten times larger. The first Dirichlet eigenfunction $\psi_1$ is simple
and can be chosen positive (Sec.~\ref{sec:bounds}). The problem is
invariant under $\theta\mapsto\theta+\pi/2$, so
$\psi_1(\rho,\theta+\pi/2)$ is again a positive first eigenfunction and
therefore equals $\psi_1(\rho,\theta)$. The mode thus lies in the block
$m\equiv0\pmod4$ of Sec.~\ref{sec:fourier} and has angular period
$\pi/2$.

\begin{figure*}[tbp]
 \centering
 \includegraphics[width=\linewidth]{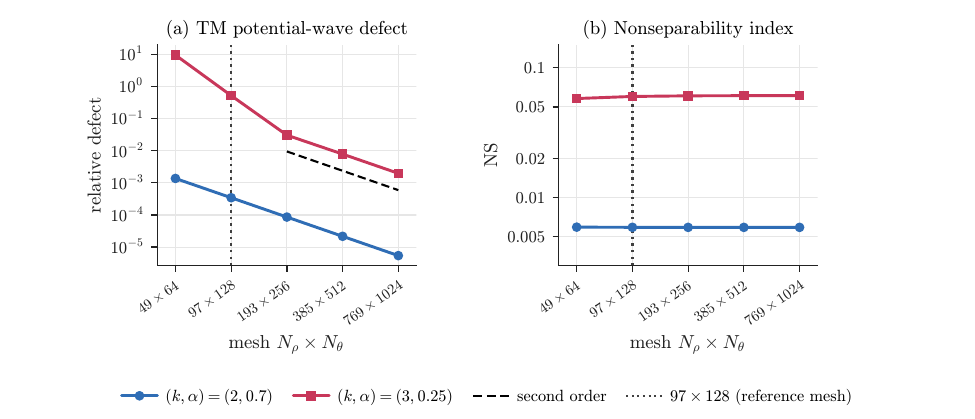}
 \caption{Mesh refinement of the first Dirichlet scalar/TM mode at
 $a/b=0.2$ for $(k,\alpha)=(2,0.7)$ and $(3,0.25)$, on
 $N_\rho\times N_\theta$ meshes from $49\times64$ to $769\times1024$,
 with both directions doubled at each step. (a) Relative TM
 potential-wave defect from the independent Maxwell diagnostics used
 for Fig.~\ref{fig:vi_fields}, with $p_zb=1$. The dashed line, offset
 for visibility, indicates the slope of second-order decrease.
 (b) Nonseparability index of Eq.~\eqref{eq:vi_ns}. The two panels
 show an asymmetry. At $(3,0.25)$, NS changes by less than $2\%$ from
 $97\times128$ onward, whereas the TM defect on the $97\times128$ mesh
 of Fig.~\ref{fig:vi_fields} exceeds the reference value by more than
 three orders of magnitude and decreases at second order only from
 $193\times256$ onward. The strong case therefore requires
 substantially finer resolution, and a stable NS does not by itself
 validate the physical field. All values are finite-mesh results,
 without extrapolation.}
 \label{fig:vi_strong_convergence}
\end{figure*}

The $97\times128$ mesh is not sufficient at this parameter point. On
that mesh the TM potential-wave defect is $0.52$, more than three
orders of magnitude above the reference value
[Fig.~\ref{fig:vi_strong_convergence}(a)].

Under refinement, the cutoff, the field, and this defect converge at
second order, with observed orders between $1.96$ and $1.99$ from
$193\times256$ onward. On $769\times1024$ the defect is $1.98\times10^{-3}$, still larger than at the
reference point, so the two calculations are not equally accurate. The mode
remains simple, with a relative gap above $7\%$ to $\lambda_2^2$ on
every mesh, and the change of the normalized field between successive
meshes decreases under refinement. By
contrast, NS is already within $2\%$ of its finest value on
$97\times128$ [Fig.~\ref{fig:vi_strong_convergence}(b)]. A stable NS
therefore does not by itself show that the physical field is resolved.

\begin{figure}[tbp]
 \centering
 \includegraphics[width=0.8\linewidth]{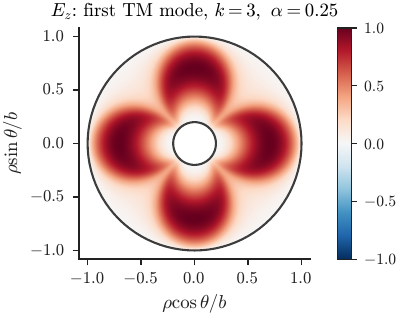}
 \caption{First Dirichlet scalar mode $\psi_1$ at $k=3$,
 $\alpha=0.25$, $a/b=0.2$, on the $769\times1024$ mesh. By
 Eq.~\eqref{eq:pec} it is also the first TM longitudinal field $E_z$,
 so the figure shows the same scalar/TM mode calculation as the TM
 panel of Fig.~\ref{fig:vi_fields}, evaluated at a stronger
 anisotropy. The real field has a fixed global phase and unit maximum
 modulus, with the same coordinates, color scale, and shading as that
 panel. The mode lies in a single Fourier block of
 Sec.~\ref{sec:fourier} (see text). Its nonseparability index
 [Eq.~\eqref{eq:vi_ns}] is $6.11\times10^{-2}$, about ten times the
 reference value $5.92\times10^{-3}$. The $97\times128$ mesh used for
 Fig.~\ref{fig:vi_fields} is insufficient at this parameter point
 (Fig.~\ref{fig:vi_strong_convergence}). On $769\times1024$ the TM
 potential-wave defect is $1.98\times10^{-3}$. The plot is a
 coordinate map, not an isometric embedding, and it does not isolate
 a curvature effect.}
 \label{fig:vi_strong_field}
\end{figure}

Figure~\ref{fig:vi_strong_field} shows the field on the finest
($769\times1024$) mesh.

At $(3,0.25)$ the lowest scalar/TM mode is thus markedly less
separable than at $(2,0.7)$, and a mesh that is adequate for the
reference case is inadequate for this one, so numerical resolution must be assessed
separately at each parameter point. Nonseparability does not measure
curvature.
For $k=0$, the transverse curvature~\eqref{eq:curvature} vanishes
identically, since it is proportional to $k(k-1)$. Yet at
$\alpha=0.7$ the confined fundamental mode remains nonseparable in
$(\rho,\theta)$, with $\mathrm{NS}=5.35\times10^{-4}$ on the
$97\times128$ mesh, a value that is stable under refinement.
Coordinate nonseparability is therefore not, by itself, evidence of
nonzero transverse curvature.
Field maps of the lowest scalar/TM, TE, and MIT modes at further
parameter points, including $k=0$ at $\alpha=0.5$, are collected in
Appendix~\ref{app:visual} as visual robustness checks.

Between $(2,0.7)$ and $(3,0.25)$, the curvature, the coefficients
$\beta_1$, $\beta_2$, and $\cM$, and the shape of the walls in
isothermal coordinates all change. The comparison therefore does not
separate curvature effects from wall-shape effects. Nor does the $k=0$
result show that the wall shape is the only source of
nonseparability for $k\ne0$.

The reported computations establish the finite-annulus low-mode
results for the specified domains and parameters. They do not take
a core limit, remove the confining walls, or define a spin-two
gravitational perturbation problem.

\section{Discussion and outlook}
\label{sec:discussion}

We started from the anisotropic family of transverse geometries and
asked how its local geometric structure and a global finite-annulus
boundary-value problem each enter the behavior of scalar, Dirac, and
electromagnetic test fields.

To answer this question, we first fixed the geometry, the behavior of
field components under frame rotations, and the angular transition
conditions, including the spin structure of the Dirac field. We then
derived the scalar, Dirac, and Maxwell equations directly. They share
a spin-weighted connection Laplacian, but the curvature term of each
field follows from its own equation: the scalar equation, the
Lichnerowicz identity, and the Lorenz-gauge Maxwell equation. Each field also retains its own
constraints. For the Maxwell field, for instance, the potential formulation makes the gauge constraint explicit, while the
physical modes are determined by the field strength. We analyzed the
local indicial problem in isothermal coordinates and posed the
finite-annulus problems with Dirichlet and Neumann data for the
scalar, the MIT bag condition for the Dirac field, and conducting
walls for the Maxwell field. For the scalar spectrum we derived
variational bounds and, from the Fourier blocks, a weak-anisotropy
selection rule. Finally, we computed the spectra, the weak-anisotropy
splittings, the reconstructed physical fields, and the TEM capacity
with finite elements, and validated these calculations against
analytic benchmarks and under independent refinement.

On the local side, the exact indicial factorization fixes the
monodromy eigenvalues for both signs of the factorizing curvature
coupling. The resulting indicial exponents depend on the spin
structure, but not on the anisotropy parameter. The holonomy analysis
shows a related limitation of local data. For the nonconical members
with $\alpha<1$ and $k\notin\{0,1\}$, whose
curvature~\eqref{eq:curvature} does not vanish identically, loops of
different shape can give different holonomies. Results computed on
coordinate circles therefore characterize those loops, not the
defect. When the frames are treated consistently, a rotating
coframe does not create a winding-dependent flat-space spectrum.

The finite-annulus spectra, by contrast, depend on the anisotropy and
on the boundary problem. In the scalar and electromagnetic cases they
are constrained by the variational bounds. The Fourier blocks and the
selection rule determine which mode couplings can produce first-order
splittings, and the computations at $k=2$ confirm this structure. The
allowed $m=\pm1$ splitting is linear in $\eta$, whereas the $m=\pm2$
splitting, which has no first-order coupling, starts at second order,
with fitted powers $1.000$ and $1.992$. The MIT bag spectrum, obtained
from recovered first-order energies, changes when the spin structure
is changed. The annulus also supports a TEM sector with zero cutoff,
which exists because the annular domain is doubly connected. Its
capacity is invariant under $k\mapsto2-k$, an exact consequence of a
radial inversion of the Dirichlet energy. The computed capacities
reproduce this identity, with discrepancies that decrease under
refinement. The finite-element results agree with the analytic
benchmarks and converge under the refinements reported in
Sec.~\ref{sec:numerics}.

Taken together, these results distinguish what the local analysis
determines from what only the confined problem determines. The
indicial exponents and the holonomy structure are fixed by the
geometry, the field equations, and the angular transition conditions,
without reference to any wall, but they do not fix the spectrum
(Sec.~\ref{sec:separable}). The global transition and the boundary
domain are therefore equally necessary parts of each field problem,
and auxiliary problems cannot replace physical ones. For example,
independent Dirichlet conditions on squared spinor components do not
represent a reflecting Dirac boundary. The same dependence on the
boundary problem means that the $\alpha$-dependence of the
finite-annulus cutoffs and of the TEM capacity cannot be attributed to
curvature alone. Within the family, only the members with $k=1$ and
$\alpha<1$ are conical, with deficit angle $2\pi(1-\alpha)$, and the
member $k=0$ is the Euclidean plane in stretched coordinates. The
$k=0$ cutoffs nevertheless change with $\alpha$, because the stretched
annulus has a different boundary geometry. The capacity depends only
on the conformal type of the annular domain, so for $k\ne1$ its
variation with $\alpha$ reflects the shape of the coordinate walls
rather than the bulk curvature. The same caution applies to the
angular structure of the confined modes. The lowest scalar/TM mode is
about ten times less separable in $(\rho,\theta)$ at
$(k,\alpha)=(3,0.25)$ than at $(2,0.7)$, while for $k=0$, where $K$
vanishes identically, it is also nonseparable. The strongly
anisotropic field was examined on meshes up to $769\times1024$, and
even there its TM potential-wave defect remains above that of the
reference point, so the adequacy of a mesh must be established
separately at each parameter point.

These conclusions concern the finite annulus with the stated walls
and boundary conditions, for a metric treated as a prescribed
geometry. They do not take a core limit, remove the confining walls,
or define a spin-two gravitational perturbation problem. Two questions lie
immediately beyond this scope. The first is the classification of
resonant indicial solutions and admissible core domains, including
their dependence on a specified core regularization. The second is a
gravitational perturbation problem, which requires a consistent source
model for the nonvacuum geometry before a physical spin-two master
equation can be claimed.

\begin{acknowledgments}
GGC is grateful to Prof. Fernando Moraes for introducing him to this line of research and for insightful discussions and encouragement throughout the years. The authors also thank Prof. Azadeh Mohammadi (UFPE) for helpful discussions during the early stages of this project. During the project, GGC was partially funded by the Flemish Government under the “Onderzoeksprogramma Artifici\"ele Intelligentie (AI) Vlaanderen” program  (Flanders AI Research). EB is partially supported by \textit{Conselho Nacional de Desenvolvimento Científico e Tecnológico} (grant N.\ 306305/2026-7) and FAPEMIG (APQ-05207-23).
\end{acknowledgments}

\appendix
\section{Explicit Fourier radial system}
\label{app:radial}

Using Eq.~\eqref{eq:fourierparameters}, one has
\begin{equation}
 \begin{gathered}
 \beta_1=b_0+d_0\cos N\theta,\qquad
 \beta_2=b_0-d_0\cos N\theta,\\
 \cM=2d_0\sin N\theta.
 \end{gathered}
\end{equation}
Multiply Eq.~\eqref{eq:master} by $\rho^2$ and substitute
Eq.~\eqref{eq:fourierexpansion}. The equation for output harmonic
$m$ is
\begin{equation}
 \mathcal D_m R_m+
 \mathcal C^+_{m-N}R_{m-N}+
 \mathcal C^-_{m+N}R_{m+N}=0,
 \label{eq:radialsystem}
\end{equation}
where the diagonal radial operator is
\begin{equation}
 \mathcal D_m=b_0\left[\rho^2\partial_\rho^2 +\rho\partial_\rho-m^2-s^2k^2\right]+\frac{2skm}{\alpha}+\lambda^2\rho^2,
\label{eq:radialdiagonal}
\end{equation}
and the operator mapping an input harmonic $m$ to $m+\varepsilon N$
is
\begin{align}
 \mathcal C_m^{\varepsilon}=\frac{d_0}{2}\big[
  &\rho^2\partial_\rho^2
    +(2k-1+2\varepsilon m)\rho\partial_\rho
       +m^2-s^2k^2\nonumber\\
  &-2c_s k(k-1)+\varepsilon mN\big],
 \qquad\varepsilon=\pm1.
 \label{eq:radialoffdiagonal}
\end{align}
These expressions use the positive-exponential convention
$\e^{+\ii m\theta}$. They apply only when $N\ne0$. For $k=1$
the conical Bessel equation applies directly. At $\alpha=1$, $d_0=0$
and $\mathcal D_m=\rho^2\partial_\rho^2+\rho\partial_\rho
-(m-sk)^2+\lambda^2\rho^2$, as required.

\section{Gauge-invariant Maxwell reconstruction}
\label{app:maxwell}

Let $\gamma=h+\dd z^2$ be the positive spatial metric, with
orientation $e^1\wedge e^2\wedge\dd z$. In terms of the full
spacetime-dependent potentials,
\begin{align}
 A_\mu\dd x^\mu
    &=A^t\dd t-\bm A^{\flat_\gamma},\nonumber\\
 \bm E^{\flat_\gamma}
    &=-\partial_t\bm A^{\flat_\gamma}-\dd_3 A^t,\qquad
 \bm B^{\flat_\gamma}=*_\gamma\dd_3\bm A^{\flat_\gamma}.
 \label{eq:potentialfields}
\end{align}
Here $\dd_3$ and $*_\gamma$ are the spatial exterior derivative
and Hodge star. Factoring out $\e^{-\ii Et+\ii p_z z}$ gives the
longitudinal amplitudes in Eq.~\eqref{eq:maxwelllongitudinal}.
The source-free Maxwell equations for the mode amplitudes are
\begin{equation}
 \begin{gathered}
 \operatorname{curl}\bm E=\ii E\bm B,\\
 \operatorname{curl}\bm B=-\ii E\bm E,\\
 \operatorname{div}\bm E=\operatorname{div}\bm B=0,
 \end{gathered}
 \label{eq:maxwellcurl}
\end{equation}
where longitudinal derivatives act as multiplication by $\ii p_z$.

The longitudinal components $u=E_z$ and $v=B_z$ obey
\begin{equation}
 \begin{gathered}
 (\Delta_0+\lambda^2)u=0,\qquad
 (\Delta_0+\lambda^2)v=0,\\
 \lambda^2=E^2-p_z^2.
 \end{gathered}
 \label{eq:maxwellscalars}
\end{equation}
For $\lambda^2>0$, solving the transverse curl equations yields
\begin{align}
 E_\perp^\flat&=\frac{\ii}{\lambda^2}
                 (p_z\,\dd_h u-E\,*_h\dd_h v),\nonumber\\
 B_\perp^\flat&=\frac{\ii}{\lambda^2}
                 (p_z\,\dd_h v+E\,*_h\dd_h u).
 \label{eq:maxwellreconstruction}
\end{align}
The transverse index is lowered with $h$. The TE sector has
$u=0$ and the TM sector has $v=0$, the geometric counterpart of
the usual waveguide reduction~\cite{MarksWilliams1992}. For example,
\begin{equation}
 \begin{gathered}
 \operatorname{div}_h E_\perp
    =\frac{\ii p_z}{\lambda^2}\Delta_0u=-\ii p_z u,\\
 *_h\dd_h E_\perp^\flat
    =-\frac{\ii E}{\lambda^2}\Delta_0v=\ii E v,
 \end{gathered}
\end{equation}
which verify the electric divergence equation and the
longitudinal Faraday equation.

For $E\ne0$, an explicit Lorenz- and temporal-gauge potential is
\begin{equation}
 \begin{gathered}
 \varphi=0,\qquad
 a_z=\frac{u}{\ii E},\\
 a=\frac{p_z\,\dd_h u-E\,*_h\dd_h v}{E\lambda^2}.
 \end{gathered}
 \label{eq:potentialreconstruction}
\end{equation}
Equations~\eqref{eq:maxwellscalars} imply both
Eq.~\eqref{eq:potentialconstraint} and the potential equations
\eqref{eq:potentialreduced}. Indeed,
$\Delta^{\rm H}_1\dd_h=\dd_h(-\Delta_0)$ on scalars, and $\Delta^{\rm H}_1$
commutes with $*_h$ on one-forms. The scalar conductor
conditions~\eqref{eq:pec} also imply the tangential potential
conditions~\eqref{eq:potentialpec}.
Conversely, let $(\varphi,a,a_z)$ solve Eqs.~\eqref{eq:potentialreduced}
and~\eqref{eq:potentialconstraint} with $\lambda^2>0$. Then
$u=\ii Ea_z-\ii p_z\varphi$ obeys $(\Delta_0+\lambda^2)u=0$, and
$v=*_h\dd_ha$ obeys it because
$*_h\dd_h\Delta^{\rm H}_1=-\Delta_0\,*_h\dd_h$ on one-forms. The fields
of the potential satisfy Eqs.~\eqref{eq:maxwellcurl}, and the transverse
curl equations form a linear system for $(E_\perp,B_\perp)$ with
determinant $(E^2-p_z^2)^2=\lambda^4\ne0$, so these fields are given by
Eq.~\eqref{eq:maxwellreconstruction}. In the gauge $\varphi=0$, the
conditions~\eqref{eq:potentialpec} give $u=\ii Ea_z=0$ and
$E_\perp(\tau)=\ii E\,a(\tau)=0$ on the walls. There $\partial_\tau u=0$,
and Eq.~\eqref{eq:maxwellreconstruction} gives
$E_\perp(\tau)=-\ii E\lambda^{-2}\partial_nv$ for the unit tangent with
$*_hn^\flat=\tau^\flat$. Hence $\partial_nv=0$, which is
Eq.~\eqref{eq:pec}.
This establishes the equivalence of the constrained potential
formulation and the TE/TM reconstruction for positive $\lambda^2$.
The $\lambda=0$ TEM sector is instead described by
Eq.~\eqref{eq:potentialtem}.

\section{Tetrad form of the transverse Dirac operator}
\label{app:tetrad}

This appendix rederives the operator $\Dslash_h$ of
Eq.~\eqref{eq:dirac2d} and its square, Eq.~\eqref{eq:diracsquared},
directly from an orthonormal tetrad, along the lines of Refs.~\cite{Soares1980FormasDiferenciais, Soares:1981kb}. It is a consistency check of
Sec.~\ref{sec:dirac} in which the curvature coupling is obtained by
explicit squaring rather than from the Lichnerowicz identity.

\emph{Frame and conventions.} We use the tetrad $e^{(0)}=\dd t$,
$e^{(3)}=\dd z$, and $e^{(1)}=e^1$, $e^{(2)}=e^2$ from
Eq.~\eqref{eq:coframe}, so that
$\dd s^2=\eta_{(A)(B)}e^{(A)}e^{(B)}$ with
$\eta_{(A)(B)}=\operatorname{diag}(1,-1,-1,-1)$. The dual frame is
$e_{(0)}=\partial_t$, $e_{(3)}=\partial_z$, and $e_{(1)}=e_1$,
$e_{(2)}=e_2$ from Eq.~\eqref{eq:inverseframe}. Transverse frame indices
are denoted $a,b\in\{1,2\}$, and repeated transverse indices are summed.
The connection one-forms and their components are defined by
\begin{equation}
 \begin{gathered}
 \dd e^{(A)}+\omega^{(A)}{}_{(B)}\wedge e^{(B)}=0,\\
 \omega_{(A)(B)}=\eta_{(A)(C)}\,\omega^{(C)}{}_{(B)},\\
 \omega_{(A)(B)(C)}=\omega_{(A)(B)}\bigl(e_{(C)}\bigr).
 \end{gathered}
 \label{eq:tetradconventions}
\end{equation}
The Dirac matrices satisfy
$\{\gamma^{(A)},\gamma^{(B)}\}=2\eta^{(A)(B)}\Id$. We use the Dirac
representation, $\gamma^{(0)}=\operatorname{diag}(\Id,-\Id)$ and
$\gamma^{(j)}=\bigl(\begin{smallmatrix}0&\sigma^j\\-\sigma^j&0
\end{smallmatrix}\bigr)$ for $j=1,2,3$. The spinor covariant derivative
and the Fock--Ivanenko connection are
\begin{equation}
 \begin{gathered}
 \nabla_{(C)}=e_{(C)}-\Gamma_{(C)},\\
 \Gamma_{(C)}=-\frac14\,\omega_{(A)(B)(C)}\,\gamma^{(A)}\gamma^{(B)},
 \end{gathered}
 \label{eq:fockivanenko}
\end{equation}
that is, $\nabla_{(C)}=e_{(C)}+\tfrac14\,\omega_{(A)(B)(C)}
\gamma^{(A)}\gamma^{(B)}$.

\emph{Connection.} The forms $e^{(0)}$ and $e^{(3)}$ are closed, and the
transverse coframe depends only on $(\rho,\theta)$. The only independent
connection component is therefore $\omega^{(1)}{}_{(2)}$. Differentiating
Eq.~\eqref{eq:coframe} gives
\begin{equation}
 \begin{gathered}
 \dd e^{(1)}=\frac{k\sin\xi}{\alpha\rho}\,e^{(1)}\wedge e^{(2)},\\
 \dd e^{(2)}=\frac{k\cos\xi}{\rho}\,e^{(1)}\wedge e^{(2)},
 \end{gathered}
 \label{eq:tetradde}
\end{equation}
and the first relation in Eq.~\eqref{eq:tetradconventions} is solved by
\begin{equation}
 \begin{gathered}
 \omega^{(1)}{}_{(2)}=-\frac{k\sin\xi}{\alpha\rho}\,e^{(1)}
                     -\frac{k\cos\xi}{\rho}\,e^{(2)},\\
 \omega_{(1)(2)}=-\omega^{(1)}{}_{(2)}=\omega .
 \end{gathered}
 \label{eq:tetradomega}
\end{equation}
Here $\omega$ is the connection one-form of Eq.~\eqref{eq:connection}.
The minus sign in the second relation comes from lowering with
$\eta_{(1)(1)}=-1$. We write $\omega_{(a)}\equiv\omega(e_{(a)})$, so that
\begin{equation}
 \omega_{(1)}=\frac{k\sin\xi}{\alpha\rho},\qquad
 \omega_{(2)}=\frac{k\cos\xi}{\rho}.
\end{equation}
Since $e^{(A)}(e_{(B)})$ is constant,
$\dd e^{(A)}(e_{(1)},e_{(2)})=-e^{(A)}([e_{(1)},e_{(2)}])$, and
Eq.~\eqref{eq:tetradde} yields
\begin{equation}
 [e_{(1)},e_{(2)}]=-\omega_{(1)}e_{(1)}-\omega_{(2)}e_{(2)} .
 \label{eq:tetradcommutator}
\end{equation}

\emph{First-order operator.} By Eq.~\eqref{eq:fockivanenko},
$\Gamma_{(0)}=\Gamma_{(3)}=0$ and
$\Gamma_{(a)}=-\tfrac12\,\omega_{(a)}\gamma^{(1)}\gamma^{(2)}$. Using
$\gamma^{(1)}\gamma^{(1)}=\gamma^{(2)}\gamma^{(2)}=-\Id$ gives
\begin{equation}
 \gamma^{(C)}\Gamma_{(C)}
 =\tfrac12\bigl(\omega_{(1)}\gamma^{(2)}-\omega_{(2)}\gamma^{(1)}\bigr).
\end{equation}
Equation~\eqref{eq:diracfirst}, written in the frame as
$(\ii\gamma^{(C)}\nabla_{(C)}-M)\Psi=0$, then becomes, for a mode
$\Psi=\e^{-\ii Et+\ii p_zz}\psi(\rho,\theta)$,
\begin{equation}
 \begin{split}
 &\bigl(E\gamma^{(0)}-p_z\gamma^{(3)}-M\bigr)\psi\\
 &+\ii\Bigl[\gamma^{(1)}\bigl(e_{(1)}+\tfrac12\omega_{(2)}\bigr)+\gamma^{(2)}\bigl(e_{(2)}-\tfrac12\omega_{(1)}\bigr)\Bigr]\psi=0 .
 \end{split}
\end{equation}
In the Dirac representation, with
$\psi=(\psi_{\mathrm u},\psi_{\mathrm d})$, the two rows of this
equation read
\begin{equation}
 \begin{gathered}
 (E-M)\psi_{\mathrm u}=T\psi_{\mathrm d},\qquad
 (E+M)\psi_{\mathrm d}=T\psi_{\mathrm u},\\
 T=\Dslash_h+p_z\sigma^3,
 \end{gathered}
 \label{eq:tetradsystem}
\end{equation}
where
\begin{equation}
 \begin{split}
 \Dslash_h
 &=-\ii\Bigl[\sigma^1\bigl(e_1+\tfrac12\omega_{(2)}\bigr)
           +\sigma^2\bigl(e_2-\tfrac12\omega_{(1)}\bigr)\Bigr]\\
 &=-\ii\sigma^a D_a,\\
 D_a&=e_a{}^iD_i^{(\sigma^3/2)}=e_a-\frac{\ii}{2}\,\omega_{(a)}\sigma^3 .
 \end{split}
 \label{eq:tetraddslash}
\end{equation}
Here $D_i^{(s)}$ is the derivative of Eq.~\eqref{eq:covderivative}, with
$s$ replaced by $\sigma^3/2$.
The second form follows from $\sigma^1\sigma^3=-\ii\sigma^2$ and
$\sigma^2\sigma^3=\ii\sigma^1$. Since $\tfrac12\omega_{(2)}=k\cos\xi/(2\rho)$
and $\tfrac12\omega_{(1)}=k\sin\xi/(2\alpha\rho)$,
Eq.~\eqref{eq:tetraddslash} is Eq.~\eqref{eq:dirac2d}, and
Eq.~\eqref{eq:tetradsystem} is the eigenvalue problem of
Eq.~\eqref{eq:dirachamiltonian}.

\emph{Squaring.} Eliminating one two-spinor from
Eq.~\eqref{eq:tetradsystem} gives
$(T^2-E^2+M^2)\psi_{\mathrm{u,d}}=0$, and
$T^2=\Dslash_h^2+p_z^2$ because $\{\Dslash_h,\sigma^3\}=0$. To compute
$\Dslash_h^2=-\sigma^aD_a\sigma^bD_b$ we use
\begin{equation}
 \begin{gathered}
 D_a\sigma^b=\sigma^bD_a+\omega_{(a)}\epsilon^{bc}\sigma^c,\\
 \sigma^a\sigma^b=\delta^{ab}\Id+\ii\epsilon^{ab}\sigma^3,\\
 [D_1,D_2]=-\omega_{(a)}D_a+\frac{\ii}{2}K\sigma^3,
 \end{gathered}
 \label{eq:tetradidentities}
\end{equation}
with $\epsilon^{12}=-\epsilon^{21}=1$ and $\epsilon^{11}=\epsilon^{22}=0$.
The first identity follows from
$[\sigma^3,\sigma^b]=2\ii\epsilon^{bc}\sigma^c$. The third follows from
Eq.~\eqref{eq:tetradcommutator} and from $\dd\omega=-K\,\dd A_h$
[Eq.~\eqref{eq:connection}] with $\dd A_h=e^{(1)}\wedge e^{(2)}$, which
gives $e_1(\omega_{(2)})-e_2(\omega_{(1)})=-K-\omega_{(a)}\omega_{(a)}$.
Moving $D_a$ to the right with the first identity and then
applying the other two gives
\begin{align}
 \Dslash_h^2
 &=-D_aD_a-\ii\sigma^3[D_1,D_2]
   +\epsilon^{ab}\omega_{(a)}D_b-\ii\sigma^3\omega_{(a)}D_a
 \nonumber\\
 &=-\bigl(D_aD_a-\epsilon^{ab}\omega_{(a)}D_b\bigr)
   +\frac{K}{2}.
 \label{eq:tetradsquare}
\end{align}
The Levi-Civita connection acts on the frame as
$\nabla_Xe_{(B)}=e_{(A)}\,\omega^{(A)}{}_{(B)}(X)$, that is,
$\nabla_Xe_1=\omega(X)e_2$ and $\nabla_Xe_2=-\omega(X)e_1$. Hence
$\nabla_{e_a}e_a=\epsilon^{ab}\omega_{(a)}e_b$. On a component with
$\sigma^3$ eigenvalue $\sigma=\pm1$, the bracket in
Eq.~\eqref{eq:tetradsquare} is therefore the connection Laplacian
$\Delta_{\sigma/2}$ of Eq.~\eqref{eq:masterdef} written in the frame. It
follows that $\Dslash_h^2=-\Delta_{\sigma/2}+K/2$, which is
Eq.~\eqref{eq:lichnerowicz}, and the squared equation becomes
\begin{equation}
 \begin{gathered}
 \bigl(L_{s,c}+\lambda^2\bigr)\psi_\sigma=0,\\
 s=\frac{\sigma}{2},\qquad c=-\frac12,\qquad
 \lambda^2=E^2-p_z^2-M^2 .
 \end{gathered}
\end{equation}
This is Eq.~\eqref{eq:diracsquared}. Writing $e_a$ in coordinates with
Eq.~\eqref{eq:inverseframe} gives Eq.~\eqref{eq:masterexpanded} with these
values of $s$ and $c$. Its zeroth-order term is Eq.~\eqref{eq:diracpotential}.
The tetrad computation thus reproduces the Dirac operator of the main text
and its curvature coupling $c_{\pm1/2}=-\tfrac12$ by direct squaring.

\section{Visual robustness across additional parameter points}
\label{app:visual}

\begin{figure*}[!htbp]
 \centering
 \includegraphics[width=\linewidth]{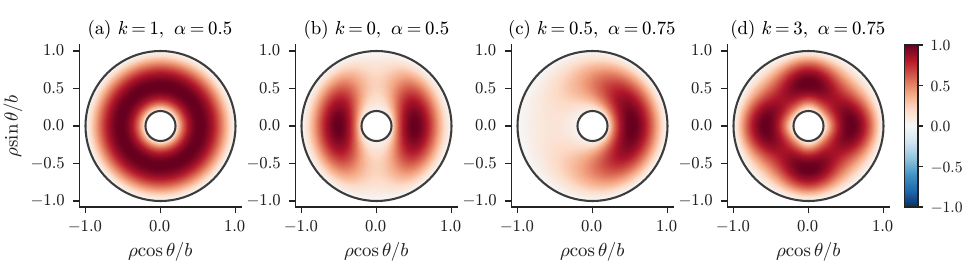}
 \caption{First Dirichlet scalar mode $\psi_1$, equivalently the first
 TM longitudinal field $E_z$, at additional parameter points:
 (a) $(k,\alpha)=(1,0.5)$, (b) $(0,0.5)$, (c) $(0.5,0.75)$, and
 (d) $(3,0.75)$, with $a/b=0.2$ on the $193\times256$ mesh. The
 representation is that of the TM panel of Fig.~\ref{fig:vi_fields}:
 fixed global phase, real part, and unit maximum modulus. The panels are
 visual robustness checks. In (b), $k=0$ and the transverse curvature
 vanishes identically, yet the confined fundamental mode is
 nonseparable in $(\rho,\theta)$, with nonzero NS
 [Eq.~\eqref{eq:vi_ns}]. This control does not show that the wall shape
 is the only source of nonseparability for $k\ne0$. The plots are
 coordinate maps, not isometric embeddings.}
 \label{fig:app_scalar_tm}
\end{figure*}

\begin{figure*}[!htbp]
 \centering
 \includegraphics[width=\linewidth]{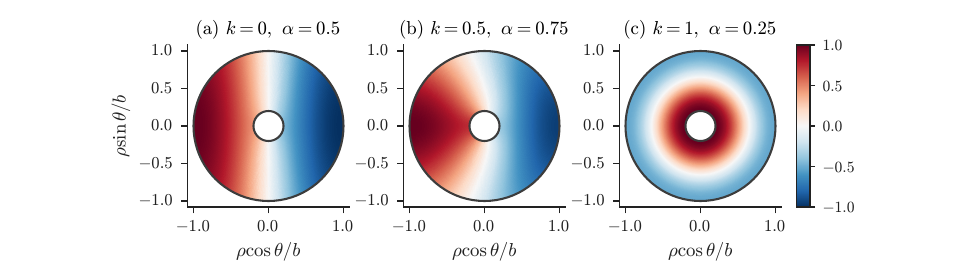}
 \caption{First TE longitudinal field $B_z$ at additional parameter
 points: (a) $(k,\alpha)=(0,0.5)$, (b) $(0.5,0.75)$, and
 (c) $(1,0.25)$, with $a/b=0.2$ on the $193\times256$ mesh, in the
 representation of the TE panel of Fig.~\ref{fig:vi_fields}. The panels
 are visual robustness checks. In (a) and (b) the first nonconstant
 Neumann mode is dominated by $|m|=1$. In (c) it is instead the second
 radial $m=0$ TE mode, with one radial node, and therefore belongs to a
 different modal branch from (a) and (b). The plots are coordinate maps,
 not isometric embeddings.}
 \label{fig:app_te}
\end{figure*}

\begin{figure*}[!htbp]
 \centering
 \includegraphics[width=\linewidth]{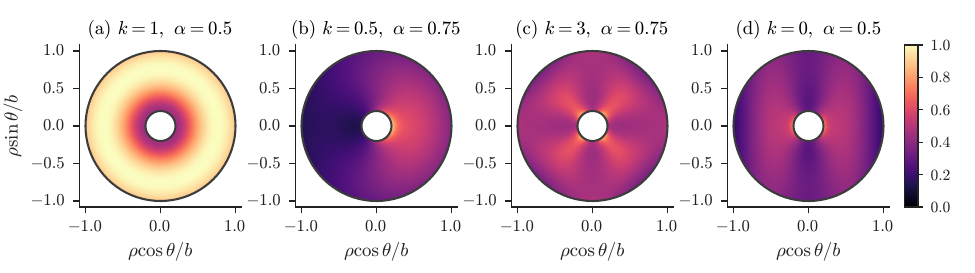}
 \caption{Full four-spinor density $\Psi^\dagger\Psi$ of the lowest
 positive MIT level, scaled to unit maximum, at additional parameter
 points: (a) $(k,\alpha)=(1,0.5)$, (b) $(0.5,0.75)$, (c) $(3,0.75)$,
 and (d) $(0,0.5)$, with $a/b=0.2$, $M=p_z=0$, and $\delta=0$ on the
 $193\times256$ mesh, in the representation of the MIT panel of
 Fig.~\ref{fig:vi_fields}. The panels are visual robustness checks.
 Panel (d) is the flat control $k=0$, whose smooth-reconstruction residual
 $\epsilon_D$ is about $1.28$ times the reference value at $(2,0.7)$. The plots are
 coordinate maps, not isometric embeddings.}
 \label{fig:app_mit}
\end{figure*}

The field maps of Figs.~\ref{fig:vi_fields} and~\ref{fig:vi_strong_field}
are shown at $(k,\alpha)=(2,0.7)$ and, for the scalar/TM mode only, at
$(3,0.25)$. Figures~\ref{fig:app_scalar_tm}--\ref{fig:app_mit} show the
same fields at additional points, to indicate how the lowest modes of the
three physical problems change across the family. The points include the
cone ($k=1$), the flat control $k=0$, for which $K$ vanishes identically,
a half-odd-integer winding, and $k=3$ at moderate anisotropy. These maps serve as visual robustness checks
that complement, but do not replace, the quantitative refinement and residual
evidence of Sec.~\ref{sec:numerics}, and their smoothness is not used as evidence of convergence.

All fields are computed with the discretization, solver settings, mode
selection, and normalization of Fig.~\ref{fig:vi_fields}, on the
$193\times256$ mesh. For each displayed field, the potential-wave and
Amp\`ere defects of the scalar/TM and TE fields and the MIT residual
$\epsilon_D$ decrease under refinement from $97\times128$ to
$193\times256$. On $193\times256$ they do not exceed the values for the
fields of Fig.~\ref{fig:vi_fields} on its $97\times128$ mesh, where the
MIT comparison uses the smooth-reconstruction residual $\epsilon_D$ of
Sec.~\ref{sec:numerics}. The exception is the MIT field at $(0,0.5)$,
whose smooth-reconstruction $\epsilon_D$ is about $1.28$ times the
reference value. Apart from this labelled exception, parameter points
whose diagnostics did not meet this standard, and TE modes in exactly degenerate eigenspaces, whose plotted
orientation is arbitrary, are not shown. The lowest MIT level is two-fold
degenerate at every displayed point, and the plotted density does not
depend on the choice within it.

\FloatBarrier
\bibliography{references}

@article{AharonovBohm,
author  = {Aharonov, Yakir and Bohm, David},
title   = {Significance of Electromagnetic Potentials in the Quantum Theory},
journal = {Physical Review},
volume  = {115},
pages   = {485--491},
year    = {1959},
doi     = {10.1103/PhysRev.115.485}
}

@article{BaerBallmann,
  author        = {B\"ar, Christian and Ballmann, Werner},
  title         = {Boundary value problems for elliptic differential
                   operators of first order},
  journal       = {Surv. Differ. Geom.},
  volume        = {17},
  pages         = {1--78},
  publisher     = {International Press},
  year          = {2012},
  doi           = {10.4310/SDG.2012.v17.n1.a1},
  eprint        = {1101.1196},
  archivePrefix = {arXiv}
}

@article{Banerjee1992,
  author  = {Banerjee, Uday},
  title   = {A note on the effect of numerical quadrature in finite element eigenvalue approximation},
  journal = {Numerische Mathematik},
  volume  = {61},
  pages   = {145--152},
  year    = {1992},
  doi     = {10.1007/BF01385502}
}

@article{BezerradeMello2008,
  author        = {Bezerra de Mello, E. R. and Bezerra, V. B. and Saharian, A. A. and Tarloyan, A. S.},
  title         = {Fermionic vacuum polarization by a cylindrical boundary in the cosmic string spacetime},
  journal       = {Physical Review D},
  volume        = {78},
  number        = {10},
  pages         = {105007},
  year          = {2008},
  doi           = {10.1103/PhysRevD.78.105007},
  eprint        = {0809.0844},
  archivePrefix = {arXiv}
}

@article{Carvalho2007,
  author        = {de M. Carvalho, A. M. and S{\'a}tiro, C. and Moraes, F.},
  title         = {{Aharonov-Bohm}-like effect for light propagating in nematics with disclinations},
  journal       = {EPL (Europhysics Letters)},
  volume        = {80},
  number        = {4},
  pages         = {46002},
  year          = {2007},
  doi           = {10.1209/0295-5075/80/46002},
  eprint        = {0709.3142},
  archivePrefix = {arXiv}
}

@article{Chodos1974,
  author  = {Chodos, A. and Jaffe, R. L. and Johnson, K. and Thorn, C. B. and Weisskopf, V. F.},
  title   = {New extended model of hadrons},
  journal = {Physical Review D},
  volume  = {9},
  number  = {12},
  pages   = {3471--3495},
  year    = {1974},
  doi     = {10.1103/PhysRevD.9.3471}
}

@article{Cortijo2007,
  author        = {Cortijo, Alberto and Vozmediano, Mar{\'\i}a A. H.},
  title         = {Effects of topological defects and local curvature on the electronic properties of planar graphene},
  journal       = {Nuclear Physics B},
  volume        = {763},
  pages         = {293--308},
  year          = {2007},
  url           = {https://www.sciencedirect.com/science/article/pii/S055032130600856X},
  eprint        = {cond-mat/0612374},
  archivePrefix = {arXiv}
}

@misc{DAncona2025,
  author        = {D'Ancona, P. and Yin, Z. and Zhang, J.},
  title         = {Dispersive and {Strichartz} estimates for {Dirac} equation in a cosmic string spacetime},
  year          = {2025},
  eprint        = {2501.10850},
  archivePrefix = {arXiv},
  primaryClass  = {math.AP},
  doi           = {10.48550/arXiv.2501.10850}
}

@article{Fumeron2021,
  author  = {Fumeron, Sébastien and Berche, Bertrand and Moraes, Fernando},
  title   = {Geometric theory of topological defects: methodological developments and new trends},
  journal = {Liquid Crystals Reviews},
  volume  = {9},
  number  = {2},
  pages   = {85--110},
  year    = {2021},
  doi     = {10.1080/21680396.2022.2163515}
}

@article{Geroch1968,
  author  = {Geroch, Robert},
  title   = {Spinor Structure of Space-Times in General Relativity. I},
  journal = {Journal of Mathematical Physics},
  volume  = {9},
  number  = {11},
  pages   = {1739--1744},
  year    = {1968},
  doi     = {10.1063/1.1664507}
}

@article{Grosse2025,
  author        = {Gro{\ss}e, Nadine and Uribe, Alejandro and van den Bosch, Hanne},
  title         = {Local Boundary Conditions for {D}irac-Type Operators},
  journal       = {Annales Henri Poincar{\'e}},
  year          = {2025},
  doi           = {10.1007/s00023-025-01634-3},
  eprint        = {2412.17396},
  archivePrefix = {arXiv}
}

@article{HindmarshKibble,
author        = {Hindmarsh, Mark B. and Kibble, T. W. B.},
title         = {Cosmic Strings},
journal       = {Reports on Progress in Physics},
volume        = {58},
pages         = {477--562},
year          = {1995},
doi           = {10.1088/0034-4885/58/5/001},
eprint        = {hep-ph/9411342},
archivePrefix = {arXiv}
}

@article{Ivrii2016,
  author        = {Ivrii, Victor},
  title         = {100 years of {W}eyl's law},
  journal       = {Bull. Math. Sci.},
  volume        = {6},
  number        = {3},
  pages         = {379--452},
  year          = {2016},
  doi           = {10.1007/s13373-016-0089-y},
  eprint        = {1608.03963},
  archivePrefix = {arXiv}
}

@book{Kato,
  author    = {Kato, T.},
  title     = {Perturbation Theory for Linear Operators},
  edition   = {2},
  series    = {Classics in Mathematics},
  publisher = {Springer},
  address   = {Berlin},
  year      = {1995},
  doi       = {10.1007/978-3-642-66282-9}
}

@book{KozlovMazyaRossmann,
  author    = {Kozlov, V. A. and Maz'ya, V. G. and Rossmann, J.},
  title     = {Spectral Problems Associated with Corner Singularities of
               Solutions to Elliptic Equations},
  series    = {Mathematical Surveys and Monographs},
  volume    = {85},
  publisher = {American Mathematical Society},
  address   = {Providence, RI},
  year      = {2001}
}

@book{LawsonMichelsohn,
  author    = {Lawson, H. B., Jr. and Michelsohn, M.-L.},
  title     = {Spin Geometry},
  publisher = {Princeton University Press},
  address   = {Princeton},
  year      = {1989},
  doi       = {10.1515/9781400883912}
}

@article{MarksWilliams1992,
  author  = {Marks, R. B. and Williams, D. F.},
  title   = {A general waveguide circuit theory},
  journal = {Journal of Research of the National Institute of Standards and Technology},
  volume  = {97},
  pages   = {533--562},
  year    = {1992},
  doi     = {10.6028/jres.097.024}
}

@book{Nakahara2003,
  author    = {Nakahara, Mikio},
  title     = {Geometry, Topology and Physics},
  edition   = {2nd},
  publisher = {Institute of Physics Publishing / CRC Press},
  year      = {2003},
  isbn      = {9780750306065}
}

@article{PhysRevD.96.105012,
  title = {Cosmology in the laboratory: An analogy between hyperbolic metamaterials and the Milne universe},
  author = {Figueiredo, David and Moraes, Fernando and Fumeron, S{\'e}bastien and Berche, Bertrand},
  journal = {Phys. Rev. D},
  volume = {96},
  issue = {10},
  pages = {105012},
  numpages = {14},
  year = {2017},
  month = {Nov},
  publisher = {American Physical Society},
  doi = {10.1103/PhysRevD.96.105012},
  url = {https://link.aps.org/doi/10.1103/PhysRevD.96.105012}
}

@article{PoissonPoundVega2011,
  author  = {Poisson, E. and Pound, A. and Vega, I.},
  title   = {The motion of point particles in curved spacetime},
  journal = {Living Reviews in Relativity},
  volume  = {14},
  pages   = {7},
  year    = {2011},
  doi     = {10.12942/lrr-2011-7}
}

@article{Satiro2009,
  author  = {S{\'a}tiro, C. and de M. Carvalho, A. M. and Moraes, F.},
  title   = {An asymmetric family of cosmic strings},
  journal = {Modern Physics Letters A},
  volume  = {24},
  number  = {18},
  pages   = {1437--1442},
  year    = {2009},
  doi     = {10.1142/S0217732309029995}
}

@incollection{Soares1980FormasDiferenciais,
  author    = {Soares, I. Dami{\~a}o},
  title     = {O C{\'a}lculo de Formas Diferenciais e a Equa{\c c}{\~a}o de {Dirac} em Espa{\c c}os Curvos},
  booktitle = {Anais da II Escola de Cosmologia e Gravita{\c c}{\~a}o},
  editor    = {Novello, M{\'a}rio},
  volume    = {1},
  pages     = {479--546},
  year      = {1980},
  publisher = {Centro Brasileiro de Pesquisas F{\'i}sicas},
  address   = {Rio de Janeiro}
}

@article{Soares:1981kb,
  author  = {Soares, I. D.},
  title   = {Gravitational coupling of neutrinos to matter vorticity: Microscopic asymmetries},
  journal = {Physical Review D},
  volume  = {23},
  number  = {2},
  pages   = {272--286},
  year    = {1981},
  doi     = {10.1103/PhysRevD.23.272}
}

@book{Teschl2009,
  author    = {Teschl, Gerald},
  title     = {Mathematical Methods in Quantum Mechanics: With Applications
               to {Schr\"odinger} Operators},
  series    = {Graduate Studies in Mathematics},
  volume    = {99},
  publisher = {American Mathematical Society},
  address   = {Providence, RI},
  year      = {2009},
  isbn      = {978-0-8218-4660-5},
  note      = {Online edition, version of February 12, 2009}
}

@book{TeschlODE,
  author    = {Teschl, Gerald},
  title     = {Ordinary Differential Equations and Dynamical Systems},
  series    = {Graduate Studies in Mathematics},
  volume    = {140},
  publisher = {American Mathematical Society},
  address   = {Providence, RI},
  year      = {2012}
}

@article{Teukolsky1972,
  author  = {Teukolsky, S. A.},
  title   = {Rotating black holes: separable wave equations for gravitational and electromagnetic perturbations},
  journal = {Physical Review Letters},
  volume  = {29},
  pages   = {1114--1118},
  year    = {1972},
  doi     = {10.1103/PhysRevLett.29.1114}
}

@article{Weyl1912,
  author  = {Weyl, Hermann},
  title   = {Das asymptotische {V}erteilungsgesetz der {E}igenwerte linearer
             partieller {D}ifferentialgleichungen (mit einer {A}nwendung auf
             die {T}heorie der {H}ohlraumstrahlung)},
  journal = {Math. Ann.},
  volume  = {71},
  pages   = {441--479},
  year    = {1912}
}

@article{bini2015,
doi = {10.1088/0264-9381/32/21/215010},
url = {https://doi.org/10.1088/0264-9381/32/21/215010},
year = {2015},
volume = {32},
number = {21},
pages = {215010},
author = {Bini, Donato and Bittencourt, Eduardo and Geralico, Andrea},
title = {Massless {Dirac} particles in the vacuum C-metric},
journal = {Classical and Quantum Gravity}
}

@book{brezis2010functional,
  author    = {Brezis, Haim},
  title     = {Functional Analysis, Sobolev Spaces and Partial Differential Equations},
  publisher = {Springer Science \& Business Media},
  year      = {2010},
  edition   = {1st},
  isbn      = {978-0387709130},
  note      = {Completely revised, updated, and expanded English edition of Analyse fonctionnelle (1983)}
}

@book{docarmo2016differential,
  title     = {Differential Geometry of Curves and Surfaces},
  author    = {do Carmo, Manfredo P.},
  edition   = {2},
  year      = {2016},
  publisher = {Dover Publications},
  address   = {Mineola, New York},
  isbn      = {9780486806990}
}

@book{evans1998partial,
  author    = {Evans, Lawrence C.},
  title     = {Partial Differential Equations},
  series    = {Graduate Studies in Mathematics},
  volume    = {19},
  publisher = {American Mathematical Society},
  address   = {Providence, RI},
  year      = {1998}
}

@article{moraes06,
  author  = {S{\'a}tiro, C. and Moraes, F.},
  title   = {Lensing effects in a nematic liquid crystal with topological defects},
  journal = {The European Physical Journal E},
  volume  = {20},
  number  = {2},
  pages   = {173--178},
  year    = {2006},
  doi     = {10.1140/epje/i2005-10127-2}
}

\end{document}